\documentclass[prd,preprint,tightenlines,floatfix,showpacs,preprintnumbers,nofootinbib,eqsecnum]{revtex4-1}

\pdfoutput=1

\usepackage[dvips,final]{graphicx}
\usepackage{color,xcolor}
\usepackage{amssymb}
\usepackage{amsmath}
\usepackage{mathtools}
\usepackage{amsfonts}
\usepackage{epsfig}
\usepackage{bm}
\usepackage{ulem}
\usepackage{comment}
\usepackage{float}
\usepackage[utf8]{inputenc}
\usepackage{comment}

\DeclareMathOperator\arctanh{arctanh}
\def\Pom{\mathrm{I\!P}}

\def\fm{\,\mbox{fm}}
\def\GeV{\,\mbox{GeV}}

\newcommand{\zvtx}{$z_{\mbox{\tiny vtx}}$}

\DeclareSymbolFont{myletters}{OML}{ztmcm}{m}{it}
\DeclareMathSymbol{\uplambda}{\mathord}{myletters}{"15}

\begin{document}

\title{Multi-dimensional hadron structure\\ 
through the lens of gluon Wigner distribution}

\author{Roman Pasechnik$^{1}$}
\email{Roman.Pasechnik@cern.ch}

\author{Marek Ta\v{s}evsk\'{y}$^{2}$}
\email{Marek.Tasevsky@cern.ch}

\affiliation{
\\
{$^1$\sl
Department of Physics, Lund
University, SE-223 62 Lund, Sweden
}
\\
{$^2$\sl 
Institute of Physics of the Czech Academy of Sciences, 
Na Slovance 1999/2, 18221 Prague 8, Czech Republic\vspace{0.5cm}
}
}
\begin{abstract}
\vspace{0.5cm}
In this review, we present the current status of phenomenological research on constraining the multi-dimensional proton (and nucleus) structure at high energies through studies of the so-called gluon Wigner distributions. We provide a brief pedagogical introduction into the corresponding theoretical definitions and modelling of exclusive and diffractive scattering observables in terms of the Wigner distribution. Also, we present a detailed outlook into the existing and planned experimental measurements that attempt to constrain the Wigner distribution. We briefly overview possible interconnections between various manifestations of the gluon Wigner distribution emerging, for instance, in azimuthal-angle correlations in (semi)exclusive reactions and elliptic flow measurements in inclusive processes. We also summarise the current knowledge on the most important processes that would potentially enable one to constrain the elliptic gluon density in the proton and to separate it from the genuine effect of hydrodynamic evolution in the flow measurements.
\end{abstract}

\maketitle

\section{Introduction}
\label{Sect:intro}

Over a few decades, a substantial effort of the particle physics community has been devoted to unveiling the proton structure at high energies in both longitudinal and transverse dimensions relative to the collision axis. Since a long ago, it has been well-known that the most detailed information on a given compound quantum system can be unwound through the knowledge of kinematical distributions of its constituents over phase space. In the case of the nucleon target, for instance, this information is contained in the Wigner parton distribution in Quantum Chromodynamics (QCD)~\cite{Ji:2003ak,Belitsky:2003nz,Lorce:2011kd,Lorce:2011ni,Mukherjee:2014nya,Mukherjee:2015aja,Liu:2015eqa,Chakrabarti:2016yuw} which represents a comprehensive visualisation of the partonic structure of the nucleon in five dimensions, also referred to as the multi-dimensional parton imaging or tomography. Such an imaging has grown into a paradigm in contemporary studies of hadron structure in high-energy particle collisions \cite{Boer:2011fh}. For this reason, the Wigner distribution is also referred to as the ``mother of all distributions'' \cite{Ji:2003ak,Belitsky:2003nz,Lorce:2011kd} as it is connected to other well known lower-dimensional parton distributions through its integration over one or more dimensions. Its Fourier transform known as the Generalized Transverse Momentum Dependent Distribution (GTMD) \cite{Meissner:2009ww,Hatta:2011ku,Lorce:2013pza,Echevarria:2016mrc} is also widely used for modelling of the nucleon structure (for earlier reviews on this topic, see e.g.~Refs.~\cite{Boer:2011fh,Accardi:2012qut,Boussarie:2023izj}). A comprehensive review of forward small-$x$ physics phenomenology, experimental developments and theoretical models addressing the challenges of the largely yet unknown nucleon structure can be found in Ref.~\cite{LHCForwardPhysicsWorkingGroup:2016ote}.

While the Wigner distribution (and GTMD) encodes all the non-perturbative QCD dynamics of parton constituents inside the nucleon (such as their spatial and momentum distributions as well as the orbital angular momentum \cite{Lorce:2011ni,Leader:2013jra}), it is not calculable in the framework of perturbative QCD and represents the main non-perturbative ingredient of QCD factorization. Consequently, it has been argued in Refs.~\cite{Elze:1994qa,Calabrese:2004eu,Kutak:2011rb,Peschanski:2012cw,Kovner:2015hga,Kharzeev:2017qzs,Hagiwara:2017uaz,Berges:2017zws,Kovner:2018rbf} that the Wigner distribution can be used to quantify entanglement entropy in the proton wave function. Ongoing developments of phenomenological methods and techniques enabling to constrain the Wigner distribution directly from the experimental data (the co-called ``partonometry'') is necessary to make further significant advances in nucleon imaging  \cite{Courtoy:2013oaa,Kanazawa:2014nha,Rajan:2016tlg}. A direct measurement of the Wigner distribution appears to be a challenging problem as it generally requires the most detailed knowledge of particle kinematics in the final state in a clean environment (i.e.~with the maximal degree of exclusivity).

A large amount of data on various scattering processes are being currently collected by the measurements in hadron and nuclear collisions at the LHC, and it would be desirable to utilise these data to reconstruct the proton structure encoded in the Wigner distribution. Such processes as inclusive DIS or semi-inclusive DIS (SIDIS) are used to probe the integrated (spin-averaged and spin-dependent) collinear Parton Distribution Functions (PDFs) or Transverse Momentum Dependent Distributions (TMDs), respectively, whereas the Generalized Parton Distributions (GPDs) are constrained by the measurements of exclusive reactions, such as the Deeply Virtual Compton Scattering (DVCS) (for a recent review, see Ref.~\cite{Diehl:2023nmm}). While these distributions are effectively connected to the Wigner distribution by integrating it over one or more dimensions, none of the above reactions can be utilized as a direct probe for the 5D Wigner distribution itself. Indeed, SIDIS and DVCS are sensitive to the parton longitudinal momentum fraction $x$ and to either the impact parameter $\bm{b}$ of a given parton inside the nucleon (or nucleus) or to its transverse momentum $\bm{q}$. 

Diffractive processes such as those in Deep Inelastic Scattering (DIS) of leptons off nucleons or large nuclei play an important role in extracting the crucial information on the structure of the target particle in both coordinate and momentum spaces \cite{Wusthoff:1999cr}. All of this information is encoded in the parton Wigner function $W(x,\bm{q},\bm{b})$ effectively representing the (quasi)probability to find a parton with a given transverse momentum $\bm{q}$ at the impact parameter $\bm{b}$ (its transverse separation from the center of the nucleon or nucleus), and with a given momentum fraction $x$. Among important examples, diffractive vector meson production effectively probes the spatial profile as well as fluctuations of the gluon field in the target \cite{Mantysaari:2016ykx,Mantysaari:2016jaz,Mantysaari:2018zdd}.

Throughout this review, we are focused mainly on the gluon Wigner distribution\footnote{The quark GTMDs can be accessed for any values of $x$ e.g.~through exploring the exclusive double Drell-Yan process as has been elaborated upon in Ref.~\cite{Bhattacharya:2017bvs}.} relevant for both hadro- and photoproduction reactions at high energies, i.e. at $x\ll 1$. Once the collision axis is fixed and both 2D vectors $\bm{b}$ and $\bm{q}$ are in the plane transverse to this axis, we deal with the non-trivial 5D gluon Wigner distribution, $W(x,\bm{q},\bm{b})$, which is the subject of phenomenological studies of gluon tomography of the nucleon.

As an important example of this research, the pioneering study of Ref.~\cite{Hatta:2016dxp} has proposed that the gluon Wigner distribution at a given momentum fraction $x\ll 1$ can be probed experimentally in exclusive dijet photoproduction in Deep Inelastic Scattering (DIS), in particular, by measuring the correlation in azimuthal angle between the produced dijet transverse momentum and the recoiled nucleon transverse momentum (see also Ref.~\cite{Altinoluk:2015dpi}). A follow-up work of Refs.~\cite{Hagiwara:2017fye,ReinkePelicer:2018gyh} (see also Ref.~\cite{Linek:2023kga}) has elaborated on this possibility more quantitatively considering exclusive light-quark dijet and heavy quark-pair photoproduction in ultraperipheral collisions (UPCs). A comprehensive analysis of diffractive dijets production in $ep$ collisions in the Color Glass Condensate (CGC) framework in the context of the Electron-Ion Collider (EIC) has been performed in Ref.~\cite{Mantysaari:2019csc}. Theoretical uncertainties in the existing models for phase space distributions that actually predict non-trivial angular correlations in impact-parameter space, namely, between the impact parameter $\bm b$ and the dipole separation $\bm r$, remain large \cite{Linek:2023kga}. Hence, it is mandatory to experimentally access such correlations that would enable us to understand the nucleon structure in greater detail. 

Given the multi-dimensional character of the Wigner function, rather large-statistics data samples would be needed if one wanted to map all (five) dependencies. Since very low pile-up data\footnote{Pile-up corresponds to typically soft independent interactions in the same bunch crossing whose average number per bunch crossing, $\mu$, rises with increasing instantaneous luminosity.} conditions are preferable, collecting large statistics may require larger integrated luminosities and, hence, longer data-taking periods, depending on the actual average pile-up rate. Furthermore, considering for instance $pA$ UPCs, in the ideal case both, the proton and nucleus on opposite sides should be tagged. If tagging nucleus is problematic or not possible, one could also require a veto in Zero-Degree-Calorimeter (if it exists). If the efficiency of such a veto turns out to be low, one can loosen this requirement by imposing an energy threshold in the forward direction -- its value may depend on a given integrated luminosity. 

Especially in the case that the double-tagging mentioned above is not possible or inefficient, one could also require a large rapidity gap to accompany the intact nucleon/nucleus on both sides from the interaction point (IP). The rapidity gap is an extremely interesting quantity from both, the theory and experiment points of view. Experimentally, the size of the rapidity gap inevitably depends on the noise level of individual subdetectors and energy thresholds of objects we wish to veto (tracks, calorimeter clusters, jets). A typical lowest $p_{\rm T}$ threshold of jets reconstructed from calorimeter clusters/towers at ATLAS or CMS below which the jet reconstruction is significantly less reliable is 20~GeV, while it can go down to 10~GeV, if tracks are considered. But vetoing jets with such $p_{\rm T}$ thresholds is not considered to be sufficient in reaching exclusivity. One should also be aware that in the context of measurements in UPCs, the incoming quasi-real photon can reveal its structure and hence its remnants then can spoil the gap(s) representing one of the major challenges in precision measurements of exclusive photoproduction processes. 

In this review, we summarise the basic properties of the gluon Wigner distribution (and the corresponding GTMD) and the existing methods to access it through detailed measurements of exclusive processes, including the existing phenomenological models and the current experimental situation, as well as their role in our understanding of the nucleon structure. Theoretical modelling of these processes is most conveniently performed in the framework of color dipole approach which has been thoroughly formulated together with the key results from the literature. Besides, we briefly overview the corresponding challenges and published measurements which can be sensitive to the gluon Wigner function, with a potential to probe it either with the existing collected data or with future data.

The review is structured as follows. In Sect.~\ref{Sect:Wigner}, we provide key definitions of the phase space parton distributions in QCD and a brief discussion of fundamental properties of the Wigner distributions, with a particular focus on the gluon case, connecting it to the fundamental ingredient of QCD scattering, the dipole $S$-matrix. Sect.~\ref{Sec:correlations} focuses on observable effects of the dipole orientation with respect to the color background field of the target such as azimuthal angle correlations quantified by the elliptic flow. In Sect.~\ref{Sec:modelling}, we outline the basics of the phenomenological dipole approach and the dipole orientation effects on phenomenological grounds. Namely, the basic properties of the dipole $S$-matrix and its widely used phenomenological modelling are discussed in connection to the azimuthal angle correlations. More elaborate QCD evolution based approaches are briefly presented in Sect.~\ref{Sect:QCD-evolution}. Sect.~\ref{Sect:pheno} reviews a few most relevant processes as suitable probes for dipole orientation effects and the associated elliptic gluon Wigner distribution, as well as provides some of the key insights into the corresponding experimental measurements. A brief overview of future measurements that could offer new possibilities for constraining the gluon Wigner distribution is given in Sect.~\ref{sec:Future}. Finally, Sect.~\ref{Sect:summary} concludes the review summarizing its main takeaways.

\section{Phase space distributions in QCD}
\label{Sect:Wigner}

A remarkable property of QCD is a dramatic difference in complexity of its UV and IR regimes. While in the UV regime one enjoys the simplicity of perturbative description stemming directly from the QCD Lagrangian given in terms of fundamental degrees of freedom, quarks and gluons (partons), in the IR regime one encounters an enormous complexity associated with dynamics of the QCD ground state (quark and gluon condensates), bound states, their interactions and the corresponding nonperturbative QCD phenomena such as confinement and mass gap generation. At short distances, i.e.~very far from the ``confinement boundary'' in the IR regime, an approximation of relatively weakly interacting partons as asymptotic states of the theory works well, at spacetime separations close or larger than this boundary such an approximation breaks down as the asymptotic states can no longer be coloured partons but colorless composite structures of different types (hadrons). A consistent transition between the parton and the hadron levels remains a long-standing theoretical challenge being so far exclusively addressed by lattice QCD simulations in Euclidean spacetime. No single consistent theoretical approach that bridges and unifies both the sub-confined (partonic) and super-confined (hadronic) regimes in a consistent framework in Minkowski spacetime has been developed yet (for a recent review on the status of nonperturbative QCD and confinement research, see Ref.~\cite{Pasechnik:2021ncb}). One particularly important aspect of this problem is the hadron structure highly motivated by the practical need for interpretation of collider measurements (such as those at the LHC) and actively explored in ample literature over past few decades, from both theoretical and phenomenological points of view \cite{Boussarie:2023izj}.

In scattering of high-energy hadrons, a fast-moving hadron whose internal partonic structure is being probed (or which plays a role of a probe itself) sets the longitudinal, or Light-Cone (LC) direction, such that the motion of its constituents is effectively separated into ``longitudinal'' and ``transverse'' parts. At short distances, distributions of asymptotically free partons in their longitudinal momenta in a hadron are described in terms of collinear parton distributions --  the basic non-perturbative ingredients of collinear factorisation --  defined in a single space (momentum) dimension as functions of parton's dimensionless longitudinal momentum fraction. The collinear factorisation concept works out extremely well in a myriad of different inclusive processes in lepton-hadron and hadron-hadron collisions, and has highlighted the triumph of QCD-improved parton model. However, in certain cases the collinear parton evolution picture does not capture all relevant phenomena, and a more detailed information on parton motion in the transverse plane (to the direction of the parent hadron) is required. 

\subsection{Basic definitions}
\label{Sect:definitions}

As a suitable choice of the reference frame in studies of a lepton-nucleon or nucleon-nucleon scattering process, one typically considers a frame where both colliding particles move fast, e.g.~center-of-mass frame, or where one of them is at rest (the target) while the other one moves fast (the projectile), known as the target rest frame. Either way, longitudinal and transverse directions play very different roles enabling the LC decomposition of an incident 4-vector $l$ in terms of its LC components: $+/-$ LC projections $l^\pm=(l^0\pm l^3)/\sqrt{2}$ and 2D transverse momentum $\bm{l}=(l^1,l^2)$, with its absolute value $l_\perp \equiv |\bm{l}|$. Then, a parton carrying 4-momentum $k$ inside a high-energy nucleon is characterized by its longitudinal momentum fraction $x\equiv k^+/P^+$ that it takes from the parent nucleon momentum $P^+$, transverse momentum $\bm{k}$ and the transverse position with respect to the center of the nucleon, $\bm{b}$. The phase space distributions separately in $\bm{k}$ and $\bm{b}$ are known in the literature as the transverse momentum dependent distribution (TMD) $T(x,\bm{k})\equiv T(x,k_\perp)$ and the generalized parton distribution (GPD) $G(x,b_\perp)$, respectively. The former quantifies transverse momentum of a parton in the considered parent hadron affecting the distribution of produced particles, whereas the latter describes the spatial distribution of partons in the transverse plane which can be found via a Fourier transform over the transverse momentum transferred to the parent hadron (causing, in particular, its deflection at a small angle). While being important ingredients of exclusive cross sections, these distributions encode crucial information on the two-dimensional structure of the nucleon, quantifying it in $(x,k_\perp)$ and $(x,b_\perp)$ planes of the phase space. In each such case, one probes dynamics of partons at typical hadronic length (or momentum) scales being sensitive to non-perturbative QCD phenomena (such as confinement). 

However, in several situations an even more detailed information about distribution of partons in both the $\bm{k}$ and $\bm{b}$ variables is required i.e.~accounting also for relative (azimuthal) angle between these 2D vectors. One of the most important examples is the orbital angular momentum $\propto \bm{b} \times \bm{k}$ particularly relevant for understanding of how the total nucleon spin emerges from the angular momenta of its constituents and their spin-orbital correlations, hence being of significant phenomenological interest.
\begin{figure}[hb]
\begin{center}
  \includegraphics[width=0.4\textwidth]{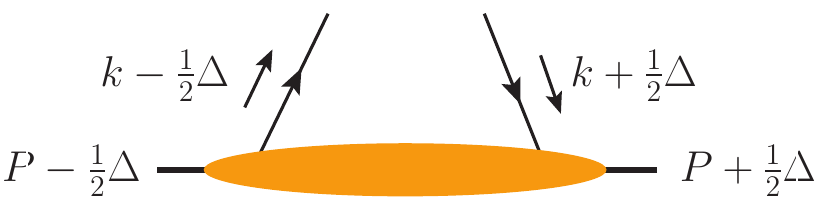}
\end{center}
\caption{Two-quark correlation function (\ref{two-quark_CF}) with the corresponding kinematical variables. From Ref.~\cite{Diehl:2015uka}.}
\label{fig:2q-correlator}
\end{figure}

The most generic object in QCD describing the parton dynamics in the nucleon is the so-called two-parton correlation function. For instance, the manifestly Lorentz-covariant two-quark correlation function is represented as a matrix element of the bilinear quark operator,
\begin{eqnarray}
\label{two-quark_CF}
    H(k, P, \Delta) = \frac{1}{(2\pi)^4}
    \int d^4z\; e^{izk} \bigl\langle P + \Delta/2 |
    \bar{\psi}(-z/2)\Gamma {\mathcal L} \psi(z/2) | P - \Delta/2 \bigr\rangle \,,
\end{eqnarray}
given in terms of the nucleon state $|P\rangle$ and its momentum $P^\mu$, the momentum transfer in the $t$-channel $\Delta^\mu=(0,0,\bm{\Delta})$ ($|\bm{\Delta}| \equiv \Delta_\perp = \sqrt{-t}$), the quark-antiquark relative separation $z^\mu=(0,z^-,\bm{z})$, the staple-shaped Wilson line along $z^-$ ${\mathcal L}$ introduced to ensure gauge invariance of the corresponding operator, and some Dirac matrix $\Gamma=\gamma^+,\gamma^+\gamma_5,\dots$ defining the spin structure and the relevant twist \cite{Jaffe:1996zw}. The incident momenta satisfying the on-shell nucleon condition, $P \Delta = 0$ and $4 P^2 + \Delta^2 = 4 m^2$, with $m$ being the nucleon mass, are depicted in Fig.~\ref{fig:2q-correlator}. For a pedagogical discussion of properties of $H(k, P, \Delta)$ and its phenomenological significance in different kinematical domains, see e.g.~Ref.~\cite{Diehl:2015uka} (see also Ref.~\cite{Boussarie:2023izj} and references therein).
\begin{figure*}
  \includegraphics[width=0.85\textwidth]{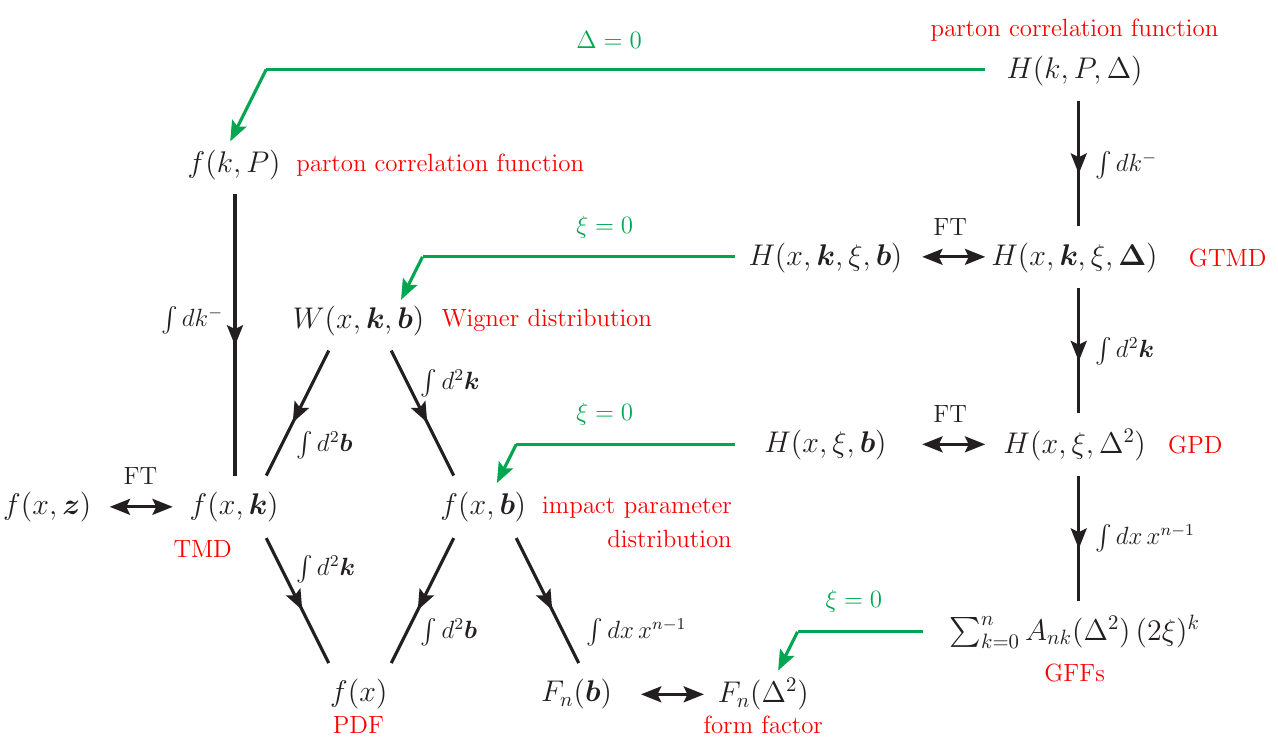}
  \caption{\label{fig:PDF-tree} Relations between different distributions originating from the two-parton correlation function. From Ref.~\cite{Diehl:2015uka}.}
\end{figure*}

Considering particular scattering processes, one typically integrates $H(k, P, \Delta)$ over one or more components of the parton 4-momentum $k$ giving rise to distribution functions in lower number of dimensions as schematically illustrated in Fig.~\ref{fig:PDF-tree}. Starting with integration over $k^-$, one notices that the virtuality of partons is not definite any longer as they are in general off-shell in consistency with confinement. Moreover, the parton evolution in this case is naturally consistent with the parton model in the LC quantisation picture where asymptotically free parton fields are quantised at LC time $z^+=0$, i.e.~just before being probed in scattering. In this case, the $k^-$-integrated parton correlation function gives rise to the six-dimensional GTMD $G(x,\bm{k},\xi,\bm{\Delta})$ describing the emission and absorption of the (anti)quark as well as the emission and/or absorption of a quark-antiquark pair as shown in Fig.~\ref{fig:2q-correlator}. Here, the second LC fraction $\xi = -\Delta^+/2P^+$ known as skewness variable determines the total momentum transfer squared as
\begin{eqnarray}
    \Delta^2 = - \frac{(4\xi^2 m^2 + \bm{\Delta}^2)}{1-\xi^2} \,, \qquad \bm{\Delta}^2 \equiv \Delta_\perp^2 \,.
\end{eqnarray}
In the forward limit suitable e.g.~for exclusive reactions in high-energy lepton-nucleon (lepton-nucleus) or nucleon-nucleon (nucleus-nucleus) collisions, $\Delta^+ \to 0$ (hence, vanishing $\xi$), we arrive at the five-dimensional parton GTMD $G(x,\bm{k},\bm{\Delta})$ where the dependence on the skewness parameter $\xi$ is suppressed. Thus, its Fourier-transform in the mixed (impact parameter/momentum space) representation $W(x,\bm{b},\bm{k})$ describes evolution in both the particle transverse momentum $\bm{k}$ and its average position in the transverse plane $\bm{b}$ and is known as the Wigner distribution \cite{Ji:2003ak} whose notion has been used in physics in many different contexts. In particular, in the context of exploring the nucleon substructure, the quark Wigner distribution takes the following form \cite{Belitsky:2003nz,Lorce:2011kd}
\begin{eqnarray}
W_q(x, \bm{k}, \bm{b}) &=&\int \frac{dz^- d^2\bm{z}}{16\pi^3} \frac{d^2\Delta}{(2\pi)^2}e^{i(xP^+z^- -\bm{k} \cdot \bm{z})}e^{-i\bm{\Delta} \cdot \bm{b}} \nonumber \\
&& \qquad  \times\langle P+\Delta/2| \bar{\psi}(-z/2)\Gamma {\mathcal L} \psi(z/2)|P-\Delta/2\rangle \nonumber \\
&=& \int \frac{dz^- d^2\bm{z}}{16\pi^3} \frac{d^2\Delta}{(2\pi)^2}e^{i(xP^+z^- -\bm{k} \cdot \bm{z})} \nonumber \\ && \qquad \times \langle P+\Delta/2| \bar{\psi}(b -z/2)\Gamma {\mathcal L}\psi(b + z/2)|P-\Delta/2\rangle \,, 
\label{wigner-q}
\end{eqnarray}
in terms of the two-quark correlation function defined above. In analogy to quarks, following Ref.~\cite{Bomhof:2006dp} (see also Refs.~\cite{Meissner:2009ww,Lorce:2013pza}), the gluon Wigner distribution in the proton can formally be written as follows
\begin{eqnarray}
xW_g(x,\bm{k}, \bm{b}) &=& \frac{2}{P^+(2\pi)^3 }\int dz^+ 
d^2\bm{z} \int \frac{d^2\bm{\Delta}}{(2\pi)^2}  
e^{i\bm{k} \cdot \bm{z}  -ixP^-z^+}  \nonumber \\ 
&& \times \left\langle P+\tfrac{\bm{\Delta}}{2} \left|  {\rm Tr}\, 
\left[U_+F_a^{+i}\left(\bm{b}+\tfrac{z}{2}\right) U_-
F_a^{+i}\left(\bm{b}-\tfrac{z}{2}\right)\right] \right|P -
\tfrac{\bm{\Delta}}{2}\right\rangle \,, \label{wigner-g}
\end{eqnarray}
in terms of the staple-shaped Wilson line $U_{\pm}$ that goes to $z^+=\pm \infty$ on the LC, and then returns. For further convenience, denoting initial and final proton momenta as $p=P-\Delta/2$ and $p'=p+\Delta$, here and below, the brackets $\langle\dots\rangle$ denote the off-forward proton matrix element implicitly normalised as $\langle p'|\dots|p\rangle/\langle p|p \rangle$.

\subsection{Properties of the gluon Wigner distribution}
\label{Sect:gluon-Wigner}

The parton Wigner distributions contain the most detailed and complete information about the quark and gluon evolution in five-dimensional phase space, and hence encode the parton structure of the nucleon wave function \cite{Ji:2003ak,Belitsky:2003nz,Lorce:2011kd} (see also Refs.~\cite{Meissner:2009ww,Lorce:2013pza}), being of significant phenomenological interest. In quantum theory, however, such a definition immediately contradicts the uncertainty principle $\delta x \delta p\ge \hbar/2$ as the particles' position and momentum cannot be determined simultaneously with arbitrary precision. In practice, this fact causes the Wigner distribution to violently oscillate and to become negative in some domains of the phase space. The Wigner distributions (\ref{wigner-q}), (\ref{wigner-g}) have been evaluated in various models, in particular, for quark one see Refs.~\cite{Belitsky:2003nz,Lorce:2011kd,Lorce:2011ni,Kanazawa:2014nha,Mukherjee:2014nya,Liu:2014vwa,Muller:2014tqa,Courtoy:2014bea}. In simple models without gluons, it turns out to be a positively-definite function. However, once gluons are included in the full QCD description, the Wigner distribution becomes negative for some phase space configurations \cite{Mukherjee:2014nya}. For that reason, in QCD the parton Wigner distribution does not have an interpretation of a joint probability distribution in both $\bm{k}$ and $\bm{b}$, but rather of a quasi-probability one, due to not being manifestly positively-definite. 

This situation has motivated the authors of Refs.~\cite{Hagiwara:2014iya,Hagiwara:2016kam} to turn to the so-called Husimi distribution as a close analogue to the classical phase-space distribution. In quantum mechanics, it has the meaning of a positively-definite probability density to locate a particle not at precise values of its momentum and coordinate but within an uncertainty band $(x\pm \delta x,p\pm \delta p)$, with $\delta x$ and $\delta p$ satisfying $\delta x \delta p\ge \hbar/2$. In QCD, a transition to Husimi distribution implies a Gaussian smearing of the Wigner distribution in both $\bm k$ and $\bm b$ as follows \cite{Hagiwara:2014iya} 
\begin{eqnarray}
{\cal W}(x,\bm{k}, \bm{b}) = \frac{1}{\pi^2}\int d^2 \bm{b}^{\prime} d^2 \bm{k}^{\prime} e^{-\frac{1}{l^2}(\bm{b} - \bm{b}^{\prime})^2 - l^2(\bm{k} - \bm{k}^{\prime})^2} W(x,\bm{k}^{\prime}, \bm{b}^{\prime}) \,,
\label{husimi-QCD}    
\end{eqnarray}
where the Gaussian widths are related inversely to each other in consistency with the uncertainty relation, hence, ensuring that the resulting distribution is positive semi-definite.

Interestingly enough though, by integrating $W$ over one of the transverse vectors yields a proper probability distribution in the remaining one. Indeed, integrating $W$ over $\bm{k}$ (with an appropriate regularisation procedure) gives rise to the parton distribution $f(x,b_\perp)$ in impact parameter space 
\begin{eqnarray}
&& f(x,b_\perp) \equiv \int d^2\bm{k} \, W(x, \bm{k}, \bm{b}) \nonumber\\ && =\int \frac{d^2\bm{\Delta}}{(2\pi)^2}e^{-i\bm{\Delta} \cdot \bm{b}}\int \frac{dz^-}{4\pi} e^{ixp^+z^-} \langle P+\Delta/2| \bar{\psi}(-z^-/2)\gamma^+ {\mathcal L} \psi(z^-/2)|P-\Delta/2\rangle\,,
\end{eqnarray}
representing a probability density for a parton carrying longitudinal fraction $x$ to be localised at impact parameter $\bm{b}$ from the nucleon center. A 2D Fourier-transform of $f(x,b_\perp)$ represents a GPD in momentum space, $G(x,k_\perp)$, for which in general $\xi\not=0$. On the other hand, integrating $W$ over $\bm{b}$ results in the corresponding TMD, i.e.
\begin{eqnarray}
{\cal T}(x,k_\perp) \equiv \int d^2\bm{b}\, W(x,\bm{k}, \bm{b}) &=& \int \frac{dz^- d^2\bm{z}}{16\pi^3} e^{i(xp^+z^- -\bm{k} \cdot \bm{z})}\langle P| \bar{\psi}(-z/2)\gamma^+ {\mathcal L} \psi(z/2)|P\rangle \,. \label{tmd}
\end{eqnarray}
From this point of view, the Wigner distribution is often dubbed as the `mother distribution' for both TMDs and GPDs. Further possible connections to lower-dimensional distributions are briefly summarised in Fig.~\ref{fig:PDF-tree}. As was mentioned above, for a longitudinally polarized nucleon, the Wigner distribution is also related to the canonical orbital angular momentum \cite{Lorce:2011kd,Hatta:2011ku}
\begin{eqnarray}
L_W = \int dx d^2\bm{b} d^2\bm{k} \, (\bm{b}\times \bm{k}) \, 
W(x, \bm{k}, \bm{b})\,, 
\label{orbital}
\end{eqnarray}
which is an important ingredient in the nucleon spin decomposition (for more details, see also Refs.~\cite{Aschenauer:2015ata,Hatta:2016aoc}).

\subsection{Gluon Wigner distribution in the dipole picture}
\label{Sec:Wigner-dipole}

Traditionally, diffractive particle production processes, for instance, with $q\bar q$ di-jets in the final state, are often regarded as a convenient probe for hadron structure (see e.g.~Ref.~\cite{Ashery:2006zw} and references therein). As was demonstrated in Ref.~\cite{Hatta:2016dxp}, the gluon Wigner distribution at small momentum fractions of the gluons in the target, i.e.~$x\ll 1$, gets dramatically simplified and can be accessed experimentally in diffractive dijet (electro)production (see also Ref.~\cite{Altinoluk:2015dpi,Hagiwara:2017fye}). Indeed, through its connection to the gluon GTMD,
\begin{equation}
xW_g(x, \bm k, \bm b) = \int \frac{d^2 \bm{\Delta}}{(2\pi)^2} e^{i\bm{\Delta}\cdot \bm b}\, xG_g(x,\bm{k},\bm{\Delta}) \,,
\end{equation}
which, following Refs.~\cite{Bomhof:2006dp,Belitsky:2002sm,Dominguez:2011wm}, reads
\begin{eqnarray}
xG_g(x,\bm{k},\bm{\Delta}) & \simeq & \frac{2N_{c}}{\alpha_s}\int
\frac{d^2\bm{R}\, d^2\bm{R}^{\prime}}{(2\pi)^4} \, e^{i {\bf k}\cdot \left(
{\bm R} - {\bm R}^{\prime}\right) +i \frac{\bm \Delta}{2}\cdot({\bm R}+{\bm R}^\prime)} \left(\bm{\nabla}_{R}\cdot
\bm{\nabla}_{R^{\prime}}\right) \\
&\times& \frac{1}{N_{c}}\left\langle{\rm Tr}\left[
U({\bm R}) U^{\dagger}({\bm R}^{\prime})
\right]\right\rangle_x\,, \qquad x\ll 1 \,,
\label{op}    
\end{eqnarray}
in terms of the coupling constant $\alpha_s$ and the number of colors $N_c=3$ in QCD, one straightforwardly identifies the impact parameter dependent off-forward dipole $S$-matrix in the eikonal approximation, which determines the dipole survival probability, 
\begin{eqnarray}
\frac{1}{N_c}\left\langle {\rm Tr}\,  U\left(\bm b+\frac{\bm r}{2}\right) U^\dagger\left(\bm b-\frac{\bm r}{2}\right)\right\rangle_x \equiv S_Y\Big(\bm b+\frac{\bm r}{2},\bm b-\frac{\bm r}{2}\Big) \,,
\label{S-matrix}
\end{eqnarray}
as a product of two light-like Wilson lines in the fundamental representation $U$ of a quark at $\bm b+\frac{\bm r}{2}$ and an antiquark at $\bm b-\frac{\bm r}{2}$ transverse positions, respectively. The Wilson line
\begin{eqnarray}
U(x_\perp) \equiv P\exp\left(-ig\int_{-\infty}^\infty dx^- A^+(x^-,x_\perp)\right) \,,
\end{eqnarray}
is an important ingredient of the dipole scattering describing the eikonal propagation of the quark in the colored medium.
\begin{figure*}[!t]
 \centerline{\includegraphics[width=0.4\textwidth]{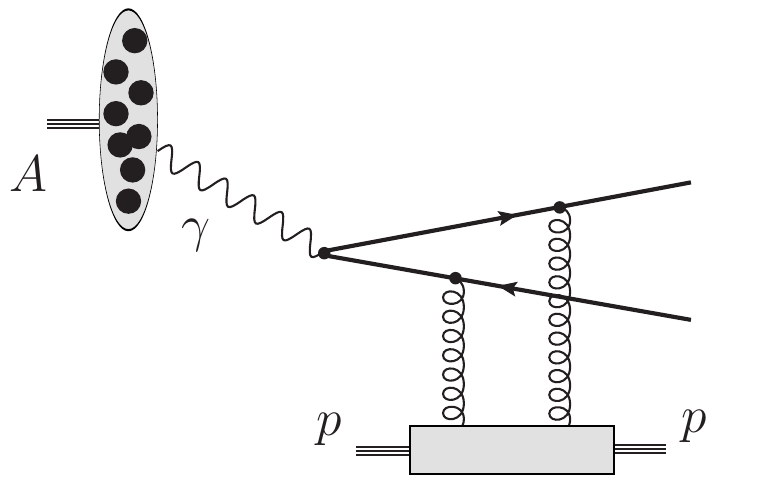}}
   \caption{
   A schematic illustration of the $\gamma p \to (q\bar q) p$ scattering in the dipole picture. 
   From Ref.~\cite{Hagiwara:2017fye}.}
 \label{fig:dijet-UPC}
\end{figure*}

In order to see this more explicitly, let us consider the $\gamma p$ scattering in the dipole picture \cite{Kopeliovich:1981pz,Nikolaev:1991et}, where the real photon fluctuates into a $q\bar q$-dipole that scatters off the target proton $p$ as illustrated in Fig.~\ref{fig:dijet-UPC}. Here, let us fix the frame in which both initial $\gamma$ and $p$ are collinear, with the fast proton propagating in the positive $z$-direction. Then, the process is most conveniently described in the impact parameter representation where the transverse coordinates of the quark and antiquark are set to be $\bm{x}_{1}=\bm{b} + (1-z)\bm{r}$ and $\bm{x}_{2} = \bm{b} -z\bm{r}$, respectively. Here, $z$ stands for the longitudinal momentum fraction of the quark taken from the incoming real photon. The lowest Fock state of the photon is described by $\gamma \to q\bar q$ wave function \cite{Kovchegov.2012}, with an emerging $q\bar q$ dipole, whose ``center of gravity'' is located at impact parameter $\bm{b} \equiv z\bm{x}_{1} + (1-z) \bm{x}_{2}$, which is the same as that of the incoming photon. The transverse separation of the dipole is then found as $\bm{r} \equiv \bm{x}_{1} - \bm{x}_{2}$ defining the strength of the dipole-target scattering,
\begin{eqnarray}
S_Y(\bm{x}_{1}, \bm{x}_{2}) \equiv \frac{1}{N_c} \left\langle \textrm{Tr} \left[U(\bm{x}_{1}) U^\dagger (\bm{x}_{2})\right] \right\rangle_x\,, 
\label{qqbar_S-matrix}
\end{eqnarray}
In the above expressions, the averaging $\langle\dots\rangle_x$ in the target is defined in the CGC formalism \cite{McLerran:1993ni,Gelis:2010nm} (for further details on the gluon Wigner distribution in the CGC framework, see Sect.~\ref{Sect:BK} below) yielding the dipole $S$-matrix as a function of ``rapidity'' $Y\equiv \ln(x_0/x)$, $x\ll 1$, with some adjustable parameter $x_0$ typically fixed to $x_0 = 0.01$ unless noted otherwise. Hence, the energy dependence of the dipole $S$-matrix is effectively encoded through its $Y$ dependence. In the particularly important example of diffractive DIS process, $\gamma p \to X p$, $x$ is identified with $x_\Pom$ being the LC momentum fraction of the projectile dipole carried by a collective $t$-channel exchange (often regarded as the Pomeron exchange) between the dipole and the target, and is found as
\begin{eqnarray}
    x\equiv x_\Pom = \frac{M_\perp^2}{\hat{s}} \,, \qquad \hat{s} \equiv s_{\gamma p} \,, \label{xPom}
\end{eqnarray}
in terms of the transverse mass $M_\perp$ of the produced diffractive system $X$. In case of dijet production in diffractive DIS off the proton target, for instance, one has
\begin{equation}
    x_\Pom = \frac{M_{jj}^2 + Q^2 - t}{W^2 + Q^2 - m_p^2} \,,
\end{equation}
in terms of photon virtuality, $Q^2$, the proton mass $m_p$, the c.m.~energy squared of the photon-target system, $W^2$, and the momentum transfer squared given by the Mandelstam $t$ variable.

Then, it is instructive to switch to momentum space by means of double Fourier transform, $\widetilde{S}_Y$, as follows \cite{Hatta:2017cte}
\begin{eqnarray}
&\quad& \int \frac{d^2\bm{r} d^2\bm{b}}{(2\pi)^4} e^{i\bm{\Delta} \cdot \bm{b}  + i \tilde{\bm{q}} \cdot \bm{r}} S_Y\big(\bm{b} + (1-z) \bm{r}, \bm{b} - z\bm{r}\big)
\nonumber\\
&=& \int\frac{d^2\bm{x}_{1} d^2\bm{x}_{2}}{(2\pi)^4}\, e^{i\bm{k}_{1}\cdot \bm{x}_{1} - i\bm{k}_{2}\cdot \bm{x}_{2}} S_Y(\bm{x}_{1}, \bm{x}_{2})
\nonumber\\
&=& \int \frac{d^2\bm{r} d^2{\bm b}^{\prime}}{(2\pi)^4} e^{i\bm{\Delta} \cdot {\bm b}^{\prime}  + i\tilde{\bm{q}} \cdot \bm{r}} e^{-i\delta_\perp\cdot \bm{r}} S_Y\left({\bm b}^{\prime} +\frac{\bm{r}}{2}, {\bm b}^{\prime} - \frac{\bm{r}}{2}\right)  \nonumber\\
&\equiv& \widetilde{S}_Y(\bm{q}\equiv\tilde{\bm{q}} -\bm{\delta},\bm{\Delta})\,, \qquad \bm{\delta} \equiv \frac{1-2z}{2}\bm{\Delta} \,,
\label{S_Y-Fourier}
\end{eqnarray}
where the exchanged gluon transverse momenta, $\bm{k}_1\equiv \tilde{\bm{q}} + z\bm{\Delta}$ and $\bm{k}_2\equiv \tilde{\bm{q}} - (1-z)\bm{\Delta}$, flow in opposite directions and are conjugate to the impact parameters, $\bm{x}_{1}$ and $\bm{x}_{2}$, respectively. The phase factor $e^{-i\bm{\delta} \cdot \bm{r}}$ recovered in Eq.~(\ref{S_Y-Fourier}) is invariant under interchange of quark and antiquark, i.e. $z\to 1-z$ and $\bm{r} \to -\bm{r}$, being the exact symmetry of the dipole amplitude \cite{Bartels:2003yj}\footnote{As has been pointed out in Ref.~\cite{Hatta:2017cte}, this symmetry has not been respected when using a different phase factor e.g.~in Refs.~\cite{Kowalski:2006hc,Goncalves:2007sa,Berger:2012wx,Rezaeian:2013tka,Goncalves:2014wna,Xie:2016ino}.}. This phase factor becomes relevant, for instance, in the DVCS with the longitudinally polarized virtual photon \cite{Hatta:2017cte}, while it does not play any role in diffractive dijet production \cite{Hatta:2016dxp}. One then arrives at the gluon GTMD 
\begin{eqnarray}
\widetilde{S}_Y(\bm{q},\bm{\Delta}) &=&  \int \frac{ d^2\bm{r} d^2\bm{b} }{(2\pi)^4} e^{i\bm{b} \cdot \bm{\Delta} + i\bm{r} \cdot \bm{q}}  S_Y\left(\bm{b} + \frac{\bm{r}}{2}, \bm{b} - \frac{\bm{r}}{2}\right) \,, 
\label{gluon-GTMD}
\end{eqnarray}
as a Fourier transform of the dipole $S$-matrix defined in Eq.~(\ref{S-matrix}). Finally, one finds \cite{Hatta:2016dxp}
\begin{eqnarray}
x\widetilde{W}_g(x,{\bm q},\bm{\Delta}) \simeq \frac{2N_c}{\alpha_s}\left({\bm q}^2 - \frac{{\bm \Delta}^2}{4}\right)\, \widetilde{S}_Y(\bm q, \bm{\Delta}) \,.
\label{xWg-F}
\end{eqnarray}
Hence, the Wigner distribution is the basic ingredient in the description of the forward cone in various diffractive processes as it naturally incorporates dependence on the transverse momentum transfer between the diffractively produced system and the target, $\bm{\Delta}$. Coming back to the Wigner distribution, one writes
\begin{eqnarray}
\label{xWg}
xW_g(x,\bm{q},\bm{b}) = -\frac{2 N_c}{\alpha_s} \int \frac{d^2\bm{r}}{(2\pi)^2} e^{i\bm{q} \cdot \bm{r}} \left(\frac{1}{4}\bm{\nabla}^2_b +\bm{q}^2 \right) T_Y(\bm{r},\bm{b})\,, 
\end{eqnarray}
in terms of the scattering matrix $T_Y$ found conventionally as $S_Y \equiv 1 - T_Y$. Due to the fact that the colorless dipole of a vanishing size ceases to interact with the target -- the so-called color transparency property -- no scattering off the target gluons occurs leading to $T_Y\to 0$, hence, $S_Y(\bm{r},\bm{b}) \to 1$ as $r_\perp\equiv |\bm{r}| \to 0$. This implies the `sum rule'
\begin{eqnarray}
\int d^2\bm{q} d^2 \bm{\Delta} e^{-i\bm{\Delta}\cdot \bm{b}} \, \widetilde{S}_Y(\bm{q}, \bm{\Delta}) = 1 \,.
\label{unitarity}
\end{eqnarray}
while $\int d^2\bm{q} \widetilde{S}_Y(\bm{q},\bm{\Delta})=0$ for $\Delta_\perp \neq 0$, in consistency with unitarity of the $S$-matrix.

The above derivation highlights the proof of Ref.~\cite{Hatta:2016dxp} showing that the gluon Wigner distribution at small $x$ can be expressed in terms of a Fourier-integral containing the dipole $S$-matrix defined in the impact parameter space, $S_Y(\bm r,\bm b)$. Hence, one may conclude that the framework of Wigner distributions effectively unifies three common descriptions based on the color dipole approach \cite{Kopeliovich:1981pz,Nikolaev:1991et} as well as on the TMDs and GPDs in the parton distributions approach. It not only gives us access to the transverse momentum distribution of the considered parton at a given rapidity (or $x$), but also provides an average position of it relative to the center of the target particle.

Such an important connection between the quasi-real Weisz$\ddot{{\rm a}}$cker-Williams (WW) type gluon distribution in the nucleon in its rest frame and the dipole scattering matrix probed by a projectile $q\bar q$ dipole propagating through the color background field of the nucleon at small $x$ represents a universality between the corresponding two topologically different gauge-invariant operators \cite{Hatta:2016dxp}. The partial dipole amplitude is an important ingredient of the color dipole picture of QCD scattering where dipoles are considered to be eigenstates of the scattering operator \cite{Kopeliovich:1981pz} (see below) such that practically any QCD scattering process can potentially be represented as a superposition of different ``elementary'' dipole scatterings. This concept has been widely explored in a vast literature starting originally from the inclusive DIS in $ep$ collisions, then applying it in inclusive hadron production in $pp$, $pA$ and $AA$ collisions as well as for a variety of exclusive and diffractive reactions. Typically, the dipole amplitude gets averaged over possible dipole orientations with respect to the background field as most of the measured processes are not sensitive to azimuthal-angle correlations between $\bm r$ and $\bm b$. From the theory point of view, considerations of Refs.~\cite{Hatta:2016dxp,Hagiwara:2017fye,ReinkePelicer:2018gyh} have pointed out that the non-trivial correlations in angle between the impact parameter and the dipole separation should be sensitive to possible correlations in angle between the gluon's impact parameter and its transverse momentum in the corresponding gluon Wigner distribution. The dependence on the dipole orientation in inclusive processes, for instance, is known to induce flow-like azimuthal angle correlations such as those in production of prompt photons \cite{Kopeliovich:2007fv}, or in two-particle production processes \cite{Iancu:2017fzn}. Below, we will elaborate on the possible origin and significance of such correlations in the small-$x$ regime relevant for high-energy collisions in the framework of the color dipole picture.

\subsection{Elliptic gluon Wigner distribution}
\label{Sec:elliptic}

The angular correlations of the differential cross sections are connected, in particular, to a correlation of the partial dipole amplitude in azimuthal angle between the impact parameter $\bm{b}$ and the dipole separation $\bm{r}$ in the course of dipole propagation through the colored medium of the target. One often distinguishes two possible sources of such correlations. The first one associated with the initial-state structure arises from the gluon Wigner distribution (or the corresponding GTMD) and represents a correlation in the angle between the transverse momentum and impact parameter vectors in the transverse plane. The second source is related to dynamics of the produced states in the medium and, to the leading order, is associated with the relative angular dependence of the (anti)quark plane-waves in the $q\bar q$ dipole which probes a gluon in the target nucleon. While both sources of dipole orientation effects may have an impact on multi-partilce correlation observables in a scattering process, a precise prediction of their effect in different kinematical regimes remains a challenging problem under intense studies in the literature. For a comprehensive discussion on the possible origins of azimuthal correlations in small systems, see e.g.~Refs.~\cite{Dusling:2015gta,Schlichting:2016sqo,Hagiwara:2017ofm,Dusling:2017dqg,Schenke:2017bog,Mantysaari:2017cni,Zhao:2017rgg,Schenke:2019pmk,Mantysaari:2020axf,Zhao:2022ayk,Linek:2023kga} and references therein. 

When the initial-state effects encoded in the gluon Wigner distribution are concerned, the dipole orientation effects give rise to a characteristic correlation in azimuthal angle between $\bm{q}$ and $\bm{b}$ determined by the so-called ``elliptic'' gluon Wigner distribution \cite{Hatta:2016dxp,Hagiwara:2016kam,Zhou:2016rnt}. Indeed, assuming that the elliptic $\propto \cos 2\theta$ dependence of the partial dipole amplitude is dominant, the gluon GTMD decomposition into the angular-independent (isotropic) ${\widetilde S}_{Y,0}$ and elliptic ${\widetilde S}_{Y,\epsilon}$ components can be written as~\cite{Hatta:2016dxp,Hagiwara:2016kam,Zhou:2016rnt,Hagiwara:2017ofm}
\begin{eqnarray}
{\widetilde S}_Y(\bm{q}, \bm{\Delta}) = {\widetilde S}_{Y,0}(q_\perp, \Delta_\perp) + 2\cos 2(\phi_q-\phi_\Delta) {\widetilde S}_{Y,\epsilon}(q_\perp, \Delta_\perp) + \dots 
\label{GTMD-decomposition}
\end{eqnarray}
ignoring small higher harmonics. For hard scattering processes, the elliptic Wigner distribution ${\widetilde S}_{Y,\epsilon}$ is relatively small (of the order a few percent, see below) compared to the isotropic one, ${\widetilde S}_{Y,0}$, but exhibits very different dependencies on $x$ and $q_\perp$ \cite{Hagiwara:2016kam}. Indeed, their asymptotic behavior is generically expected as follows \cite{Hagiwara:2017fye}
\begin{eqnarray}
{\widetilde S}_{Y,0}(q_\perp,\Delta_\perp) \sim 
\begin{cases} {\rm const} & q_\perp\to 0 \\ 
\left(\frac{1}{q_\perp}\right)^4 & q_\perp \to \infty\,,
\end{cases} \,, \qquad 
{\widetilde S}_{Y,\epsilon}(q_\perp,\Delta_\perp) \sim 
\begin{cases} q_\perp^2 & q_\perp\to 0 \\ 
\left(\frac{1}{q_\perp}\right)^6 & q_\perp \to \infty\,.
\end{cases}
\end{eqnarray}
Thus, the presence of the elliptic density may lead to distinct experimental signatures such as elliptic flow etc being the subject of intense explorations in the literature (see e.g.~Refs.~\cite{Hatta:2016dxp,Zhou:2016rnt,Hagiwara:2017ofm}). The elliptic gluon distribution ${\widetilde S}_{Y,\epsilon}$ provides crucial information on small-$x$ gluon tomography in the proton and can potentially be constrained through measurements of different processes discussed below, in particular, at the future EIC (for a comprehensive review of the EIC concepts and research, see e.g.~Refs.~\cite{Boer:2011fh,Accardi:2012qut,Aschenauer:2017jsk}). 

The above decomposition (\ref{GTMD-decomposition}) is typically used in analysis of such observables as, for instance, the differential cross section of exclusive di-jet photoproduction offering interesting opportunities for a direct measurement of the elliptic gluon GTMD. Still, a more generic analysis of feasibility of direct probes for the elliptic gluon Wigner distribution in differential observables accounting for both the initial- and final-state effects on the same footing is still missing, and it remains yet unclear to what extent the elliptic component ${\widetilde S}_{Y,\epsilon}(q,\Delta)$ can actually be accessed in a measurement. For a more detailed discussion on this topic, see Sect.~\ref{Sect:dijet-photo} below. Let us first elaborate in more detail on physical origins of such correlations and, further on, how the azimuthal angle correlations can be modelled in the dipole picture of QCD scattering from theoretical and phenomenological points of view.

\section{Azimuthal correlations and the elliptic flow}
\label{Sec:correlations}

The experimentally observed fact that the hadron production processes in both proton-proton and proton-nucleus collisions at high multiplicities exhibit unexpectedly large azimuthal asymmetries (see e.g.~Refs.~\cite{CMS:2010ifv,CMS:2012qk,CMS:2013jlh,ALICE:2014dwt,ATLAS:2012cix,ATLAS:2013jmi,STAR:2015kak}) has prompted continuous debates in the literature on the origins of such asymmetries. The long-standing theoretical problem is to consistently explain why small systems emerging in $pp$ or $pA$ collisions reveal a very similar collective behavior as order-of-magnitude larger systems formed in $AA$ collisions. Such a universality between small and large systems has been found in various observables differential in rapidities, transverse momenta and masses of final-state hadrons produced in such reactions \cite{ALICE:2016fzo,ALICE:2012eyl,ALICE:2013snk,ATLAS:2012cix,ATLAS:2014qaj}. 

Two different approaches are typically put forward to interpret such an universality and confronted to each other. One of them resides on a picture of collectivity due to the final-state interactions, also known as the hydrodynamic flow -- an effect typically considered to be natural in the hydrodynamic description of large systems, at least, for small transverse momenta (see e.g.~Refs.~\cite{dEnterria:2010xip,Bozek:2012gr,Bozek:2013ska,Qin:2013bha,Werner:2013ipa,Bzdak:2014dia,Weller:2017tsr}). The second approach relies on interpretation of collective phenomena as due to universality in the initial state connected to parton densities in the incoming hadron or nuclear wave functions before they collide (see e.g.~Refs.~\cite{Teaney:2002kn,Kovchegov:2002nf,Gavin:2008ev,Dumitru:2010iy,Avsar:2010rf,Kovner:2011pe,Levin:2011fb,Iancu:2011ns,Schenke:2012hg,Dusling:2015rja,
Kovchegov:2013ewa,Kovner:2015rna,Lappi:2015vta,Rezaeian:2016szi,Schenke:2016lrs}.)

In order to measure an asymmetry in the azimuthal angle $\phi$ reflecting a spontaneous breakdown of the rotational symmetry in the plane transverse to the collision axis, it is of practical relevance to consider an inclusive single-particle distribution as a function of $\phi$ on event-by-event basis. Experimentally, such an angle is determined in the so-called reaction plane as an angle between the momentum of a produced state and its impact factor, both measured in the transverse plane. Then, the Fourier $\cos(n\phi)$-moments of such a single-particle distribution characterise, in particular, an ellipticity of the overlapping (interaction) domain of the two non-centrally colliding states. These moments are commonly known as the ``flow coefficients'' denoted as $v_n(p_\perp)$ \cite{Snellings:2011sz} and are traditionally considered as an effective measure of the azimuthal asymmetry. The latter can be relevant even for central collisions and is induced by fluctuations in transverse momentum distributions of incident particles in the compound colliding states such as nucleons in a nucleus or partons in a nucleon \cite{Alver:2010gr}. In this sense, $v_n$ is sensitive to inhomogeneities in the transverse-plane parton distributions effectively probing variation in the gluon transverse momentum distribution in the target. In the standard hydrodynamic simulations in the case of heavy-ion collisions, the particle number fluctuations are naturally present in the initial conditions, hence, triggering an asymmetry even without final-state interactions. Such an `initial-state approach' is typically formulated in the effective framework of CGC \cite{Iancu:2003xm,Gelis:2010nm}.

The high energy limit of QCD (i.e.~at very small $x$) is described in terms of the CGC effective theory which provides a suitable framework for analysis of diffractive processes. According to the CGC picture of the fast moving target, one distinguishes the gluon fields with large occupation numbers (``slow modes''), described by classical Yang-Mills equations of motion, and their sources (``fast modes''). Consequently, the target nucleon in this picture has a sub-structure at very high energies describing the formation of regions of a typical size $\sim 1/Q_s$, in terms of the semi-hard saturation scale $Q_s$, with large occupation numbers of coherent gluons. Each such region, also known as the ``saturation domains'', can be interpreted as being in the ground-state (condensate) of strong chromo-electric fields. The orientation of these fields is not symmetric setting up a preferred direction in the transverse plane, thus breaking the rotational invariance and inducing azimuthal correlations in the final-state particles' distribution. The transverse momentum dependence of the flow parameters $v_n(p_\perp)$ appears to have a maximum at around $p_\perp\sim Q_s$ as expected from the domain-like substructure of the target. This phenomenon demonstrates the relevance of a realistic saturation model incorporating the correct $b_\perp$-dependence in the inhomogeneous transverse-plane distribution of soft gluons in the target that should be tested against the experimental measurements.

Indeed, consider the CGC dipole picture of a very forward particle production (and/or a heavy nucleus in the target) when a dilute projectile probes a dense target at a very low $x$. In this case, if incoming particles interact with the same saturation domain at a given impact parameter, they would get similar ``kicks'' giving rise to final states emerging at similar angles. This effect would be washed away in the case of inclusive single-particle production when averaging over scatterings off different domains since their color fields' orientations are assumed to be random. However, very interesting residual effects of such correlations in the CGC approach remain in observables corresponding to production of, at least, two particles in the final state with ``semi-hard'' spectra  decreasing with $Q_s$. Indeed, as $Q_s$ grows, $v_n$ decrease with energy, while the scaling $v_n\sim 1/N_c^2$ is valid for a large number of colors $N_c\to \infty$. Interestingly enough, in the case of non-Guassian correlations the effects of fluctuations in the CGC approach can be reproduced by the color field domain model of Refs.~\cite{Kovner:2011pe,Dumitru:2014dra,Dumitru:2014vka,Dumitru:2014yza}.

For the scattering cross section to have a non-trivial correlation in azimuthal angle between transverse momentum $\bm{p}$ of the produced hadron, acquired due to scattering, and its impact parameter $\bm{b}$, there must be an inhomogeneity in the small-$x$ gluon density in the transverse plane \cite{Kopeliovich:2007fv,Kopeliovich:2007sd,Kopeliovich:2008nx,Levin:2011fb,Zhou:2016rnt,Hagiwara:2017ofm} (see also Ref.~\cite{Iancu:2017fzn}). The latter is precisely characterised by the elliptic part of the gluon Wigner distribution in the target in the high-energy (dipole) factorisation valid for `dilute-dense' collisions. Hence, the flow coefficients $v_n$, more precisely the elliptic flow $v_2$, should be sensitive to (or rather determined by) the presence of the elliptic Wigner density of the target which may adopt a finite value even in the large-$N_c$ limit due to not relying on non-planar gluon exchanges only. Just like the elliptic density itself, the azimuthal asymmetry is directly connected to the dipole orientation phenomena encoded in the dipole $S$-matrix.

The simplest illustration of how to probe the azimuthal angle correlations in the target rest frame is via the inclusive single particle production at forward rapidities such as quark production in collinear quark-target scattering, $qT\to qX$, where $T=p(A)$ is the proton (nucleus) target. The scattered quark has impact parameter relative to the center of the projectile proton, $\tilde{\bm{b}}\equiv \bm{b} - \bm{b}'$, and acquires a transverse momentum $\bm{p}$ due to multiple gluon exchanges with the dense target. The corresponding cross section, differential in both $\bm{p}$ and the projectile proton impact parameter relative to the center of the target $\bm{b}'$, can be written in terms of the dipole $S$-matrix ${\bar S}_Y(\bm{p},\bm{b})\equiv {\bar S}(x_g,\bm{p},\bm{b})$ in the mixed (transverse momentum/impact parameter) representation as
\begin{eqnarray}
\frac{\partial \sigma^q(qT\to qX)}{\partial \eta\,
\partial^2\bm{p}\, \partial^2\bm{b}'} = 
\frac{1}{(2\pi)^2}\, \int {d^2\bm{b}} \ x_q q(x_q,\bm{b} - \bm{b}')\,
{\bar S}_Y(\bm{p},\bm{b})\,,
\label{qCS}
\end{eqnarray}
where $\eta$ is the (pseudo)rapidity of the final-state quark, $xq(x,\bm{b})$ is the quark GPD incorporating the quark impact parameter dependence in the dilute projectile proton. In the soft regime, one typically adopts the factorised form of the quark GPD, with an isotropic profile of the projectile,
\begin{eqnarray}
xq(x,\bm{b}) \simeq xq(x) f(\bm{b}) \,, \qquad \int {d^2\bm{b}} \,f(\bm{b}) = 1 \,, \qquad f(\bm{b}) \equiv f(b_\perp) \,,
\end{eqnarray}
where $xq(x)$ is the collinear quark density in a dilute projectile. The dependence on the transverse size of the projectile proton becomes irrelevant in case of nearly homogeneous large targets in central collisions, the projectile proton can be considered point-like such that $f(\bm{b})=\delta^{(2)}(\bm{b})$ to a good approximation. Above, $x_{q,g}$ are the longitudinal momentum fractions of the projectile quark and exchanged gluon given by
\begin{eqnarray}
x_{q} = \frac{p_\perp}{\sqrt{s}}\,e^{\eta}\,, \qquad x_{g} = \frac{p_\perp}{\sqrt{s}}\,e^{-\eta}\,, \qquad x_g \ll x_q \,,
\end{eqnarray}
respectively, in terms of the total c.m.~energy of the (projectile-target) scattering, $\sqrt{s}$. Provided that the incoming quark propagates at a fixed transverse position (although different one in the amplitude and its conjugate), multiple gluon exchanges are viewed as effectively resummed to all perturbative QCD (pQCD) orders using the eikonal approximation in the impact parameter space. The latter, after squaring the scattering amplitudes, gives rise to the $q\bar q$ dipole $S$-matrix $S_Y$ at an impact parameter $\bm{b}$ and the dipole separation $\bm{r}$. The transverse-plane positions of the quark and antiquark in the dipole relative to the target can then be written as 
\begin{eqnarray}
 \bm{x} = \bm{b} + \frac{\bm{r}}{2} \,, \qquad \bm{y} = \bm{b} - \frac{\bm{r}}{2} \,,
\end{eqnarray}
respectively, while the rapidity separation reads $Y = \ln(x_0/x_g) \gg 1$. The unitarity condition (\ref{unitarity}) in this case implies that the total probability for the final-state quark to acquire some momentum $\bm{p}$ is equal to unity, with the probability density ${\bar S}_Y(\bm{p},\bm{b})$ for the quark to emerge at the impact parameter $\bm{b}$ with the transverse momentum $\bm{p}$. 

Due to a finite size of the dipole, the differential cross section depends on the azimuthal angle $\phi$ between $\bm{p}$ and $\bm{b}'$, thus being sensitive to the dipole orientation with respect to the target color field. In order to measure an anisotropy in $\phi$, one introduces a set of flow coefficients,
\begin{equation} 
v_n(p_\perp,b'_\perp) \equiv \int_0^{2\pi}\!\!\!\!d\phi \cos(n\phi)\, \frac{\partial\sigma^q(qT\to qX)}{\partial \eta\, \partial^2\bm{p}\, \partial^2\bm{b}'} \Big/
\int_{0}^{2\pi}\!\!\!\!d\phi \,
\frac{\partial \sigma^q(qT\to qX)}{\partial\eta\, \partial^2\bm{p}\, \partial^2\bm{b}'} \,, \quad v_{2} \gg v_{n>2} \,.
\label{vn}
\end{equation}
Using the relation (\ref{qCS}) between the cross section and the dipole $S$-matrix, the dominant (``elliptic'') flow coefficient can be represented as
\begin{eqnarray} \nonumber
v_2(p_\perp,b'_\perp) = -\frac{ \int b_\perp db_\perp d\alpha\,\cos (2\alpha)\,f(|\bm{b} - \bm{b}'|)
\int r_\perp dr_\perp d\theta\,\cos(2\theta)\,
J_2(p_\perp r_\perp)\,S_Y(r_\perp,b_\perp,\theta) } 
{\int b_\perp db_\perp d\alpha\,f(|\bm{b} - \bm{b}'|) 
 \int r_\perp dr_\perp d\theta\,J_0(p_\perp r_\perp) \,S_Y(r_\perp,b_\perp,\theta)}\,, \\
 \label{v2-gen}
\end{eqnarray}
in terms of the dipole $S$-matrix in the impact parameter representation, $S_Y(\bm{r},\bm{b})\equiv S_Y(r_\perp,b_\perp,\theta)$. Here, the angle $\alpha$ is made by the vectors ${\bf b}$ and ${\bf b'}$ while $\theta$ -- by vectors $\bm{r}$ and $\bm{b}$. In the case of a point-like projectile, this expression simplifies to
\begin{eqnarray}
v_2(p_\perp,b_\perp) = -\frac{\int r_\perp dr_\perp d\theta\,\cos(2\theta)\,
J_2(p_\perp r_\perp)\,S_Y(r_\perp,b_\perp,\theta)}{ 
\int r_\perp dr_\perp d\theta\,J_0(p_\perp r_\perp)\,S_Y(r_\perp,b_\perp,\theta)} \,, \label{v2-pointlike}
\end{eqnarray}
thus, the dependence on the projectile quark distribution is conveniently cancelled in the ratio such that the elliptic flow is determined by the elliptic gluon density in the target.

\section{Phenomenological modelling of the dipole $S$-matrix}
\label{Sec:modelling}

Being inspired by intricate connections between the Wigner distribution and the partial dipole amplitude discussed above, let us elaborate on the main aspects and theoretical basics of the dipole formulation of QCD scattering processes. Such processes are considered in the so-called dipole (infinite-momentum) frame sometimes regarded simply as the target rest frame (with the target typically being a nucleon or nucleus) corresponding to ``dilute-dense'' collisions. This is due to the fact that the target gluon density is much higher than that of the projectile hadron as caused by a small-$x$ evolution. The final-state particles in this case originate from a dipole or a system of dipoles and are produced at very forward rapidities. The dipoles consist of fast, (nearly) collinear, partons in a ``dilute'' projectile that undergo multiple scatterings off gluons in a dense target, thus, acquiring some non-zero transverse momenta. In the eikonal approximation, such gluon exchanges can be effectively resummed to all orders in perturbation theory in the impact parameter representation \cite{Pasechnik:2010cm,Pasechnik:2010zs}. At the parton level, such a scattering in the leading order may be considered in perturbative QCD as a probe for very high gluon density in the target -- subject to continued theoretical and experimental investigations.

\subsection{Universality of the dipole description and DIS}
\label{Sec:universality}

The universality of the dipole picture of QCD scattering resides on the basic argument of Ref.~\cite{Kopeliovich:1981pz} suggesting that the color dipoles are eigenstates of QCD interactions. This promotes the dipole scattering matrix as the universal fundamental ingredient of {\it any} QCD scattering process such that any inclusive and diffractive process can be represented as a superposition of ``elementary'' elastic dipole scatterings. Indeed, color dipoles in this approach are considered to experience elastic scattering only and the corresponding partial amplitudes are characterized only by a transverse dipole separation $\bm{r}$ at a fixed small value of Bjorken $x\lesssim 0.01$, where the dipole approach is considered to be valid. For illustration, one often considers a projectile hadron state $|h\rangle$ as a subject of excitations due to QCD interactions that can be decomposed into the orthonormal complete set of eigenstates of interactions $|\alpha\rangle$ labeled by a continuous $\alpha$ variable \cite{Kopeliovich:1978qz,Miettinen:1978jb,Kopeliovich:1981pz}
\begin{eqnarray}
|h\rangle = \sum_\alpha C^h_\alpha\,|\alpha\rangle \,, \qquad 
\hat f_{\rm el}|\alpha\rangle = f_\alpha |\alpha\rangle\,,
\end{eqnarray}
in terms of the universal elastic dipole scattering operator $\hat f_{\rm el}$ and its eigenstates $f_{\alpha}$. Then, elastic $h\to h$ and single diffractive $h\to h'$ scattering amplitudes are represented as, respectively,
\begin{eqnarray}
f^{hh}_{\rm el}=\sum_\alpha |C_\alpha^h|^2\,f_\alpha\,, \qquad 
f^{hh'}_{\rm sd}=\sum_\alpha (C^{h'}_\alpha)^*C^h_\alpha\, f_\alpha \,,
\end{eqnarray}
such that a diffractive excitation is realised as a result of quantum fluctuations in projectile hadron state \cite{Good:1960ba,Miettinen:1978jb}. Finally, the forward single-diffractive and total hadron-proton scattering cross sections read
\begin{eqnarray}
\sum_{h'\not=h}\frac{d\sigma_{\rm sd}^{hp}}{dt}\Big|_{t=0}&=&
\frac{1}{4\pi}\Big[\sum_{h'}|f_{\rm sd}^{hh'}|^2-|f_{\rm el}^{hh}|^2\Big]=
\frac{\langle f_\alpha^2 \rangle_h - \langle f_\alpha \rangle_h^2}{4\pi}=
\frac{\langle \sigma_\alpha^2 \rangle_h - \langle \sigma_\alpha \rangle_h^2}{16\pi} \,, \\
\sigma_{\rm tot}^{hp}&=&\sum_\alpha |C_\alpha^h|^2\sigma_\alpha \equiv
\langle\sigma_\alpha\rangle_h \,,
\end{eqnarray}
respectively, with the former being expressed as a dispersion of the eigenvalues distribution, $\sigma_\alpha$ is the eigenvalue of the latter known as the universal dipole cross section \cite{Kopeliovich:1981pz}. Turning to a continuous set of kinematical variables in the dipole (impact parameter) representation, the phase space of dipole scattering is characterised by intrinsic separation between its constituents (quark and antiquark), $\bm{r}$, such that
\begin{eqnarray}
\sigma_\alpha \to \sigma_{q\bar q}(r_\perp,x) \,, \qquad
\langle\sigma_\alpha\rangle_h &\to& \int d^2\bm{r} \,\int_0^1 \frac{dz}{4\pi} |\Psi_h(r_\perp,z)|^2\sigma_{q\bar q}(r_\perp,x) \,, \label{sig-h} \\
\int d^2\bm{r} \,\int_0^1 \frac{dz}{4\pi} |\Psi_h(r_\perp,z)|^2 &=& 1 \,, \nonumber
\end{eqnarray}
in terms of the normalised $h\to q\bar q$ transition LC wave function $\Psi_h(\bm{r},z)$, with $z$ being the incident quark momentum fraction in the projectile hadron $h$. For small dipoles (compared to the typical hadron size $R_{\rm had}\sim \Lambda_{\rm QCD}^{-1}$), $\sigma_{q\bar q}$ is a function of momentum fraction $x$ of the gluon exchange (i.e. the Bjorken variable of the target gluons). An important example to mention is a heavy system (with total momentum $P$) production in $pp$ collisions such as virtual photon in the Drell-Yan process studied in the dipole picture in Refs.~\cite{Kopeliovich:1999am,Kopeliovich:2006tk,Basso:2015pba} (see also Ref.~\cite{Pasechnik:2015fxa,Kopeliovich:2016rts} and references therein). Another important example is the inclusive heavy flavour production in $pp$ and $pA$ collisions successfully treated in the dipole picture beyond QCD factorisation \cite{Nikolaev:1994de,Kopeliovich:2001ee,Kopeliovich:2002yv,Kopeliovich:2005ym,Goncalves:2017chx}. The exchanged gluon momentum fraction is found in terms of the phase space variables as 
\begin{eqnarray}
x\equiv x_2 = \frac{P^-}{p_2^-}\ll 1 \,, \quad x_1 = \frac{P^+}{p_1^+} \,, \quad x_1x_2=\frac{M^2+k_T^2}{s} \,, \quad x_1 - x_2 \equiv x_F \,,
\label{x12}
\end{eqnarray}
in terms of the $pp$ center-of-mass energy $s$, and $M$, $k_T$ and $x_F$ being the invariant mass, transverse momentum and Feynman variable of the produced system, respectively. In the very high energy limit and for soft dipole scattering the dipole-proton collision center-of-mass energy, $\sqrt{\hat s}$ (with $\hat s=x_1s$), is a more appropriate variable in the dipole cross section than $x$ as advocated e.g.~in Refs.~\cite{Kopeliovich:1999am,Golec-Biernat:1998zce}. Remarkably, it has been demonstrated in Ref.~\cite{Raufeisen:2002zp} that in the case of the Drell-Yan process the dipole model yields the differential cross sections fully consistent with the results of the next-to-leading order (NLO) collinear factorisation framework. This observation provides a solid argument that the dipole approach effectively incorporates the dominant NLO QCD corrections. Besides, as ample literature suggests, the dipole formalism provides a universal way to quantify such effects as saturation, initial state interactions, nuclear coherence as well as the gluon shadowing (see e.g.~Refs.~\cite{Nikolaev:1990ja,Nikolaev:1994de,Brodsky:1996nj,Kopeliovich:1998nw,Kopeliovich:2001ee,Kopeliovich:2002yv,Basso:2016ulb} and references therein).

\begin{figure*}[!h]
\begin{minipage}{1.0\textwidth}
 \centerline{\includegraphics[width=0.7\textwidth]{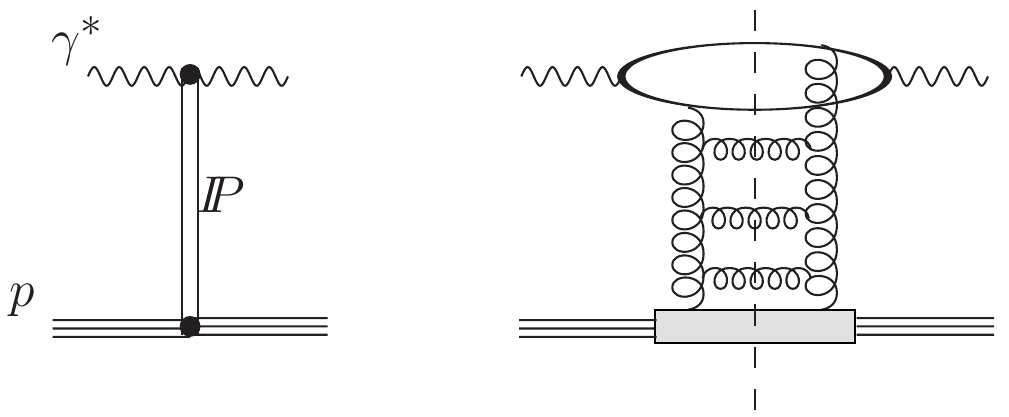}}
\end{minipage}
   \caption{The cross section of the DIS process $\gamma^* p \to X$ in the Regge picture, via a Pomeron exchange (left), and in the dipole picture, via a perturbative QCD gluonic ladder (right). From Ref.~\cite{Pasechnik:2015fxa}.}
 \label{fig:DIS}
\end{figure*}

Let us now consider an important example of the inclusive DIS process in $\gamma^* p$ collisions being one of the most precise and rich source of phenomenological information on the dipole scattering in QCD which historically came about due to HERA measurements \cite{H1:2009pze,H1:2012xnw}. The optical theorem suggests that the DIS cross section is given in terms of the imaginary part of the amplitude of the elastic virtual photon $\gamma^* N \to \gamma^* N$ (with virtuality $q^2=-Q^2$ and with longitudinally (L) and transversely (T) polarised photon, $\lambda_\gamma=L,T$) scattering off the nucleon target $N$ in the forward limit, i.e.
\begin{eqnarray}
  \sigma_{\lambda_\gamma}^{\gamma^*N}(x,Q^2) = 
  \mathrm{Im}\,\mathcal{A}_{\lambda_\gamma}^{\gamma^* N}(x,Q^2,\Delta_\perp=0) \,, \quad \sum_{\lambda_\gamma=L,R}\sigma_{\lambda_\gamma}^{\gamma^*N}(x,Q^2) = \frac{4\pi^2\alpha_{\rm em}}{Q^2}\,F_2(x,Q^2) \,,
  \label{sigma-DIS}
\end{eqnarray}
where the transverse momentum of the outgoing nucleon is denoted by $\bm{\Delta}$, and the DIS structure function -- by $F_2(x,Q^2)$. Then, a standard illustration of the DIS $\gamma^* N \to X$ cross section at small Bjorken $x$ can be seen in Fig.~\ref{fig:DIS} in both Regge (left) and dipole (right) pictures.  Indeed, the virtual photon reveals its partonic structure represented as its lowest Fock component to the leading order $\gamma^* \to q\bar q$ -- the colorless $q\bar q$ dipole characterised by its transverse separation $\bm{r}$ as introduced above.

Such a dipole undergoes scattering off the color field of the target at impact parameter $\bm{b}$ through a small-$x$ color-singlet compound state exchange, often regarded as a Pomeron exchange in Fig.~\ref{fig:DIS} (left) (see e.g.~Refs.~\cite{Levin:1998pk,Donnachie:2002en}, as well as Ref.~\cite{ParticleDataGroup:2022pth} and references therein). In QCD, such a Pomeron is viewed as a colorless multi-gluon exchange in the $t$-channel such as a di-gluon ladder in Fig.~\ref{fig:DIS} (right), thus, connecting to the unintegrated gluon density (the small-$x$ gluon GTMD, in general kinematics) in the target. The corresponding dipole $S$-matrix (and hence the gluon Wigner distribution) carries an important information about the gluon saturation effects relevant at small $x$~\cite{Gelis:2010nm}, as well as the dipole orientation effects due to its interactions with the medium.

In such a dipole (also known as Light-Front) representation, the elastic $\gamma^* N \to \gamma^* N$ amplitude at high energies \cite{Kowalski:2003hm,Kowalski:2006hc},
\begin{eqnarray}
  \mathcal{A}_{\lambda_\gamma}^{\gamma^*N}(x,Q^2,\Delta_\perp) = \sum_f \sum_{\lambda,\bar \lambda} \int d^2\bm{r}\,\int_0^1 \frac{dz}{4\pi}\,(\Psi^{\lambda_\gamma})^\dagger_{\lambda\bar \lambda}(r_\perp,z,Q^2)\,\mathcal{A}_{q\bar q}(x,\bm{r},\bm{\Delta})\,\Psi^{\lambda_\gamma}_{\lambda\bar \lambda}(r_\perp,z,Q^2) \,,\nonumber\\
  \label{gammap-el}
\end{eqnarray}
can be viewed as a convolution (over the dipole separation $\bm{r}$ and a fraction $z$ of the projectile photon LC momentum carried by a quark in the dipole) of three basic ingredients specified as follows. The elastic amplitude of the $q\bar q$ dipole scattering off the proton is denoted as $\mathcal{A}_{q\bar q}(x,\bm{r},\bm{\Delta})$, and the amplitudes (also known as the light-front wave functions) of $\gamma^* \to q\bar q$ and $q\bar q \to \gamma^*$ transitions are labelled as $\Psi^{\lambda_\gamma}_{\lambda\bar \lambda}$ and $(\Psi^{\lambda_\gamma})^\dagger_{\lambda\bar \lambda}$, for $q,\bar q$ helicities $\lambda,\bar \lambda$, respectively (with the quark flavour index $f$). The latter can be found in abundant literature, e.g.~in Refs.~\cite{Bjorken:1970ah,Kowalski:2006hc} (see also Ref.~\cite{Kovchegov.2012}). Then, combining Eq.~(\ref{sigma-DIS}) with Eq.~(\ref{gammap-el}) one writes the DIS cross section (c.f.~Eq.~(\ref{sig-h})),
\begin{eqnarray}
  \sigma_{\lambda_\gamma}^{\gamma^* N}(x,Q^2) = \sum_f \int d^2\bm{r} \int_0^1 \frac{dz}{4\pi}
  (\Psi^\dagger\Psi)_{\lambda_\gamma}^f \, \sigma_{q\bar q}(r_\perp,x) \,.
  \label{sigma-DIS-dipoleCS}
\end{eqnarray}
Such a convolution of the kinematical dependence of the LC wave function with the dipole cross section gives rise to the mean dipole size squared
\begin{eqnarray}
\langle r_\perp^2 \rangle \sim \frac{1}{\epsilon^2}=\frac{1}{Q^2z(1-z)+m_q^2} \,,
\end{eqnarray}
determined as an inverse to the quark energy squared, $\epsilon^2$, depending on the photon virtuality, $Q^2$, momentum sharing within the dipole, $z$, and a constituent quark mass, $m_q\sim \Lambda_{\rm QCD}$. The dipole size $r\sim 1/Q$ is then considered to be ``frozen'' in the target rest frame in the course of dipole propagation through the color background field of the target. Analogically, one gets the following expression for diffractive DIS $\gamma^*N\to XN$ cross section in the forward limit
\begin{eqnarray}
16\pi\, \frac{d\sigma_{\rm sd}^{\gamma^*N}(x,Q^2)}{dt}\Big|_{t=0}=
\int d^2 \bm{r} \int_0^1 \frac{dz}{4\pi}\, (\Psi^\dagger\Psi)_{\lambda_\gamma}^f \,\sigma_{q\bar q}^2(r_\perp,x)\,, \quad
t \equiv \Delta_\perp^2 \,.
\end{eqnarray}
The above well-known results of the dipole model are expressed in terms of the universal dipole cross section which, by means of the optical theorem, is equal to the imaginary part of the amplitude for the elastic $q\bar q$ dipole scattering off the proton, i.e. 
\begin{eqnarray}
  \sigma_{q\bar q}(r_\perp,x) \equiv \mathrm{Im}\,\mathcal{A}_{q\bar q}(x,r_\perp,\Delta_\perp=0) \,.
  \label{sigma-qq}
\end{eqnarray}
The latter is determined by the dipole $S$-matrix in the impact parameter representation, $S_Y(\bm{r},\bm{b})$, introduced above
\begin{eqnarray}
\mathcal{A}_{q\bar q}(x,\bm{r},\bm{\Delta}) = \int d^2\bm{b} e^{-i\bm{b}\cdot\bm{\Delta}}\,\mathcal{A}_{q\bar q}(x,\bm{r},\bm{b}) = i\,\int d^2\bm{b}e^{-i\bm{b}\cdot\bm{\Delta}}\,2\left[1-S_Y(\bm{r},\bm{b})\right] \,,
\label{Aqq-SY}
\end{eqnarray}
yielding finally the universal dipole cross section in the form
\begin{eqnarray}
    \sigma_{q\bar q}(r_\perp,x) = 2\int d^2\bm{b}\, {\rm Im} f^N_{q \bar q}(x,\bm{r},\bm{b}) \,,
\label{sigma-qqbar_universal}
\end{eqnarray}
where
\begin{eqnarray}
    {\rm Im} f^N_{q \bar q}(x,\bm{r},\bm{b}) \equiv {\cal N}_Y (\bm{r}, \bm{b}) = 1 - {\rm Re}\,S_Y(\bm{r},\bm{b})\,,
\label{Imfel-qq}
\end{eqnarray}
is known in the literature as the partial (elastic) dipole amplitude off a nucleon (e.g.~proton) target $N$ \cite{Bartels:2003yj}. Then, one deduces the dipole $T$-matrix in the Glauber approximation for a nucleus with mass number $A$ ($A=1$ for a nucleon),
\begin{equation}
T_Y(\bm{r}, \bm{b}) = 1 - e^{-A\,{\cal N}_Y (\bm{r}, \bm{b})} \,.
\end{equation}
For a given interaction potential of a quark with the medium $V(\bm b,z)$, this amplitude adopts an eikonal form, nearly imaginary at high energies \cite{Pasechnik:2012ac}. In analogy to Eq.~(\ref{GTMD-decomposition}), the Fourier expansion of the partial dipole amplitude reads
\begin{eqnarray}
{\cal N}_Y(\bm{r}, \bm{b}) = {\cal N}_0(Y, r_\perp, b_\perp) + 2 \, \cos (2 \phi_{br}) \, {\cal N}_\epsilon(Y,r_\perp, b_\perp) + \dots \,, \label{N-Y_elliptic}
\end{eqnarray}
where the elliptic part ${\cal N}_\epsilon$ quantifies the dipole orientation effect leading to a non-trivial correlation in azimuthal angle $\phi_{br}$ between the dipole separation $\bm{r}$ and its impact parameter $\bm{b}$.

For small dipoles, i.e.~when the dipole separation $r_\perp$ is small compared to the hadron scale $r_\perp\ll R_{\rm had}$, the universal dipole cross section $\sigma_{q\bar q}$ realises the color transparency property vanishing quadratically with separation i.e.~$\sigma_{q\bar q}(r_\perp,x)\propto r_\perp^2$ \cite{Kopeliovich:1981pz,Bertsch:1981py,Brodsky:1988xz,Nikolaev:1990ja,Nikolaev:1991et}. This is an important consequence of the gauge invariance property of the fundamental QCD theory as well as nonabeliance of strong interactions. Indeed, a colorless object of vanishing size propagating through the target does not interact with its color field, i.e. the color medium is transparent for a point-like color-neutral object. For large dipole sizes, instead, the dipole cross section levels off at the so-called saturation scale $Q_s^2$ which depends either on Bjorken $x$ or on total energy $\sqrt{\hat{s}}$ in the $\gamma^* p$ c.m.~frame. 

To summarise this part, the universality of the dipole cross section (hence, the dipole $S$-matrix, $S_Y$, or the related partial dipole amplitude, ${\cal N}_Y$) is its most important property making it the main ingredient of a given QCD scattering process, for both nucleon and nuclear targets. Such a colour dipole framework has enjoyed a lot of developments over last few decades, both in theoretical modelling and in phenomenological approaches. Equipped with the universal dipole picture of QCD scattering, one can utilize the same fundamental ingredient, $S_Y(\bm{r},\bm{b})$, as a building block for a vast majority of high-energy scattering phenomena covering the particle production reactions in nucleon-nucleon $NN$, lepton-nucleon $lN$ (such as the DIS), lepton-nucleus $lA$, nucleon-nucleus $NA$, as well as heavy nuclei $AA$ collisions, in case of both inclusive and diffractive topologies of final states. On the other hand, an intricate connection between the gluon Wigner distribution and the dipole scattering enables us to probe factorisation mechanisms at small $x$ which are being actively developed in the literature over past decades.

\subsection{From impact parameter to transverse momentum representation}
\label{sec:transverse-mom}

It is often more convenient and instructive to turn to momentum-space representation for the dipole amplitude in studies of the diffractive reactions as was implemented, for instance, in Refs.~\cite{Hagiwara:2016kam,Hagiwara:2017fye,Goncalves:2017chx,ReinkePelicer:2018gyh}. Let us now establish an important connection between the dipole cross section in the impact parameter representation with the intrinsic dipole TMD in the momentum representation, ${\cal K}_{\rm dip}(x,\kappa_\perp^2)$. Following Refs.~\cite{Nikolaev:1990ja,Bartels:2002cj}, one defines
\begin{eqnarray}
\sigma_{q\bar q}(r_\perp,x) \equiv \frac{2\pi}{N_c}\int \frac{d^2\bm{\kappa}}{\kappa_\perp^4}\,
(1-e^{i\bm{\kappa}\cdot \bm{r}})(1-e^{-i\bm{\kappa}\cdot \bm{r}})\,
{\cal K}_{\rm dip}(x,\kappa_\perp^2) \,,
\label{dipole-TMD}
\end{eqnarray}
such that for large dipoles the cross section saturates to the non-perturbative quantity,
\begin{eqnarray}
    \sigma_0(x) \equiv \frac{4\pi}{N_c} \int \frac{d^2 \bm{\kappa}}{\kappa_\perp^4} {\cal K}_{\rm dip}(x,\kappa_\perp^2) \,.
    \label{sigma0}
\end{eqnarray}
The corresponding partial dipole amplitude ${\cal N}_Y$ defined in Eqs.~(\ref{sigma-qqbar_universal}), (\ref{Imfel-qq}) is connected to an unintegrated (off-forward) gluon density matrix \cite{Nemchik:1997xb} -- an off-diagonal generalization of the unintegrated gluon distribution function (UGDF) $f(Y,\bm{\kappa}_1,\bm{\kappa}_2)$ -- as follows:
\begin{eqnarray}
{\cal N}_Y(\bm{r}, \bm{b}) &=& \int d^2\bm{q} d^2\bm{\kappa} \, f\Big(Y,\frac{\bm{q}}{2} + \bm{\kappa}, \frac{\bm{q}}{2} - \bm{\kappa}\Big) e^{i \bm{q} \cdot \bm{b}}  \nonumber \\
&\times&\Big\{ e^{\frac{i}{2} \bm{q} \cdot \bm{r}} + e^{-\frac{i}{2} \bm{q} \cdot \bm{r}} - e^{i \bm{\kappa} \cdot \bm{r}} - e^{-i \bm{\kappa} \cdot \bm{r}} \Big\} \,.
\label{dipole_N-Y}
\end{eqnarray}
In the two-gluon approximation, each of the phase factors can be straightforwardly related to each of the four possible ways the two gluons can couple to the $q\bar q$ dipole, whose constituents propagate at impact parameters
\begin{eqnarray}
\bm{b}_q = \bm{b} + \frac{\bm{r}}{2} \,, \qquad 
\bm{b}_{\bar q} = \bm{b} - \frac{\bm{r}}{2} \,, \label{qqbar-impact}
\end{eqnarray}
respectively, while the $t$-channel gluons carry the transverse momenta
\begin{eqnarray}
\bm{\kappa}_1 = \frac{\bm{q}}{2} + \bm{\kappa} \,, \qquad 
\bm{\kappa}_2 = \frac{\bm{q}}{2} - \bm{\kappa} \,.
\end{eqnarray}
The genuine gluon GTMD (dipole $T$-matrix in momentum space) can then be connected to the off-forward gluon density matrix as \cite{Linek:2023kga}
\begin{eqnarray}
T_Y(\bm{k},\bm{\Delta}) &\equiv& \int \frac{d^2\bm{b}}{(2\pi)^2}
\frac{d^2\bm{r}}{(2\pi)^2} \, e^{-i\bm{\Delta} \cdot \bm{b}} \, 
e^{-i\bm{k}\cdot\bm{r}}\,{\cal N}_Y(\bm{r}, \bm{b}) = C_Y(\bm{\Delta})\left(\delta^{(2)}\Big(\bm{k} -\frac{\bm{\Delta}}{2}\Big) + 
\delta^{(2)}\Big(\bm{k} + \frac{\bm{\Delta}}{2}\Big) \right) \nonumber \\
&& - f\Big(Y,\frac{\bm{\Delta}}{2} + \bm{k}, \frac{\bm{\Delta}}{2} - \vec k_\perp\Big) 
- f\Big(Y,\frac{\bm{\Delta}}{2} - \bm{k}, \frac{\bm{\Delta}}{2} + \bm{k}\Big) \, ,
\label{T_Y-f}
\end{eqnarray}
where
\begin{eqnarray}
C_Y(\bm{\Delta}) = \int \frac{d^2\bm{k}}{2 \pi} \,f\Big(Y,\frac{\bm{\Delta}}{2} + \bm{k}, \frac{\bm{\Delta}}{2} - \bm{k}\Big) \,. 
\label{C-Y}
\end{eqnarray}
In the perturbative regime, $k_\perp \gg \Delta_\perp$, the $\delta$-functions do not play any role, so up to a factor of two $T_Y$ is the same as $f$. The color transparency property of the dipole amplitude, ${\cal N}_Y(\bm{r}, \bm{b})\to 0$ as $\bm{r}\to 0$, implies
\begin{eqnarray}
\int d^2 \bm{k}\, T_Y(\bm{k}, \bm{\Delta}) = 0 \,. \label{sumrule}
\end{eqnarray}
As was pointed out in Ref.~\cite{Linek:2023kga}, the emergence of $\delta$-functions in Eq.~(\ref{T_Y-f}) means that the Fourier integral does not converge, and hence a regularisation procedure should be advised e.g.~via introducing a Gaussian cutoff factor $e^{-\varepsilon r_\perp^2}$ preserving the color transparency \cite{Hagiwara:2016kam,Hagiwara:2017fye,ReinkePelicer:2018gyh}. Then, starting from Eq.~(\ref{N-Y_elliptic}) and utilising the Fourier-Bessel transforms one obtains the regularised isotropic and elliptic components of the gluon GTMD as \cite{Hatta:2016dxp,Hagiwara:2017fye,ReinkePelicer:2018gyh}
\begin{eqnarray}
T_0(Y,k_\perp,\Delta_\perp) &=& \frac{1}{4\pi^{2}} \int_0^\infty b_\perp db_\perp J_0(\Delta_\perp b_\perp) \int_0^\infty r_\perp dr _\perp   J_0(k_\perp r_\perp) \, {\cal N}_0(Y,r_\perp, b_\perp) e^{- \varepsilon r_\perp^2}\,, \label{eq_T-0} \\
T_\epsilon(Y,k_\perp,\Delta_\perp) &=& \frac{1}{4\pi^{2}}  \int_0^\infty b_\perp db_\perp J_2(\Delta_\perp b_\perp) \int_0^\infty r_\perp dr _\perp   J_2(k_\perp r_\perp) \, {\cal N}_\epsilon(Y,r_\perp, b_\perp) e^{-\varepsilon r_\perp^2} \,.
\label{eq_T-eps}
\end{eqnarray}

Note that the density matrix can also be recast as
\begin{eqnarray}
 f(Y,\bm{\kappa}_1,\bm{\kappa}_2) = \frac{\alpha_s}{4 \pi N_c} \, \frac{{\mathcal F}(x, \bm{\kappa}_1, \bm{\kappa}_2)}
 {\kappa_{1\perp}^2 \kappa_{2\perp}^2} \,.
 \label{f_F}
\end{eqnarray}
Then, in the perturbative QCD regime corresponding to a hard gluon transverse momentum in the target, $\kappa_\perp \gg \Lambda_{\rm QCD}$, the gluon TMD ${\cal K}_{\rm dip}$ defined in Eq.~(\ref{dipole-TMD}) approximates the diagonal (forward) UGDF ${\cal F}(x,\kappa_\perp^2)$, the standard ingredient of $k_\perp$-factorisation, times $\alpha_s$, i.e.
\begin{eqnarray} 
{\cal K}_{\rm dip}(x,\kappa_\perp^2) = \alpha_s\,{\cal F}(x,\kappa_\perp^2) \,, \qquad {\cal F}(x,\kappa_\perp^2) \equiv {\mathcal F}(x, \bm{\kappa}, -\bm{\kappa}) \,.
\label{dip-vs-kt}
\end{eqnarray}
In the forward limit, one recovers
\begin{eqnarray}
C_Y(\bm{\Delta}\to 0) \to \frac{\sigma_0(x)}{32\pi^3} \,.
\end{eqnarray}
In fact, integrating the dipole TMD over $\bm{\kappa}$ and considering the perturbative limit of hard $Q^2$ scales, one recovers in the double logarithmic approximation of the DGLAP equations \cite{Bartels:2002cj}
\begin{eqnarray}
\frac{1}{\pi}\int^{Q^2} \frac{d^2\bm{\kappa}}{\kappa_\perp^2}\,{\cal K}_{\rm dip}(x,\kappa_\perp^2)=\alpha_s(Q^2)xg(x,Q^2) \,,
\label{dip-vs-kt-integrated}
\end{eqnarray}
thus, establishing an important connection between the dipole formalism and $k_\perp$-factorisation approach. Indeed, it has been known for a long time that the forward cross section of the diffractive dissociation of projectile photons (or pions) into di-jets at a perturbatively large jet $p_T$ is directly mapped on to the target UGDF (see e.g.~Refs.~\cite{Nikolaev:1994cd,Nikolaev:2000sh}). Note, however, that Eqs.~(\ref{dip-vs-kt}) and (\ref{dip-vs-kt-integrated}) are approximate and limited to small $x$ and small dipoles (realised e.g.~for heavy quarks or high-$p_T$ jets production) only. In the soft regime corresponding to large dipole sizes where the saturation and other non-linear QCD effects become relevant, the use of conventional UGDF and the $k_\perp$-factorisation formula are not well justified, and the use of phenomenological dipole approach going beyond $k_\perp$-factorisation is more sensible due to its universality.

Following Ref.~\cite{Bartels:2002cj}, one may formally {\it define} a two-gluon amplitude in momentum space beyond $k_\perp$-factorisation -- the dipole TMD -- in terms of a known parameterisation of the universal dipole cross section, basically, by inverting Eq.~(\ref{dipole-TMD}) yielding
\begin{eqnarray}
\frac{1}{\kappa_\perp^4}{\cal K}_{\rm dip}(x,\kappa_\perp^2)=\frac{3}{8\pi^2}\int^\infty_0 dr_\perp\,r_\perp\,J_0(\kappa_\perp r_\perp)\, 
\Big[ \sigma_{q\bar q}^\infty(x) - \sigma_{q\bar q}(r_\perp,x) \Big] \,,
\label{dipole-UGDF}
\end{eqnarray}
in terms of the Bessel function of the first kind, $J_0(x)$, and $\sigma_{q\bar q}^\infty(x) \equiv {\rm lim}_{r_\perp\to\infty}\sigma_{q\bar q}(r_\perp,x)$. As was elaborated in detail in Ref.~\cite{Goncalves:2017chx}, in the framework of dipole approach the dipole TMD is a very convenient object to use for practical calculations directly in momentum representation.

\subsection{Applied dipole parameterisations}
\label{sec:sigma_qq-models}

A first-principle computation for the dipole cross section (and the corresponding dipole $S$-matrix) remains a major theoretical challenge and its exact theoretical modelling is still far from complete understanding despite of a three-decade long community effort. A remarkable progress has been made on reconstruction of kinematic and energy dependence of ${\cal N}_Y$ strongly boosted by a vast amount of precision experimental data from the HERA collider. Among important QCD-based advancements, the rapidity (or energy) evolution at high energies can be understood in the CGC formalism \cite{McLerran:1993ka,McLerran:1994vd} and is encoded in the Renormalization Group equations represented by an infinite hierarchy of the so-called Balitsky-Jalilian-Marian–Iancu–McLerran–Weigert–Leonidov–Kovner (Balitsky-JIMWLK) equations \cite{Balitsky:1995ub,Jalilian-Marian:1996mkd,Jalilian-Marian:1997qno,Jalilian-Marian:1997jhx,Jalilian-Marian:1997ubg,Balitsky:1998kc,Kovner:2000pt,Weigert:2000gi,Iancu:2000hn,Iancu:2001md,Ferreiro:2001qy,Iancu:2001ad}. The latter determines how the separation of hard and soft modes evolves towards smaller momentum fractions effectively resuming terms enhanced by large $\ln(1/x)$ in leading logarithmic accuracy. In the mean-field approximation, these equations yield the Balitsky-Kovchegov (BK) equation \cite{Balitsky:1995ub, Kovchegov:1999yj} whose $b_\perp$-dependent solutions for the partial dipole amplitude ${\cal N}_Y (\bm{r}, \bm{b})$ are very difficult to obtain \cite{Golec-Biernat:2003naj}. As the $b_\perp$-slope is driven essentially by non-perturbative QCD effects, one often looks for translationally invariant solutions ignoring the $b_\perp$-dependence in numerical solutions and then introducing it ``by hands'' through an extra soft exponential factor to match the observed $t$-dependence of differential observables for exclusive processes.

The universality of the dipole scattering discussed above enables one to extract it from one process (typically, inclusive DIS) and then to use it in many other hadronic and photon-induced processes in $p/l+p$, $p/l+A$ and $AA$ collisions. Starting from the fundamental work of Ref.~\cite{Mueller:1989st}, many different phenomenological dipole parameterisations for the partial dipole amplitude have been advised in the literature attempting to account for the saturation phenomena, the QCD-inspired hard-scale and Bjorken $x$ evolution, as well as for impact parameter dependence and the dipole orientation effects. For some of the widely used parameterisations relying on fits to the HERA data \cite{H1:2009pze,H1:2012xnw}, see e.g.~Refs.~\cite{Golec-Biernat:1998zce,Bartels:2002cj,Kowalski:2003hm,Iancu:2003ge,Kharzeev:2004yx,Dumitru:2005gt,deSantanaAmaral:2006fe,Kowalski:2006hc,Goncalves:2006yt,Boer:2007ug,Watt:2007nr,Soyez:2007kg,Basso:2012nb,Rezaeian:2012ji,Rezaeian:2013tka}. 

One of the simplest and well-known phenomenological parametrizations of the saturated dipole cross section satisfying the color transparency property,
\begin{eqnarray}
&& \sigma_{q\bar q}(r_\perp,x) = \sigma_0\left(1-e^{-r_\perp^2/R_0^2(x)}\right)\,, \quad \sigma_0\simeq 23.03\,{\rm mb}\,, \nonumber \\
&& R_0(x) \simeq 0.4\,{\rm fm}\times\left(\frac{x}{x_0}\right)^{0.144} \,, \quad x_0\simeq 3.04\times 10^{-4} \,,
\label{GBW-dipole}
\end{eqnarray}
represents the simplest and phenomenologically consistent way to incorporate the saturation effects and provides a very good description of a number of different observables in $ep$ and $pp$ collisions with both inclusive and exclusive topologies of final states at high energies (with $x\lesssim 0.01$), as well as for processes with nuclear targets. Such a simple parametrization reminds the Glauber model of multiple interactions and is known in the literature as the Golec-Biernat--Wuestoff (GBW) model \cite{Golec-Biernat:1998zce,Golec-Biernat:1999qor}. The GBW parameter values in Eq.~(\ref{GBW-dipole}) are found by a fit to the inclusive $F_2$ data from the HERA collider \cite{H1:2009pze,H1:2012xnw} (initially, without accounting for charm quarks) such that the model is found to simultaneously describe the diffractive DIS data (and is denoted in the literature as ``GBWold'' parameterization). The saturation scale related to the gluon density in the target nucleon is then conveniently given by
\begin{eqnarray}
    Q^2_s(x) \equiv \frac{4}{R^2_0(x)} = \left(\frac{x_0}{x}\right)^{\lambda_{\rm GBW}}\;{\rm GeV}^2 \,, \quad \lambda_{\rm GBW} \simeq 0.288 \,.
\end{eqnarray}
Incorporating the charm quark contribution (with $m_c=1.5$ GeV) in the fit yields somewhat different parameter values: $\sigma_0\simeq 29$ mb, $x_0\simeq 4.01\times 10^{-5}$ and $\lambda_{\rm GBW}\simeq 0.277$ (``GBWnew'' parameterization). It is rather surprising that such a simple phenomenologically inspired parametrization, well justified for small dipoles only, has turned out to be surprisingly successful in giving an acceptable description of HERA data \cite{H1:2009pze,H1:2012xnw} in a wide range of $Q^2\sim [0.1\dots 30]$ GeV$^2$ for both inclusive and diffractive DIS, as well as for a large variety of hard and semi-hard inclusive and diffractive processes in $pp$, $pA$ and $AA$ collisions (see e.g.~Ref.~\cite{Kowalski:2006hc} and references therein).

For sufficiently hard scales $\mu^2=\mu^2(r_\perp)$, it was understood in Refs.~\cite{Blaettel:1993rd,Frankfurt:1993it,Frankfurt:1996ri} that the saturation scale $Q_s(x)$ must depend on the hard scale $\mu^2$ as a manifestation of the QCD evolution and the corresponding scaling violation. Connecting the universal dipole cross section to the leading-order (LO) DGLAP-evolved gluon density in the target, one writes \cite{Nikolaev:1994cn}
\begin{eqnarray}
\sigma_{q\bar q} \simeq \frac{\pi^2}{3}\alpha_s\Big(\frac{\Lambda}{r_\perp^2}\Big)\,r_\perp^2\,xg\Big(x,\frac{\Lambda}{r_\perp^2}\Big)\,, \qquad \Lambda\approx 10 \,.
\end{eqnarray}
A particular realisation of the phenomenological dipole model, that explicitly incorporates the QCD DGLAP evolution of the collinear gluon density in the target $xg(x,\mu^2(r_\perp))$ in the saturation scale,
\begin{eqnarray}
Q_s^2 = Q_s^2(x,\mu^2) \equiv \frac{4\pi^2}{\sigma_0 N_c}\, \alpha_s(\mu^2)\,xg(x,\mu^2) \,, \qquad \mu^2=\frac{{\cal C}}{r_\perp^2} + \mu_0^2 \,,
\label{bgbk}
\end{eqnarray}
and utilises the same GBW saturated ansatz of Eq.~(\ref{GBW-dipole}), is known in the literature as the Bartels--Golec-Biernat--Kowalski (BGBK) model \cite{Bartels:2002cj}. The target gluon density is found as a solution of the DGLAP evolution equation accounting for the gluon splitting function $P_{gg}(z)$ only, with the following parameterisation for the starting gluon distribution at $\mu^2=\mu_0^2$:
\begin{eqnarray}
xg(x,\mu_0^2) = A_g x^{-\lambda_g} (1-x)^{5.6}\,, \quad \frac{\partial xg(x,\mu^2)}{\partial \ln \mu^2 } = 
\frac{\alpha_s(\mu^2)}{2\pi} \int_x^1 dz\, P_{gg}(z) \frac{x}{z} g\Big(\frac{x}{z}, \mu^2\Big)\,.
\label{dglap}
\end{eqnarray}
The BGBK model parameters,
\begin{eqnarray}
A_g = 1.2\,, \quad \lambda_g = 0.28\,, \quad \mu_0^2 = 0.52\, \mathrm{GeV}^2\,, \quad 
{\cal C} = 0.26\,, \quad \sigma_0 = 23\, \mathrm{mb}\,,
\end{eqnarray}
were extracted through fitting $F_2$ to the HERA data \cite{H1:2009pze,H1:2012xnw} yielding much better fits than the original GBW model, particularly, for large $Q^2$ (although omitting the charm quarks contribution). 

A generalisation of the BGBK model incorporating the $b_\perp$-dependence of the partial dipole amplitude relevant for a consistent description of the $t$-dependent observables in exclusive processes at HERA has been proposed in Refs.~\cite{Kowalski:2003hm,Kowalski:2006hc,Watt:2007nr} and reads
\begin{eqnarray}
\sigma_{q\bar q}(r_\perp,x) &=& 
2\,\int d^2\bm{b}\, 
\left[1-\exp\left(-\frac{\pi^2}{2 N_c} r_\perp^2 \alpha_s(\mu^2) xg(x,\mu^2) 
T_{\rm G}(\bm{b})\right)\right] \,, \\ 
T_{\rm G}(\bm{b}) &\equiv& T_{\rm G}(b_\perp) = 
\frac{1}{2\pi B_{\rm G}}\, e^{-\frac{b_\perp^2}{2 B_{\rm G}}} \,, \label{D-slope}
\end{eqnarray} 
where $B_{\rm G}$ is an additional exponential $b_\perp$-slope parameter that can be extracted e.g.~from the data on $t$-dependent differential elastic $ep$ scattering. 
Hence, this so-called Impact-Parameter dependent Saturation (IP-Sat) model -- also known as Kowalski-Teaney (KT) model -- does not contain any non-trivial correlation in the relative angle between $\bm{r}$ and $\bm{b}$. The IP-Sat parameters found in Ref.~\cite{Rezaeian:2012ji} by a fit to the precision (both, elastic and inelastic) $ep$ HERA data \cite{H1:2009pze,H1:2012xnw} are given by
\begin{eqnarray}
A_g = 2.373\,, \quad \lambda_g = 0.052\,, \quad \mu_0^2 = 1.428\; {\rm GeV}^2\,, \quad B_{\rm G} = 4.0\; {\rm GeV}^2 \,, \quad {\cal C} = 4.0 \,.
\end{eqnarray}

Turning to momentum-space representation and assuming that $Q_s^2(x,\mu^2(r_\perp))$ is either independent of $r_\perp$ (as in GBW model) or depends on $r_\perp$ slowly (e.g.~logarithmically as in the case of BGBK or KT models) the saturated antsatz of Eq.~(\ref{GBW-dipole}) in Eq.~(\ref{dipole-UGDF}) leads to the dipole TMD in a nearly-Gaussian form
\begin{eqnarray}
\label{dugdf}
 {\cal K}_{\rm dip}(x,\kappa_\perp^2) \simeq  \frac{3 \sigma_0}{4 \pi^2}\frac{\kappa_\perp^4}{Q_s^2(x,\mu^2)} \,
 e^{-\frac{\kappa_\perp^2}{Q_s^2(x,\mu^2)}}\,, \qquad \mu^2\equiv \mu^2(r_\perp=1/\kappa_\perp)\,.
\end{eqnarray}
and hence to the diagonal ``dipole UGDF'' ${\cal F}(x,k_\perp^2)$ via Eq.~(\ref{dip-vs-kt}). Then, accounting for the diffractive slope the off-forward gluon density matrix defined in Eq.~(\ref{f_F}) can be found as
\begin{eqnarray}
f\Big(Y,\frac{\bm{\Delta}}{2} + \bm{k}, \frac{\bm{\Delta}}{2} - \bm{k}\Big) = \frac{\alpha_s}{4 \pi N_c} \, \frac{{\cal F}(x, k_\perp^2)}{k_\perp^4} \, e^{- \frac12 B_{\rm G} \Delta_\perp^2 }\,, 
     \label{f-exp}
\end{eqnarray}
explicitly incorporating the exponential dependence on $\Delta_\perp^2\equiv -t$ (with the slope $B_{\rm G} = 4.0\, {\rm GeV}^2$) in a factorised form. The latter has been used in the analysis of diffractive vector meson production in Ref.~\cite{Ivanov:2004ax} (see also Refs.~\cite{Kopeliovich:2007fv,Cisek:2022yjj}).

Imposing translational invariance and, hence, ignoring the $b_\perp$-dependence of the partial dipole amplitude ${\cal N}_Y (\bm{r}, \bm{b})\to {\cal N} (\bm{r}, x)$, one often looks for the corresponding solution of the BK equation describing its evolution in rapidity $Y\equiv \ln(x_0/x)$ as formulated in Refs.~\cite{Albacete:2007yr,Albacete:2009fh,Albacete:2010sy}. 
\begin{equation*}
\frac{\partial {\cal N}(\bm{r},x)}{\partial Y}=\int d^2 \bm{r}_1 K(\bm{r},\bm{r}_1,\bm{r}_2)\Bigg({\cal N}(\bm{r}_1,x)+{\cal N}(\bm{r}_2,x)-{\cal N}(\bm{r},x)-{\cal N}(\bm{r}_1,x){\cal N}(\bm{r}_2,x)\Bigg)\,,
\end{equation*}
where $\bm{r}_2=\bm{r}-\bm{r}_1$. Including the Renormalization Group running of the strong coupling, the BK kernel reads \cite{Albacete:2010sy}
\begin{equation}
K(\bm{r},\bm{r}_1,\bm{r}_2)=\frac{\alpha_s(r_\perp^2) N_c}{2\pi^{2}}\Bigg(\frac{r_\perp^2}{r_{1\perp }^2 r_{2\perp }^2}+\frac{1}{r_{ 1\perp }^2}\left(\frac{\alpha_s(r_{1\perp }^2)}{\alpha_s(r_{2\perp }^2)}-1\right)+\frac{1}{r_{2\perp }^2}\left(\frac{\alpha_s(r_{2\perp }^2)}{\alpha_s(r_{1\perp }^2)}-1\right)\Bigg) \,,
\end{equation}
with
\begin{equation}
\alpha_{s}(r_{\perp}^2)=\frac{4\pi}{(11-\frac{2}{3}N_f)\ln\left(\frac{4C^2}{r_{\perp}^2\Lambda^2_{\rm QCD}}\right)}\,,
\end{equation}
in terms of the number of active quark flavours $N_f$ and an adjustable parameter $C$. In this ``rcBK'' formulation, the initial $r_\perp$-shape of the dipole amplitude at the starting value $x=0.01$ is chosen to take a McLerran-Venugopalan (MV) \cite{McLerran:1993ka,McLerran:1997fk} form,
\begin{equation}
{\cal N}(\bm{r},x=0.01)=1-\exp\Bigg(-\frac{\left(r_{\perp}^2Q^2_{s0}\right)^\gamma}{4} \ln \left(\frac{1}{r_{\perp}\Lambda_{\rm QCD}}+{\rm e}\right)\Bigg) \,,
\end{equation}
with $ \Lambda_{\rm QCD}=0.241$ MeV, $Q^2_{s0} = 0.165$ GeV$^2$, $\gamma=1.135$ and $C = 2.52$ extracted from the fit yielding $\sigma_0=32.895$ mb and assuming that $\alpha_s(r_{\perp}^2)$ freezes at $r_{\perp}>r_{0\perp}$, determined by $\alpha_s(r^2_{0\perp})=0.7$ \cite{Albacete:2010sy}.

A further variation of the dipole model can be obtained utilising the collinearly improved kernel \cite{Iancu:2015joa} given by,
\begin{equation}
K(\bm{r},\bm{r}_1,\bm{r}_2)=\frac{\bar \alpha_s N_c}{2\pi^2}\Bigg(\frac{r_{\perp}^2}{r_{1\perp}^2 r_{2\perp}^2}\left(\frac{r_{\perp}^2}{\mathrm{min}(r_{1\perp}^2,r_{2\perp}^2)}\right)^{\pm\bar\alpha_s A_1}\frac{J_1(2\sqrt{\bar\alpha_s |\ln(r_{1\perp}^2/r_{\perp}^2)\ln(r_{2\perp }^2/r_{\perp}^2)|})}{\sqrt{\bar\alpha_s |\ln(r_{1\perp}^2/r_{\perp}^2)\ln(r_{2\perp }^2/r_{\perp}^2)|}}\Bigg) \,.
\end{equation}
Here, the parameter $A_1$ is fixed to $11/12$, while the positive sign corresponds to $r_{\perp} < \mathrm{min}(r_{1\perp},r_{2\perp}) $, and the definition $\bar\alpha_s\equiv \alpha_s(\mathrm{min}(r_{\perp}^2,r_{1\perp}^2,r_{2\perp}^2)) \frac{N_c}{\pi}$ is introduced. Then, variable number of flavours scheme is applied \cite{Albacete:2010sy} such that $\Lambda_{\rm QCD}$ becomes flavour-dependent and is computed using
\begin{equation}
\Lambda_{N_f-1}=m_f^{1-\frac{\beta_{N_f}}{\beta_{N_f-1}}}\Lambda_{N_f}^{\frac{\beta_{N_f}}{\beta_{N_f-1}}} \,.
\end{equation}
with the quark mass $m_f$ of flavour $f$, and $\beta_{N_f}=(11N_c-2N_f)/3$. Starting with the measured value for the strong coupling $\alpha_s(r_{\perp}^2\equiv 4C^2/M_Z^2)=0.1189$ at the $Z^0$-boson mass scale $M_Z=91.187$ GeV (corresponding to $N_f=5$) as the initial scale, one finds
\begin{equation}
\Lambda_5 = M_Z\,e^{-\frac{2\pi}{\alpha_s(r_{\perp}^2 = 4C^2/M_Z^2)\beta_5}} \,.
\end{equation}
Similarly to the rcBK model, one employs the MV initial condition which in the collinear formulation of Ref.~\cite{Iancu:2015joa} takes the form
 \begin{equation}
{\cal N}(\bm{r},x=0.01) = \left[1 - \exp \left(-\left[\frac{r_{\perp}^2Q^2_{s0}}{4} \bar\alpha_s(r_{\perp}^2) 
\left(1 + \ln \left(\frac{\alpha_{\rm sat}}{\bar\alpha_s(r_{\perp}^2)}\right)\right)\right]^p\right)\right]^{1/p} \,,
\end{equation}
with the definition $\bar\alpha_{\rm sat}\equiv\frac{N_c}{\pi}\alpha_{\rm sat}\,, \bar\alpha_s(r_{\perp}^2)\equiv\frac{N_c}{\pi}\alpha_s(r_{\perp}^2)$, where $\alpha_{\rm sat}$ can be fixed to unity. Fitting the normalisation of the corresponding dipole cross section $\sigma_{q\bar{q}}(r_\perp,x) = \sigma_0\,{\cal N}(r_\perp,x)$ to the HERA data leads to the following parametrization, also known as ``colBK'': $\sigma_0=31.4055$ mb, $Q^2_{s0}=0.4$ GeV$^2$, $C=2.586$ and $p=0.807$ \cite{Iancu:2015joa}.

Iancu, Itakura and Munier in Ref.~\cite{Iancu:2003ge} suggested yet another parametrization (denoted as ``IIM'', in what follows),
\begin{eqnarray}
\sigma_{q\bar q}(r_{\perp},x)&=&\sigma_0 \times
\begin{cases}
N_0 \left(\frac{r_{\perp}Q_s(x)}{2}\right)^{2\gamma_{\rm eff}(r_{\perp},x)}& \qquad\text{for}\quad r_{\perp}Q_s(x) \leq 2 \,, \\
\left(1 - e^{-A\ln^{2}(B\,r_{\perp}Q_s(x))}\right)&\qquad\text{for}\quad r_{\perp}Q_s(x) > 2 
\end{cases}
\nonumber\\
\gamma_{\rm eff}(r_{\perp},x)&=&\gamma+\frac{1}{\kappa\lambda Y}\ln\left(\frac{2}{r_{\perp}Q_s(x)}\right)\,, \qquad Q_s^2(x)=\left(\frac{x_0}{x}\right)^\lambda \; \text{GeV}^2 \,,
\end{eqnarray}
in terms of an effective anomalous dimension $\gamma_{\rm eff}(r_{\perp},x)$, the saturation scale $Q_s^2(x)$ in the standard parametrization and the normalisation of the dipole cross section $\sigma_0=2\pi R_p^2$. Above, a continuity of the dipole cross section at $r_{\perp}Q_s(x)=2$ is ensured by a specific choice of the parameters,
\begin{eqnarray}
A = -\frac{N_0^{2}\gamma^{2}}{(1-N_0)^{2}\ln(1-N_0)} \,, \qquad
B = \frac{1}{2}(1-N_0)^{-\frac{1-N_0}{N_0\gamma}} \,.
\end{eqnarray}
while the remaining parameters were fitted to the HERA data in Ref.~\cite{Soyez:2007kg} yielding the values: $\gamma=0.7376$, $\lambda=0.2197$, $x_0=0.1632\times 10^{-4}$, $\kappa=9.9$, $N_0=0.7$, and $\sigma_0=27.33$ mb. A modification of the IIM model to incorporate the $b_\perp$-dependence explicitly has been achieved in Refs.~\cite{Kowalski:2006hc,Watt:2007nr} through a modified saturation scale
\begin{equation}
Q_s^2(x)\rightarrow Q_s^2(x,b_{\perp})=\left(\frac{x_0}{x}\right)^\lambda \left( e^{-\frac{b_{\perp}^2}{2B_{\rm G}}}\right)^{\frac{1}{\gamma}} \,.
\end{equation}  
The corresponding parametrization adjusted to fit the HERA data in Ref.~\cite{Rezaeian:2013tka}
\begin{eqnarray}
\gamma=0.6492\,, \quad N_0=0.3658\,, \quad \lambda=0.2023\,, \quad x_0=0.00069\,, \quad B_{\rm G}=5.5\; \mathrm{GeV}^{-2} \,.
\end{eqnarray}

For light hadron scattering at high energies off a given target driven by the soft-scale QCD dynamics (i.e. when no hard scale is involved) the total gluon-target collision energy squared in the c.m.~frame $\hat s \equiv x_1 s$ is a more appropriate variable than Bjorken $x$, such that $\sigma_{q\bar q}(r_\perp,x) \to \sigma_{q\bar q}(r_\perp,\hat s)$ is implied as was suggested by Ref.~\cite{Kopeliovich:1999am}. This is the case for instance of photoproduction such that $Q^2 \to \Lambda_{\rm QCD}^2$ where even for moderate or low energies one gets to very small values of $x$. The latter indeed becomes inappropriate for soft reactions at high energies, and the saturation scale becomes a function of rather $\hat s$ than $x$. Other examples of such processes are the pion-proton scattering, the diffractive Drell-Yan and the gluon radiation processes. Retaining the saturated ansatz of Eq.~(\ref{GBW-dipole}) and following the definition of Ref.~\cite{Kopeliovich:1999am,Kopeliovich:2007fv}, the parameterization for the dipole cross section known in the literature as the Kopeliovich-Schafer-Tarasov (KST) model is given by
\begin{eqnarray}\nonumber
 R_0(x)\to R_0(\hat s) = 0.88\,\mathrm{fm}\,\left(\frac{s_0}{\hat s}\right)^{0.14}\,, \quad
 \sigma_0\to \sigma_0(\hat s)=\sigma_{\rm tot}^{\pi p}(\hat s)
 \left(1+\frac{3R_0^2(\hat s)}{8\langle r_{\rm ch}^2 \rangle_{\pi}}\right)\,.
 \label{KST-params}
\end{eqnarray}
in terms of the pion-proton total cross section $\sigma_{\rm tot}^{\pi p}(\hat s)=23.6(\hat s/s_0)^{0.08}$ mb \cite{Barnett:1996yz}, with $s_0=1000\,{\rm GeV}^2$ and the mean pion charge radius squared $\langle r_{\rm ch}^2 \rangle_{\pi}=0.44$ fm$^2$ \cite{NA7:1986vav}. It is worth noting here that despite of its soft origin this parametrization provides an acceptable description of the pion-proton scattering cross section for scales reaching as high as $Q^2\sim 20$ GeV$^2$.
\begin{figure}[!hbt]
\begin{center}
\includegraphics[width=1.0\textwidth]{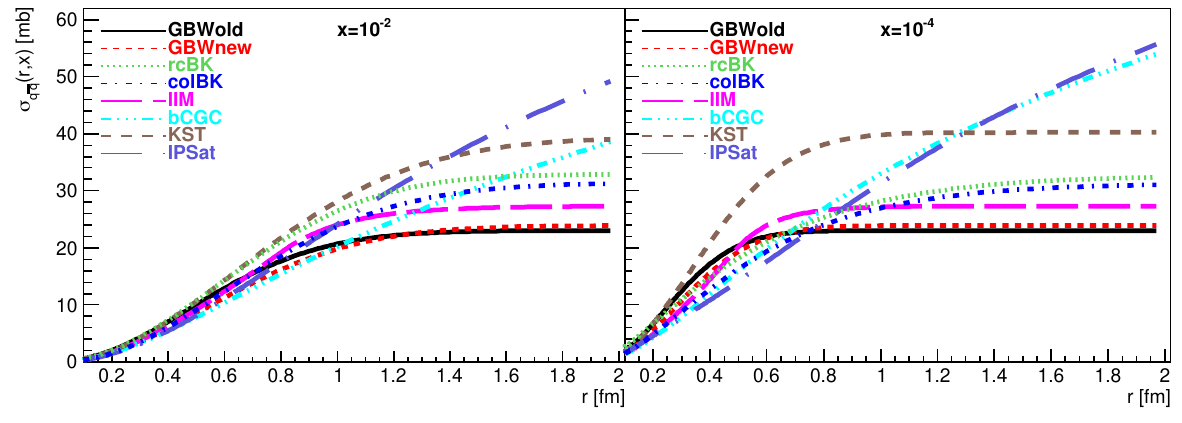}
\includegraphics[width=1.0\textwidth]{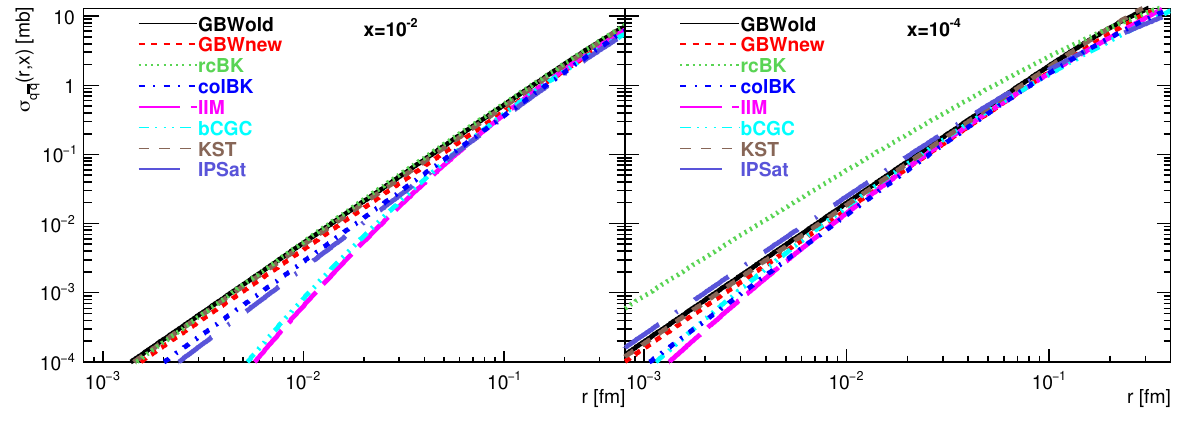}
\caption{Comparison of various parametrizations for the dipole cross section $\sigma_{q\bar q}(r\equiv r_\perp,x)$ described in the text. From Ref.~\cite{Cepila:2019skb}.}
\label{fig:sigqq-comparison}
\end{center}
\end{figure}

In order to illustrate typical uncertainties associated with a spread between different phenomenological parametrizations for the dipole cross section outlined above, in Fig.~\ref{fig:sigqq-comparison} adopted from Ref.~\cite{Cepila:2019skb} we show $\sigma_{q\bar q}(r_{\perp},x)$ for each model as a function of the dipole separation for two distinct Bjorken-$x$ values: $x=10^{-2}$ (left) and $x=10^{-4}$ (right). One notices big variations in both the magnitudes and shapes of the dipole cross section, particularly, for large dipoles in the saturation domain, as well as for very small dipole sizes in the perturbative (color transparency) domain $r_{\perp}\lesssim 0.05\div 0.06\,\fm$. The corresponding theoretical uncertainties in the soft domain grow with decreasing Bjorken $x$. So the uncertainties in the treatment of saturation effects are rather large even before incorporating angular dependence in the corresponding gluon GTMD. Other elaborate dipole parameterisations will be briefly reviewed below in connection to the dipole orientation effects.

\subsection{Elastic dipole amplitude and dipole orientation}
\label{sec:elastic-dip}

As will be discussed in more detail below, one of the most important examples of phenomenological probes for the elliptic gluon density (i.e.~the processes that are sensitive to it) is the fully differential cross section of the exclusive di-jet photoproduction \cite{Hatta:2016dxp}. Indeed, the angular dependence of it originates from a non-trivial dependence of the partial dipole amplitude on the dipole orientation. As was mentioned above, the $q \bar q$ dipole (with net color charge zero) interacts with the target only due to a non-zero relative difference (separation) $\bm r$ between the quark and anti-quark impact parameters relative to the scattering ``center of gravity''. As the ``center of gravity'' of the dipole is located at the impact parameter $\bm b$, the interaction strength encoded in the partial dipole amplitude is expected to disappear for $\bm{r} \perp \bm{b}$, while it gets maximized for $\bm{r} \parallel\bm{b}$. This is one of the generic model-independent properties of the dipole $S$-matrix. Below, we wish to summarise some of the basic theoretical and phenomenological models for dipole orientation effects employed in the literature and discuss their basic features and phenomenological consequences.

As was pointed out recently in Ref.~\cite{Linek:2023kga}, in the impact-parameter representation a non-trivial correlation of the dipole amplitude in angle between $\bm{r}$ and $\bm{b}$ emerges even if the corresponding GTMD is fully isotropic (i.e.~independent of the angle between $\bm{k}$ and $\bm{\Delta}$). This can be illustrated starting, for instance, from the isotropic gluon density matrix in the form of Eq.~(\ref{f-exp}) and substituting it into Eq.~(\ref{dipole_N-Y}). This way, one recovers the dipole amplitude as follows \cite{Linek:2023kga}:
\begin{eqnarray}
    {\cal N}(Y,\bm{r},\bm{b}) = \frac{1}{4} \Big\{ t_N\Big(\bm{b} + \frac{\bm{r}}{2} \Big)  + t_N\Big(\bm{b} - \frac{\bm{r}}{2} \Big) - 2 t_N(\bm{b}) \Big\} \sigma_0(x)  + \frac12 t_N(\bm{b}) \sigma_{q\bar q}(r_\perp,x) \,,
    \label{N_from_f}
\end{eqnarray}
where we notice an emergence of the dipole orientation effect $\propto \cos{\phi_{br}}$ controlled by $\sigma_0(x)$ being the dipole cross section in the non-perturbative limit of large dipoles given in Eq.~(\ref{sigma0}), and
\begin{eqnarray}
    t_N(\bm{b}) = \int \frac{d^2\bm{q}}{(2\pi)^2} \, e^{-i \bm{q} \cdot \bm{b}}\, e^{- \frac12 B_{\rm G} q_\perp^2} \, . 
\end{eqnarray}
Such a correlation has a straightforward geometric interpretation as coming due to contributions where only $q$ or $\bar q$ interact with the target at impact parameters $b_q$ or $b_{\bar q}$ given in Eq.~(\ref{qqbar-impact}), respectively. The corresponding elliptic contribution to ${\cal N}$ naturally vanishes in the perturbative limit of small $r_\perp\ll \Lambda_{\rm QCD}$ as the gluon density in the target can be considered constant (homogeneous) over distances $\sim r_\perp$ probed by such a very small dipole. Any genuine correlation between $\bm{k}$ and $\bm{\Delta}$ in the gluon GTMD representing the dynamical elliptic density in the target may, in principle, survive in the hard regime and comes on top of the above effect of the final-state (anti)quark wave-functions. Opportunities for disentangling both contributions unambiguously and hence for accessing the genuinely dynamic elliptic Wigner (or GTMD) gluon density experimentally constitutes yet unsolved problem being a subject of intense studies in the literature.

At Born level, considering for simplicity a symmetric dipole where quark and antiquark carry the same longitudinal momenta (i.e.~having the same distances from the dipole ``center of gravity''), the partial elastic amplitude of $q \bar q$ dipole interacting with a target quark reads \cite{Kopeliovich:2007fv} (see also Ref.~\cite{Bartels:2003yj})
\begin{eqnarray}
{\rm Im} f^q_{q\bar q}(\bm{r},\bm{b}) &=& \frac{2}{9\pi^2} \int
\frac{d^2\bm{q}\,d^2\bm{q}'\,\alpha_s(q_{\perp}^2)\alpha_s({q'}_{\perp}^2)} {(q_{\perp}^2 + m_g^2)({q'}_{\perp}^2 + m_g^2)}\, \nonumber \\ 
& \times &
\left[e^{i\bm{q}\cdot(\bm{b} + \bm{r}/2)} - e^{i\bm{q}\cdot(\bm{b} - \bm{r}/2)}\right] \left[e^{i\bm{q}'\cdot(\bm{b} + \bm{r}/2)} - e^{i\bm{q}'\cdot(\bm{b} - \bm{r}/2)}\right] \nonumber \\ 
& \simeq & 
\frac{8\alpha_s^2}{9} \left[K_0\left(m_g\left|\bm{b} + \frac{\bm{r}}{2}\right|\right) -
K_0\left(m_g\left|\bm{b} - \frac{\bm{r}}{2}\right|\right)\right]^2 \,,
\label{Born-dipole-q}
\end{eqnarray}
with the effective gluon mass $m_g$ accounting for non-perturbative (e.g.~confinement) effects, and the modified Bessel function $K_0(x)$. Indeed, as the soft gluon exchanges are dominant in the process, the gluon mass parameter encodes the exponential gluon fields' decay at large impact parameters effectively mimicking the confinement phenomenon. In Ref.~\cite{Kopeliovich:2007fv}, $m_g$ has been fixed to the pion mass enabling to reproduce typical hadronic cross sections. As anticipated above, the terms in the last line of Eq.~(\ref{Born-dipole-q}) cancel each other for $\bm{b}\cdot\bm{r}=0$, thus, yielding a non-trivial angular correlation between $\bm{r}$ and $\bm{b}$, already in the simplistic Born approximation. One should also note here that, similarly to the example above, only the first source of dipole orientation effects -- due to a combination of the phase factors representing the asymptotic plane-wave states of the dipole constituents -- emerges at Born-level, and no elliptic GTMD contribution is generated at the leading order. One therefore concludes that an angular correlation between the $\bm{k}$ and $\bm{b}$ vectors of a probed target gluon should be a higher-order (or non-perturbative) effect.

Following the footsteps of Ref.~\cite{Kopeliovich:1981pz}, one can perform an analogical calculation and derive the partial dipole amplitude off the nucleon target $N$ in the Born two-gluon exchange approximation and for a symmetric dipole resulting in
\begin{eqnarray}
{\rm Im} f^N_{q\bar q}(\bm{r},\bm{b}) &=& \frac{2}{3\pi^2} \int \frac{d^2\bm{q}\,d^2\bm{q}'\,\alpha_s(q_\perp^2)\alpha_s(q_\perp^{'2})}
{(q_\perp^2+m_g^2)(q_\perp^{'2}+m_g^2)} \;
e^{i\bm{b}\cdot(\bm{q} - \bm{q}')}\,
\Bigl(1 - e^{i\bm{q}\cdot\bm{r}}\Bigr)\Bigl(1 - e^{-i\bm{q}'\cdot\bm{r}}\Bigr)
\nonumber \\ 
& \times &
\left[F_N(|\bm{q}-\bm{q}'|) - F_N^{(2q)}(\bm{q},\bm{q}')\right] \,.
\label{2gluon-dipole-N}
\end{eqnarray}
Here, $F_N(k_\perp) \equiv \langle\Psi_N|\exp(i\bm{k}\cdot\bm{\rho}_1)|\Psi_N\rangle$ and $F_N^{(2q)}(\bm{q},\bm{q}') \equiv \langle\Psi_N|\exp[i\bm{q}\cdot\bm{\rho}_1 - i\bm{q}'\cdot\bm{\rho}_2]|\Psi_N\rangle$ are the nucleon and two-quark nucleon form factors, respectively, represented in terms of the three-quark nucleon wave function $\Psi_N(\bm{\rho}_1,\bm{\rho}_2,\bm{\rho}_3)$. 

As was extensively discussed in the previous subsection, precision measurements of the proton structure function at the HERA collider at high energies (small $x$) have enabled to reconstruct the dipole cross section $\sigma_{q\bar q}(r_\perp,x)$ in wide kinematic ranges of $r_\perp$ and $x$. The Born approximation illustrated above is, of course, too naive as it does not reproduce the experimentally well established Bjorken $x$ (or energy) dependence of $\sigma_{q\bar q}(r_\perp,x)$, particularly manifested at high energies. The experimentally observed rise of $\sigma_{q\bar q}(r_\perp,x)$ at $x\ll 1$ is due to a steep evolution of the unintegrated gluon density ${\cal F}(x,\kappa_\perp^2)$ in the target nucleon as follows from their relation given by Eq.~(\ref{dipole-TMD}) in the leading-logarithmic approximation, or equivalently,
\begin{eqnarray}
\sigma_{q\bar q}(r_\perp,x) \equiv \frac{4\pi}{3}\int \frac{d^2\bm{\kappa}}{\kappa_\perp^4}\,
(1-e^{-i\bm{\kappa}\cdot \bm{r}})\,
\alpha_s(\kappa_\perp^2)\,{\cal F}(x,\kappa_\perp^2) \,.
\label{sigma-qqbar}
\end{eqnarray}

In order to recover a correct $x$-dependence of the dipole partial amplitude in the two-gluon exchange model, in Ref.~\cite{Kopeliovich:2007fv} a generalised off-diagonal unintegrated gluon density ${\cal F}(x,\bm{q},\bm{q}')$ has been introduced, such that
\begin{eqnarray}
{\rm Im} f^N_{q\bar q}(\bm{r},\bm{b})&=&\frac{1}{12\pi} \int\frac{d^2\bm{q}\,d^2\bm{q}'}{q_\perp^2\,q_\perp^{'2}}\,\bar{\alpha}_s\,
{\cal F}(x,\bm{q},\bm{q}') e^{i\bm{b}\cdot(\bm{q}-\bm{q}')} \nonumber\\ 
&\times&
\left(e^{-i\bm{q}\cdot\bm{r}\beta} - e^{i\bm{q}\cdot\bm{r}(1-\beta)}\right)\,
\left(e^{i\bm{q}'\cdot\bm{r}\beta} - e^{-i\bm{q}'\cdot\bm{r}(1-\beta)}\right)\, \,,
\label{2gluon-dipole-N-x}
\end{eqnarray}
in terms of the longitudinal light-cone momentum fractions of anti-quark $1-\beta$ and quark $\beta$, such that the dipole ``center of gravity'' gets effectively shifted towards the fastest $q$ or $\bar q$, and $\bar{\alpha}_s\equiv \sqrt{\alpha_s(q_\perp^2)\alpha_s({q'}_\perp^2)}$. The Born approximation of Eq.~(\ref{2gluon-dipole-N}) is then recovered for
\begin{eqnarray}
{\cal F}_{\rm Born}(\bm{q},\bm{q}') =
\frac{4\bar{\alpha}_s}{\pi} \left[F_N(\bm{q} - \bm{q}') - 
F_N^{(2q)}(\bm{q},\bm{q}')\right] \,.
\label{UGDF-Born}
\end{eqnarray}
Thus, Eq.~(\ref{2gluon-dipole-N-x}) advises us that the dipole $S$-matrix should be sensitive to dipole separation not only in the transverse plane, but also in longitudinal direction due to, in general, unequal sharing of energy between $q$ or $\bar q$ in the dipole \cite{Kopeliovich:1999am,Kopeliovich:2007fv}. More specifically, we notice from the structure of Eq.~(\ref{2gluon-dipole-N-x}) that the dipole orientation effect in the partial dipole amplitude appears to be sensitive to both the angle between the transverse momenta of the exchanged gluons $\bm{q}$ and $\bm{q}'$ and the longitudinal momentum sharing between $q$ or $\bar q$ quantified by the $\beta$ variable.

Due to a remarkable simplicity of the GBW parameterisation (\ref{GBW-dipole}), it enables to analytically derive both the form of the corresponding off-diagonal unintegrated gluon density \cite{Kopeliovich:2007fv},
\begin{eqnarray}
{\cal F}(x,\bm{q},\bm{q}') &=&
\frac{3\sigma_0}{16\pi^2\bar{\alpha}_s}\ q_\perp^2\,{q'}_\perp^2\,R_0^2(x)\,
{\rm exp}\Bigl[-\frac{1}{8}\,R_0^2(x)\,(q_\perp^2+{q'}_\perp^2)\Bigr] \nonumber \\
&\times & {\rm exp}\bigl[-R_N^2(\bm{q} - \bm{q}')^2/2\bigr] \,,
\label{GBW-UGDF}
\end{eqnarray}
and the phenomenological ansatz for the related partial (elastic) dipole amplitude satisfying Eq.~(\ref{sigma-qqbar}) in the two-gluon exchange approximation (\ref{2gluon-dipole-N-x}),
\begin{eqnarray}
{\rm Im} f^N_{q\bar q}(\bm{r},\bm{b}) &=&
\frac{\sigma_0}{8\pi B_{\rm el}}\,
\Biggl\{\exp\left[-\frac{[\bm{b} + \bm{r}(1-\beta)]^2}{2B_{\rm el}}\right] +
\exp\left[-\frac{(\bm{b} - \bm{r}\beta)^2}{2B_{\rm el}}\right]
\nonumber \\ 
&-&
2\exp\Biggl[-\frac{r_\perp^2}{R_0^2} -
\frac{[\bm{b} + (1/2-\beta)\bm{r}]^2}{2B_{\rm el}}\Biggr] \Biggr\} \,.
\label{ImfN-GBW}
\end{eqnarray}
The latter contains a non-trivial dependence on azimuthal angle $\theta$ between $\bm{r}$ and $\bm{b}$ in the transverse plane. Thus, it manifestly accounts for not only the relative dipole orientation effect, but also an explicit dependence on the $q$ and $\bar q$ light-cone fractional momenta. Here, 
\begin{eqnarray}
B_{\rm el}(x) = R_N^2+R_0^2(x)/8 \,, \qquad 
R_N^2 \approx \langle r_{\rm ch}^2\rangle/3 \,,
\end{eqnarray}
are the elastic slope and its part $R_N^2$ originated from the Pomeron-proton form factor $F^p_\Pom(k_\perp^2)=\exp(-k_\perp^2 R_N^2/2)$ and determined in terms of the proton mean charge radius squared, $\langle r_{\rm ch}^2\rangle$ (the current value is $\sqrt{\langle r_{\rm ch}^2\rangle}\simeq 0.875$ fm \cite{ParticleDataGroup:2016lqr}), respectively. Note that the angular correlation in the phenomenological ansatz for the unintegrated gluon density (\ref{GBW-UGDF}) takes the following simple form,
\begin{eqnarray}
{\cal F}(x,\bm{q},\bm{q}') \propto e^{-R_N^2\,q_\perp\,q'_\perp\cos \theta } \,,
\end{eqnarray}
which can, in principle, be used to extract the elliptic gluon density by mapping the above expression to the ansatz of Eq.~(\ref{GTMD-decomposition}).

Starting from the dipole elastic amplitude off the nucleon target $f^N_{q\bar q}$, we can derive the elastic cross section of dipole-$N$ scattering -- the main building block of the dipole picture of the QCD scattering. Ignoring the real part which is typically small at high energies, one writes \cite{Kopeliovich:2007fv}
\begin{eqnarray} \nonumber
&&\frac{d\sigma^{(q\bar q)N}_{\rm el}(r_\perp)}{dk_\perp^2} = \frac{1}{4\pi}
\left|\int d^2\bm{b}\,e^{i\bm{k}\cdot\bm{b}}
{\rm Im} f^N_{q\bar q}(x,\bm{r},\bm{b})\right|^2 \\
&&\qquad \qquad \xRightarrow[]{k_\perp\to 0}
\frac{[\sigma_{q\bar q}(r_\perp,x)]^2}{16\pi}\,
\exp\left[-B^{(q\bar q)N}_{\rm el}(r_\perp) k_\perp^2\right] \,.
\label{sigma-qq_N}
\end{eqnarray}
The latter relation can be used to determine the slope of the differential elastic cross section \cite{Kopeliovich:2007fv}
\begin{equation}
B^{(q\bar q)N}_{\rm el}(r_\perp) = \frac{1}{\sigma_{q\bar q}(r_\perp,x)}
\int d^2\bm{s}\,s_\perp^2\,{\rm Im} f^N_{q\bar q}(x,\bm{r},\bm{s}) \,,
\label{slope}
\end{equation}
which has been measured at HERA collider in diffractive $\rho$-meson electroproduction process at large $Q^2$ to be \cite{ZEUS:2007iet} $B^{(q\bar q)N}_{\rm el}(r_\perp)\approx 5\GeV^{-2}$. The latter is in rough agreement with its relation to the proton-proton elastic slope, $B^{(q\bar q)N}_{\rm el}(r_\perp)\approx B^{pp}_{\rm el}/2$.

One may straightforwardly generalise the phenomenological formalism for the dipole scattering off a nucleon outlined above to the case of nuclear target, $T\equiv A$, with atomic mass number $A$. Correspondingly, the partial dipole amplitude for a $q\bar q$ dipole scattering off $A$ at impact parameter $b_\perp$ is found in the Glauber picture as
\begin{eqnarray}
{\rm Im} f^A_{q\bar q}(x,\bm{r},\bm{b}) &=& 1-\left[1-\frac{1}{2A}\sigma_{q\bar q}(r_\perp,x) {\cal T}_A(\bm{r},\bm{b})\right]^A \nonumber \\
&\simeq& 1-\exp\left[-\frac{1}{2}\sigma_{q\bar q}(r_\perp,x) {\cal T}_A(\bm{r},\bm{b})\right] \,,
\label{dipole-nuc}
\end{eqnarray}
in terms of the effective nuclear thickness \cite{Kopeliovich:2003tz}
\begin{eqnarray}
{\cal T}_A(\bm{r},\bm{b})=\frac{2}{\sigma_{q\bar q}(r_\perp,x)}\int d^2\bm{s}\,
{\rm Im} f^N_{q\bar q}(x,\bm{r},\bm{s})\,
T_A(\bm{b} + \bm{s})\,, \quad\!\!\!\!
T_A(b_\perp) \equiv \int_{-\infty}^\infty dz\rho_A(b_\perp,z) \,
\label{eff-thickness}
\end{eqnarray}
Here, the standard nuclear thickness function $T_A$ is found as an integral of the conventional nuclear density $\rho_A(b_\perp,z)$ taken along the trajectory of the particle. Provided that the mean value $\langle s_\perp^2\rangle\approx 0.4\,{\rm fm}^2$ found from Eq.~(\ref{slope}) is much smaller than the nuclear radius squared (for a heavy nucleus), i.e.~$\langle s_\perp^2\rangle\ll R_A^2$, one may expand 
\begin{eqnarray}
T_A(\bm{b} + \bm{s}) = T_A(b_\perp) + \frac{\bm{s}\cdot\bm{b}}{b_\perp}T_A^\prime(b_\perp)
+\frac{1}{2}\left(\frac{\bm{s}\cdot\bm{b}}{b_\perp}\right)^2
T_A^{\prime\prime}(b_\perp) + \dots \,,
\label{TA-expansion}
\end{eqnarray}
and apply the latter in Eq.~(\ref{dipole-nuc}) giving rise to the approximate formula for the partial dipole amplitude off a nucleus target,
\begin{eqnarray}
&& {\rm Im} f^A_{q\bar q}(x,\bm{r},\bm{b}) \simeq
1-\exp\left[-\frac{1}{2}\sigma_{q\bar q}(r_\perp,x)T_A(b_\perp)\right]
\Biggl\{1-\frac{1}{b_\perp}\,T_A^\prime(b_\perp)\,
\gamma_1(\bm{b},\bm{r})
\nonumber\\ 
&& \qquad\qquad -\,\frac{1}{2b_\perp^2}\,\left[
T_A^{\prime\prime}(b_\perp)\gamma_2(\bm{b},\bm{r}) - 
{T_A^{\prime}}^2(b_\perp)\,\gamma_1^2(\bm{b},\bm{r})
\right]\Biggr\} \,,   
\label{approx-nuc-fel}
\end{eqnarray}
where
\begin{eqnarray}
\gamma_n(\bm{b},\bm{r}) = \int d^2\bm{s}\,{\rm Im} f^N_{q\bar q}(x,\bm{r},\bm{s})(\bm{s}\cdot\bm{b})^n \,.
\end{eqnarray}
We notice that for a large nucleus, $T_A^\prime(b_\perp)\approx -2\rho_0\,b_\perp/\sqrt{R_A^2-b_\perp^2}$ and $T_A^{\prime\prime}(b_\perp)\approx 2\rho_0\,R_A^2/(R_A^2-b_\perp^2)^{3/2}$ (with $\rho_0\approx 0.16\,{\rm fm}^{-3}$ being the density at the center of the nucleus) are small for $b_\perp\ll R$ and are peaked at the nuclear periphery. Besides, in the eikonal approximation applicable for heavy nuclei one may use ${\cal T}_A(\bm{r},\bm{b})\approx T_A(b_\perp)$ such that the angular correlations are naturally small.

The formula (\ref{ImfN-GBW}) (in both GBW and KST parameterisations as well as its generalisation to nuclear targets in Eq.~(\ref{approx-nuc-fel})) represents the first known phenomenologically driven model for the 5D partial dipole amplitude that has been fit to the experimental data and that can be connected to the dipole $S$-matrix and hence to the gluon Wigner distribution using Eqs.~(\ref{Imfel-qq}) and (\ref{xWg}), respectively. The $\theta$-averaged part of the Wigner distribution is directly related to the isotropic unintegrated gluon density in the target ${\cal F}$, the basic non-perturbative ingredient of $k_\perp$-factorization. The difference between the Wigner distribution and its $\theta$-averaged part is proportional to the elliptic gluon distribution that can be extracted from the exponential parametrization of the dipole amplitude in Eq.~(\ref{ImfN-GBW}), or in more elaborate models discussed below.

In the following, we wish to illustrate how the azimuthal angle correlations of the gluon GTMD can be understood at small $x$ in the framework of QCD evolution based models, also leading to emergent effects of saturation and elliptic flow.

\section{QCD evolution based models for dipole scattering}
\label{Sect:QCD-evolution}

Both angular-dependent and angular-independent components of the Wigner distribution are highly sensitive to the soft and saturation physics, and thus are a subject of intense research in the literature \cite{Hagiwara:2016kam,Iancu:2017fzn}. Here, we would like to overview the widely used models for the Wigner distribution (dipole $S$-matrix) based upon QCD evolution.

\subsection{Azimuthal angle correlation in the BFKL model}
\label{Sect:BFKL}

In the BFKL approximation valid at $x\ll 1$ (or high energies), the $T$-matrix of a dipole scattering off another dipole at transverse position ${\bm x}$ in the impact parameter space reads \cite{Lipatov:1985uk,Navelet:1997tx,Navelet:1997xn} (see also Ref.~\cite{Hatta:2016dxp})
\begin{eqnarray}
T_Y(\bm{r},\bm{b}) &=& 2\pi\alpha_s^2\sum_n\int\frac{d\nu}{(2\pi)^3}
\frac{(1+(-1)^n)\left(\nu^2+\frac{n^2}{4}\right)}{\left(\nu^2+\left(\frac{n-1}{2}\right)^2\right)
\left(\nu^2+\left(\frac{n+1}{2}\right)^2\right)}e^{\chi(n,\nu)Y} \nonumber \\
&& \qquad \times
\int d^2\bm{\omega} E^{1-h,1-\bar{h}}\Big(\bm{b} + \frac{\bm{r}}{2} - \bm{\omega}, \bm{b} -\frac{\bm{r}}{2} - \bm{\omega}\Big) \nonumber \\ 
&& \qquad \qquad \qquad \times\; E^{h,\bar{h}}
\Big(\frac{\bm{x}}{2} - \bm{\omega},-\frac{\bm{x}}{2} - \bm{\omega}\Big) \,, 
\label{Tmatrix-BFKL}
\end{eqnarray}
valid for a symmetric energy sharing between the quark and antiquark in the dipole. Here, the holomorphic eigenfunction of the BFKL operator and the corresponding characteristic function read
\begin{eqnarray}
&& E^{h,\bar{h}}(a-\omega,b-\omega) = (-1)^n\left(\frac{z_{ab}}{z_{a\omega}z_{b\omega}}\right)^h \left(\frac{\bar{z}_{ab}}{\bar{z}_{a\omega}\bar{z}_{b\omega}}\right)^{\bar{h}} \,, \\
&& \chi(n,\nu) = \frac{2\alpha_sN_c}{\pi} \left[\psi (1) - 
\textrm{Re}\,\psi\Big(\frac{|n|+1}{2}+i\nu\Big)\right] \,,
\end{eqnarray}
respectively, with $h=\frac{1-n}{2}+i\nu$ and $\bar{h}=\frac{1+n}{2}+i\nu$. At $x\ll 1$, the dominant contribution comes from the $n=0$ term. In order to disentangle a non-trivial correlation in azimuthal angle $\phi_{br}$ between $\bm{r}$ and $\bm{b}$, one can apply the saddle point approximation in Eq.~(\ref{Tmatrix-BFKL}) in a vicinity of $\nu=0$ and consider the limit of $x_\perp\ll r_\perp,b_\perp$, yielding the result for the dipole $T$-matrix \cite{Hatta:2016dxp}
\begin{eqnarray}
T_Y(\bm{r},\bm{b}) \simeq \frac{\alpha_s^2 \rho_\perp}{\sqrt{\pi} } \frac{\ln\frac{16}{\rho_\perp}} {\left(\frac{7}{2}\bar{\alpha}_s\zeta(3)Y\right)^{3/2}}\exp\left(4\bar{\alpha}_sY\ln 2 -\frac{\ln^2\frac{16}{\rho_\perp}}{14\bar{\alpha}_s\zeta(3) Y}\right) \,, 
\label{Tmatrix-BFKL_approx}
\end{eqnarray}
in terms of
\begin{eqnarray}
\rho_\perp^2 \equiv \frac{x_\perp^2 r_\perp^2}{\left(\bm{b}+\frac{\bm{r}}{2}-\frac{\bm{x}}{2}\right)^2 \left(\bm{b}-\frac{\bm{r}}{2}+\frac{\bm{x}}{2}\right)^2} \simeq \frac{x_\perp^2 r_\perp^2}{b_\perp^4+\frac{r_\perp^4}{16} - \frac{b_\perp^2r_\perp^2}{2}\, \cos(2\phi_{br})}\,.    
\end{eqnarray}
Thus, one observes that, in consistency with the discussion in the previous section and with Ref.~\cite{Kopeliovich:2007fv}, the partial dipole amplitude is enhanced for a dipole configuration with parallel $\bm{b}$ and $\bm{r}$ (i.e.~$\phi_{br}=0$) compared to the case of $\bm{b}\perp \bm{r}$. Analogically, beyond the saddle point approximation, i.e.~close to the nonlinear saturation domain where the partial dipole amplitude reaches a constant value $T(\bm{r},\bm{b}, Y)|_{r_\perp \to 1/Q_s} \to {\rm const}$ at the saturation momentum scale $Q_s$, one finds
\begin{eqnarray}
\frac{1}{\rho_\perp^2}\Big|_{r_\perp \simeq 1/Q_s}\sim e^{\frac{\chi(\gamma_s)}{\gamma_s}Y} \,, \qquad \gamma_s\equiv \frac{1}{2}+i\nu_s=0.628 \,.
\end{eqnarray}
Resolving the latter with respect to $Q_s$ at large impact parameters (small momentum transfers) corresponding to $b_\perp \gg r_\perp \simeq 1/Q_s$, one gets \cite{Hatta:2016dxp}
\begin{eqnarray}
Q_s^2 \sim \frac{x_\perp^2}{b_\perp^4}e^{\frac{\chi(\gamma_s)}{\gamma_s}Y} + 
\frac{\cos(2\phi_{br})}{2b_\perp^2} \,,
\end{eqnarray}
in overall agreement with a numerical analysis of the nonlinear small-$x$ dynamics of Refs.~\cite{Golec-Biernat:2003naj,Berger:2011ew}. Note that a non-trivial angular correlation is observed also for configurations with $r_\perp\sim b_\perp$ and hence is a generic phenomenon that should be inherent for realistic saturation models.

Switching to momentum-space amplitude by means of a Fourier transform $T_Y\xRightarrow[]{\cal F} \widetilde{T}_Y$ over $\bm{r}$ and $\bm{b}$ and averaging the result over all possible target dipole $\bm{x}$ orientations, one obtains \cite{Hatta:2016dxp}
\begin{eqnarray}
\widetilde{T}_Y(\bm{q},\bm{\Delta}) &\simeq& \frac{\alpha_s^2 x_\perp}{(2\pi)^2 \Delta_\perp^3} \frac{e^{4\bar{\alpha}_s\,Y\ln 2}}
{\left(\frac{7}{2}\bar{\alpha}_s\zeta(3)Y \pi\right)^{3/2}}\, \int_0^{\pi/2} d\theta J_0\left(\frac{\sin \theta \Delta_\perp x_\perp }{2}\right)K_{0}\left(\frac{\cos \theta \Delta_\perp x_\perp }{2} \right) \notag\\
&& \times \int_0^1\frac{d\alpha}{\alpha^2(1-\alpha)^2}\,_2F_1\left(\frac{3}{2},\frac{3}{2},1,-\frac{|\bm{q} + (1/2-\alpha)\bm{\Delta}|^2}{\Delta_\perp^2 \alpha(1-\alpha)} \right) \,.
\end{eqnarray}
Thus, the angular correlation is manifested also in momentum space, particularly, for even harmonics, starting from a sizable elliptic $\sim\cos 2\phi$ part, while the contributions of the odd harmonics vanish. Furthermore, in the linear (saddle-point) approximation and for finite momentum transfers $\bm{\Delta}$, one may write the relation between the partial dipole amplitude and the Wigner function in momentum space following Ref.~\cite{Hatta:2016dxp} as
\begin{eqnarray}
x\widetilde{W}_g(x,\bm{q},\bm{\Delta}) = -\frac{2N_c}{\alpha_s}\left(q_\perp^2 - \frac{\Delta_\perp^2}{4}\right)\, \widetilde{T}_Y(\bm{q},\bm{\Delta}) \,.
\label{xWg-linear-BFKL}
\end{eqnarray}

\subsection{Elliptic gluon distribution in the Color Glass Condensate}
\label{Sect:BK}

Let us turn now to a discussion of the gluon Wigner distribution in a high-energy nucleon or nucleus in the saturation regime at small-$x$ accounting for the gluon saturation phenomenon. The latter can be incorporated in the dipole $S$-matrix formalism through small-$x$ QCD evolution described by the Balitsky-Kovchegov (BK) equation with impact parameter dependence \cite{Balitsky:1995ub,Kovchegov:1999yj},
\begin{eqnarray}
\label{BK_equation}
\partial_Y S_Y(\bm{x},\bm{y}) = \frac{N_c\alpha_s}{2\pi^2} \, 
\int d^2\bm{z} \frac{(\bm{x}-\bm{y})^2}{(\bm{x}-\bm{z})^2 (\bm{z}-\bm{y})^2} \{S_Y(\bm{x},{\bm z}) S_Y(\bm{z},\bm{y}) - S_Y(\bm{x},\bm{y}) \}\,,
\end{eqnarray}
providing a rapidity evolution for $S_Y$ in the leading-log accuracy and large-$N_c$ limit. Here, $\bm{x}\equiv \bm{b}+\frac{\bm{r}}{2}$ and $\bm{y}\equiv \bm{b}-\frac{\bm{r}}{2}$ are the transverse positions of a quark and an antiquark in the dipole, respectively. 

A computation of both the isotropic and elliptic components of the gluon Wigner distribution (and the corresponding GTMD) through numerical solution of the BK equation (\ref{BK_equation}) has been performed by Hagiwara, Hatta, and Ueda in Ref.~\cite{Hagiwara:2016kam} (we denote it as HHU model, in what follows). Here, a simplifying assumption that the considered BK solutions are invariant under an $SO(3)$ subgroup of the maximal symmetry of the BK equation, the conformal M$\ddot{\text{o}}$bius group in the transverse plane, has been applied \cite{Gubser:2011qva}. Namely, the BK solutions were sought in the following $SO(3)$-symmetric form,
\begin{eqnarray}
    S_Y(\bm{x},\bm{y}) \equiv S_Y(d^2(\bm{x},\bm{y})) \,, \quad
    d^2(\bm{x},\bm{y}) \equiv \frac{R^2(\bm{x}-\bm{y})^2}{(R^2+x_\perp^2)(R^2+y_\perp^2)} \,, \quad 0 \le d^2 \le 1 \,,
    \label{S_Y-SO3}
\end{eqnarray}
and with symmetric initial condition,
\begin{eqnarray}
S_{Y=0}(\bm{x},\bm{y}) = e^{-d^2(\bm{x},\bm{y})} \,,
\end{eqnarray}
in terms of an arbitrary length-scale parameter $R$. The above ansatz for the considered solution relies on a conjecture that the BK equation may dynamically restore $SO(3)$ symmetry at the level of solutions despite its breaking by realistic initial conditions as has been observed numerically at small or large $r_\perp$ in Ref.~\cite{Golec-Biernat:2003naj} (see also Ref.~\cite{Hatta:2007fg}). Note, the ansatz (\ref{S_Y-SO3}) is symmetric with respect to the change of variables, $r_\perp\to r_m^2/r_\perp$, with $r_m=2\sqrt{b_\perp^2+R^2}\simeq 2b_\perp$ for $b_\perp\gg R$, thus, recovering the observation of Ref.~\cite{Golec-Biernat:2003naj} that the partial dipole scattering amplitude $T_Y$ is maximised for the dipole separation of $r_\perp=2b_\perp$. 

\begin{figure}[t]
\includegraphics[width=80mm]{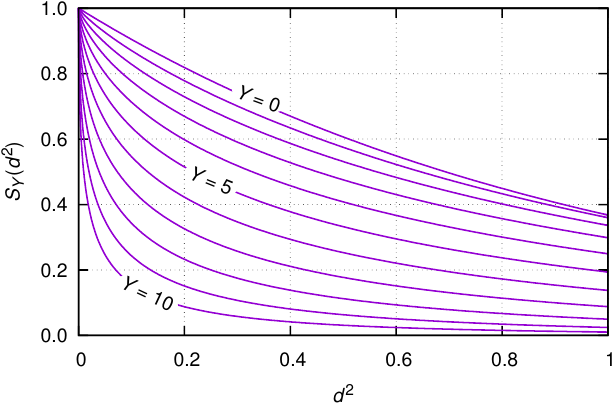}
\includegraphics[width=80mm]{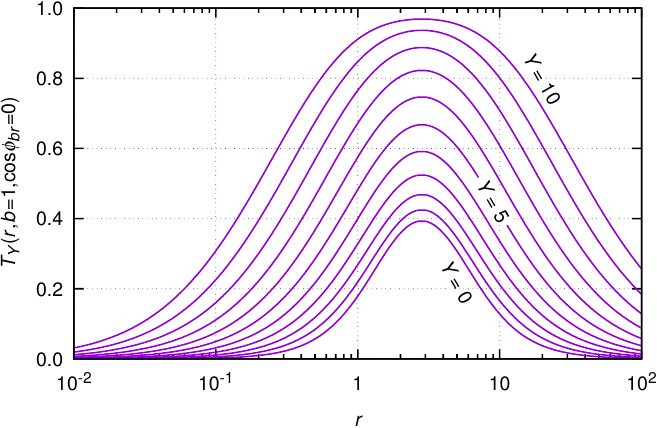}
\caption{The dipole $S$-matrix $S_Y(d^2)$ in the HUU model as a function of $d^2$ (left panel) and the partial dipole amplitude $T_Y=1-S_Y$ as a function of $r\equiv r_\perp$ (in logarithmic scale) for fixed $b\equiv b_\perp=1$ and $\cos(\phi_b-\phi_r)=0$ (right panel) for different rapidities $Y=[0,\dots,10]$. From Ref.~\cite{Hagiwara:2016kam}.} 
\label{fig:S_Y-T_Y}
\end{figure}

With these considerations, the BK equation (\ref{BK_equation}) can be recast into the following $SO(3)$-symmetric form \cite{Hagiwara:2016kam},
\begin{eqnarray}
\partial_Y g(x_\perp) &=&  \frac{N_c\alpha_s}{\pi} \, \int_0^{2\pi} \frac{d\phi}{2\pi} \int_0^\infty \frac{dz_\perp}{z_\perp} \frac{x_\perp^2}{(x_\perp^2+z_\perp^2-2x_\perp z_\perp\cos\phi)} \nonumber \\ && \qquad\times \left\{g_Y\left(\sqrt{\frac{R^4(x_\perp^2+z_\perp^2-2x_\perp z_\perp\cos\phi)}{R^4+x_\perp^2z_\perp^2+2R^2x_\perp z_\perp \cos\phi}}\right)g_Y(z_\perp)-g_Y(x_\perp)\right\}\,, \label{BK_equation-SO3}
\end{eqnarray}
in terms of $g_Y(x_\perp)\equiv S_Y({\bm x},0)$, such that the $SO(3)$-symmetric dipole $S$-matrix is reconstructed as
\begin{eqnarray}
S_Y(d^2)=g_Y\left(\sqrt{\frac{R^2d^2}{1-d^2}}\right)\,, \label{S_Y-SO3_result}
\end{eqnarray}
with $d^2$ given in Eq.~(\ref{S_Y-SO3}). The numerical results of Ref.~\cite{Hagiwara:2016kam} for $S_Y(d^2)$ and $T_Y(r_\perp,b_\perp=1,\cos(\phi_b-\phi_r)=0)$ are shown in Fig.~\ref{fig:S_Y-T_Y} in left and right panels, respectively, fixing the units by setting $R=1$, as well as taking $N_c\alpha_s/\pi=0.2$. As expected above, the peak in $T_Y$ appears at $r_\perp=r_m$ at any value of $Y$.

\begin{figure}[t]
\includegraphics[width=77mm]{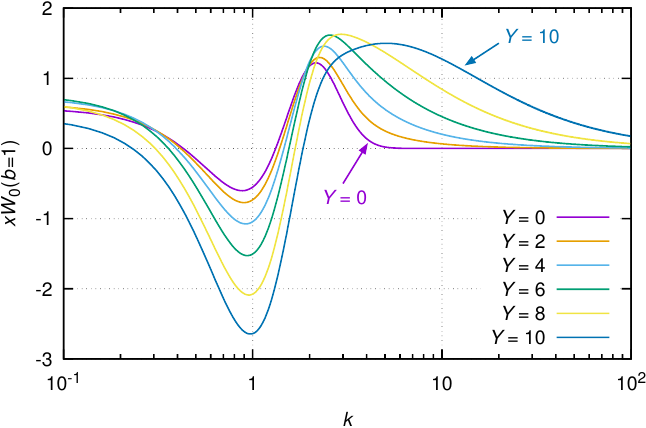}
\includegraphics[width=82mm]{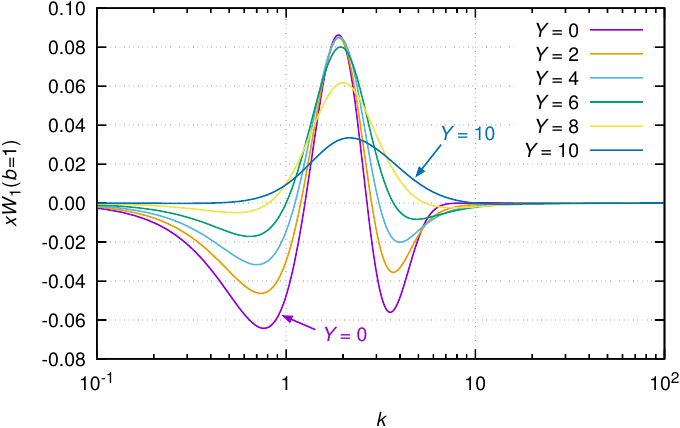}
\caption{The isotropic $xW'_{g,0}$ (left) and elliptic $xW'_{g,1}$ (right) Wigner distributions in the HHU model as functions of $k\equiv k_\perp$ at fixed $b\equiv b_\perp=1$ and for different rapidities $Y=[0,\dots,10]$. From Ref.~\cite{Hagiwara:2016kam}.} 
\label{fig:Wigner0-1}
\end{figure}

As was further pointed out in Ref.~\cite{Hagiwara:2016kam}, there is an artefact of the BK equation whose kernel maintains the perturbative Coulomb interaction at large separations which is a non-physical behavior due to confinement. In real QCD, however, the effect of confinement is expected to lead to the black disc limit corresponding to $T_Y(r_\perp\gg R)\to 1$ such that the contribution to the Fourier integral in the Wigner distribution (\ref{xWg}) from the perturbative tail at large $r_\perp\gg r_m$ domain should vanish and cannot affect the observables. A natural way to eliminate non-physical contributions is to use the Husimi distribution (\ref{husimi-QCD}) where the integral over $r$ would be cut off at large distances by means of a Gaussian factor. A similar dumping factor has been introduced in Ref.~\cite{Hagiwara:2016kam} also in the Wigner distribution (\ref{xWg}) such that
\begin{eqnarray}
\label{xWg-corrected}
xW'_g(x,\bm{k},\bm{b}) = -\frac{2N_c}{\alpha_s} \int \frac{d^2\bm{r}}{(2\pi)^2} e^{i \bm{k} \cdot \bm{r}} e^{-\epsilon r_\perp^2} \left(\frac{1}{4}\bm{\nabla}^2_{b} +\bm{k}^2 \right) T_Y(\bm{r},\bm{b})\,, 
\end{eqnarray} 
where a natural choice of $\epsilon = 1/4$ has been made, thus, effectively cutting off the non-physical contribution from large $r_\perp\gtrsim 2$. As follows from Eq.~(\ref{S_Y-SO3}), the Fourier transform of $W'_g$ contains only even harmonics such that the first $\propto \cos 2\phi$ (elliptic) term provides the leading angular dependence \cite{Hatta:2016dxp}. Isolating the isotropic and elliptic contributions in Eq.~(\ref{xWg-corrected}), one finds
\begin{eqnarray}
xW'_{g,0}(x,k_\perp,b_\perp) &=& -\frac{N_c}{2\alpha_s \pi^2}\left(\frac{1}{4} \frac{\partial^2}{\partial b_\perp^2} + \frac{1}{4b_\perp}\frac{\partial}{\partial b_\perp} + k_\perp^2 \right)  \int_0^{\infty}\, r_\perp\, e^{-\epsilon r_\perp^2}J_0(k_\perp r_\perp)dr_\perp  \nonumber \\ 
&& \qquad \times \int_0^{2\pi} d\phi_{br}\, T_Y(r_\perp,b_\perp,\cos2\phi_{br}) \,,    
\end{eqnarray}
and
\begin{eqnarray}
xW'_{g,1}(x,k_\perp,b_\perp) &=& \frac{N_c}{2\alpha_s \pi^2} \left(\frac{1}{4} \frac{\partial^2}{\partial b_\perp^2} + \frac{1}{4b_\perp}\frac{\partial}{\partial b_\perp} - \frac{1}{b_\perp^2} + k_\perp^2 \right) \int_0^{\infty}\, r_\perp\, e^{-\epsilon r_\perp^2} J_2(k_\perp r_\perp)  dr_\perp  \nonumber \\ 
&& \times \int_0^{2\pi} d\phi_{br}\, \cos(2\phi_{br})\, T_Y(r_\perp,b_\perp,\cos2\phi_{br}) \,,
\end{eqnarray}
shown as functions of $k_\perp$ (at fixed $b_\perp=1$) for different rapidities $Y=[0,\dots,10]$ in Fig.~\ref{fig:Wigner0-1}, left and right panels, respectively, and $\phi_{br}$ is as usual an angle between $\bm{b}$ and $\bm{r}$ vectors. We notice that the position of the peak in $xW'_{g,0}$ found at the saturation momentum $k_{\rm peak} = Q_s(Y,b_\perp)$ grows with rapidity $Y$ and falls with impact parameter $b_\perp$. The latter behavior is in consistency with the geometric scaling, $T_Y(\bm{r},\bm{b}) \propto (r_\perp Q_s)^{2\gamma}$ for small $r_\perp$. Here, $\gamma$ is weakly dependent on $r_\perp$ and is found to be between $\gamma = 1$ at $r_\perp\to 0$ and $\gamma \simeq 0.63$ at $r_\perp \lesssim 1/Q_s$ \cite{Gribov:1983ivg,Iancu:2002tr}. The peak value in the elliptic part $xW'_{g,1}$ is only a few percent of that of $xW'_{g,0}$, and evolves much slower with $Y$ since there is no geometric scaling in the elliptic contribution \cite{Hagiwara:2016kam}.

\subsection{Azimuthal correlations in the McLerran-Venugopalan model}
\label{Sect:MV}

As was mentioned earlier, the color dipole orientation with respect to its impact parameter $\bm{b}$ probes the transverse inhomogeneity of the gluon distribution in the target. A robust semi-analytical computation of the partial dipole amplitude incorporating the dipole orientation effects has been performed in Ref.~\cite{Iancu:2017fzn}. For this purpose, in the latter work an effective McLerran-Venugopalan (MV) model \cite{McLerran:1993ka} has been employed and extended to incorporate inhomogeneities in the transverse-plane distribution of soft gluons in the target. Such an extension procedure is in overall consistency with the phenomenological dipole models that feature a non-trivial impact-parameter dependence and saturation such as the bCGC  \cite{Iancu:2003ge,Kowalski:2006hc,Rezaeian:2013tka} and IP-Sat \cite{Kowalski:2003hm,Kowalski:2008sa,Rezaeian:2012ji} parameterisations being successfully matched to the HERA data as was discussed above (see also Refs.~\cite{Munier:2001nr,Berger:2011ew,Berger:2012wx,Armesto:2014sma}). However, they do not contain a non-trivial azimuthal angle dependence essentially assuming the rotational invariance of the dipole $S$-matrix. 

Being equipped with the MV model embedding an inhomogeneous Gaussian-like valence quark distribution of the color-field sources in the impact parameter $b_\perp$, in Ref.~\cite{Iancu:2017fzn} Iancu and Rezaeian have computed explicitly the angular dependence of the partial amplitude for dipole scattering off such sources in the target (this model will then be called MV-IR in what follows). This has been worked out for both smooth and dense targets such as nucleon and for a lumpy target like a nucleus. An important observation here is that the elliptic gluon density (and hence the azimuthal asymmetry) has been shown to originate as a genuinely non-perturbative phenomenon significantly impacted by soft or semi-hard gluon multiple scattering effects at characteristic momentum scales of order of the scale of inhomogeneities in the target (i.e.~the saturation scale). By focusing on valence quark distributions as a starting point, the MV model relies on a semi-classical pQCD-motivated approach that may not reflect the full picture of the soft QCD phenomena. It is believed, however, that it, at least, captures their most pronounced characteristics signifying the relevance of the realistic $b_\perp$-dependence in the dipole scattering process. Important physical parameters of the MV-IR model \cite{Iancu:2017fzn} are the effective gluon mass setting up the gluon confinement scale, the width of the Gaussian source distribution in the transverse plane as well as the saturation scale characterising the transverse inhomogeneity of the target. A significant model dependence and variations among predictions (e.g. for $v_2$) existing in the literature typically come from different modelling of these intrinsically non-perturbative ingredients.

It has been demonstrated in Ref.~\cite{Iancu:2017fzn} that the inhomogeneity in the transverse plane is largest at the edge of the target, hence, causing a large $v_2(p_\perp)$ (peaked at $p_\perp\sim Q_s$) for peripheral collisions, while multiple scatterings strongly impact $v_2$ causing a change in its sign. Let us elaborate on this phenomenon in more detail as it would help us to shed light on the common origin of elliptic flow in nucleon and nuclear collisions.

In the original MV model \cite{McLerran:1993ka} formulated for the case of very large (hence, homogeneous) target nucleus with atomic mass number $A$, the nuclear gluon density is assumed to be saturated in the high-energy limit, such that its small-$x$ QCD evolution is ignored. In this regime (which can, in principle, be considered as an initial condition for the high-energy QCD evolution), one views the target as a collection of `valence' color-field sources $\rho^a(\bm{b})$ whose two-point correlation function reads
\begin{eqnarray}
\langle \rho^a(\bm{b})\rho^b(\bm{b}')\rangle = 
\delta^{ab}\delta^{(2)}(\bm{b} - \bm{b}')\mu(\bm{b})\,,
\label{rho-corr}
\end{eqnarray}
in terms of the uniform color-charge squared of the valence quarks per unit area $\mu(\bm{b})\to \mu_0={\rm const}$ within a disk of radius $R\propto A^{1/3}$. As was advocated in Ref.~\cite{Iancu:2017fzn}, inhomogeneities in the transverse color-charge distribution are responsible for azimuthal asymmetries in particle production. To this end, a phenomenologically inspired impact-parameter dependence of the isotropic color-charge distribution $\mu(\bm{b}) \equiv \mu(b_\perp)$ has been introduced in the MV-IR model in a simple Gaussian form
\begin{eqnarray}
\mu(b_\perp) = \mu_0 \,e^{-b_\perp^2/4R^2} \,,\qquad 
\mu_0 = \frac{\alpha_s}{2R^2} \,,
\label{mub-inhom}
\end{eqnarray}
suggested by successful fits to HERA data on diffractive production processes as well as on vector meson production in DIS of Refs.~\cite{Kowalski:2003hm,Kowalski:2008sa,Rezaeian:2012ji,Rezaeian:2013tka}. Such a generalised MV-IR model has been formulated for both the proton and nuclear targets as an attempt of a unified description of flow-type effects in events with high-multiplicity final states in $pp$ and heavy-ion collisions. While $\mu_0$ scale characterises the saturation momentum squared at $b_\perp=0$, in Ref.~\cite{Iancu:2017fzn} the Gaussian width $R\simeq 0.3$ fm (setting up the soft scale $1/R\sim \Lambda_{\rm QCD}$) and the semi-hard saturation scale squared, 
\begin{eqnarray}
Q^2_s(\bm{b})\equiv Q^2_s(b_\perp) = \alpha_s C_F \mu(b_\perp) = Q_{0s}^2\, e^{-b_\perp^2/4R^2}\,, \qquad Q^2_{0s} \equiv \alpha_s C_F\mu_0 \,,
\label{Qsb}
\end{eqnarray}
(here, $C_F=(N^2-1)/2N$ for $SU(N)$) have been taken from the phenomenological fits to the HERA data in the framework of IP-Sat model \cite{Kowalski:2003hm,Rezaeian:2012ji}. 

Employing the eikonal approximation, in the MV model it is assumed that the color dipole scatters off the sources of the color field independently. For a single such scattering, the forward amplitude corresponds to an exchange of two gluons in the $t$-channel and reads,
\begin{eqnarray}
N_{2g}(\bm{x},\bm{y}) = \frac{g^2C_F}{2} \Big[\gamma(\bm{x},\bm{x}) + 
\gamma(\bm{y},\bm{y}) - 2\gamma(\bm{x},\bm{y})\Big]\,,
\label{N2g}
\end{eqnarray}
in the impact parameter representation, where $\gamma(\bm{x},\bm{y})=\gamma(\bm{y},\bm{x})$ represents a partial elastic amplitude, with one of the two gluons coupled to the quark at transverse position $\bm{x}$ and another one -- to the antiquark at position $\bm{y}$. The latter can be found in terms of the two-point correlation function of the two-dimensional Coulomb-like color fields $A^-_a(\bm{x})$ of the ultrarelativistic color charges,
\begin{eqnarray}
\langle A^-_a(\bm{x})A^-_b(\bm{y})\rangle = \delta^{ab}\gamma(\bm{x},\bm{y})\,, \qquad
A^-_a(\bm{x}) = \int d^2\bm{z} \,G(\bm{x}-\bm{z})\rho^a(\bm{z})\,,
\label{AA}
\end{eqnarray}
in terms of the two-dimensional effective Coulomb propagator,
\begin{eqnarray}
G(\bm{b}) = \int \frac{d^2\bm{q}}{(2\pi)^2} \, 
\frac{e^{i\bm{q}\cdot\bm{b}}}{q_\perp^2+m_g^2} = \frac{1}{2\pi}
{\rm K}_0(m_g\,b_\perp)\,,
\label{Coul-prop}
\end{eqnarray}
where the modified Bessel function of the second kind ${\rm K}_0$ exponentially falls at large separations $m_g\,b_\perp\gg 1$. One should notice here that in the case of $m_g\to 0$ the small-$x$ gluon field of the color source would extend far beyond the typical size of the source itself due to a power-like decay of the Coulomb propagator, in immediate contradiction to confinement. Hence, the presence of the effective gluons mass $m_g\sim \Lambda_{\rm QCD}$ manifestly introduces confinement into the picture which is an important physical property of the gluon propagator even though the amplitude $N_{2g}(\bm{x},\bm{y})$ is well-behaved in the infrared limit in perturbative $m_g\to 0$ regime of the theory. Using (\ref{rho-corr}) and (\ref{AA}), one obtains
\begin{eqnarray}
\gamma(\bm{x},\bm{y}) = \int \frac{d^2\bm{q} }{(2\pi)^2}\frac{d^2\bm{q}' }{(2\pi)^2} \,
e^{i\bm{q}' \cdot \bm{x} + i\bm{q} \cdot \bm{y}}\, \frac{\tilde{\mu}(\bm{q}' + \bm{q})}{(q_\perp^{'2} + m_g^2) (q_\perp^2 + m_g^2)}\,,
\end{eqnarray}
in terms of the Fourier transform of the color-charge distribution, $\tilde{\mu}(\bm{q})\equiv \tilde{\mu}(q_\perp)$, with $\tilde{\mu}(0)=4\pi R^2\mu_0$, such that the area of valence charge distribution of the proton is $4\pi R^2$ and hence the transverse proton size is $2R$. Introducing the average transverse momentum transfer between the quark and the target, $\bm{k}=(\bm{q}' - \bm{q})/{2}$, and a difference between the momentum transfers between the amplitude and its conjugate characterising the inhomogeneity of the target, $\bm{\Delta}=\bm{q}' + \bm{q}$, one finally arrives at \cite{Iancu:2017fzn}
\begin{eqnarray} \nonumber
N_{2g}(\bm{r},\bm{b}) &=& \frac{g^2C_F}{2} \int \frac{d^2\bm{\Delta} }{(2\pi)^2}\frac{d^2\bm{k}}{(2\pi)^2} \,
\frac{\tilde{\mu}(\bm{\Delta})}{((\bm{k} + \bm{\Delta}/2)^2 + m_g^2) ((\bm{k} - \bm{\Delta}/2)^2 + m_g^2)}\\
&\times& e^{i\bm{\Delta} \cdot \bm{b}} \left[ e^{i\bm{\Delta} \cdot \bm{r}/2} + e^{-i\bm{\Delta} \cdot \bm{r}/2} - 2 e^{i\bm{k} \cdot \bm{r}}  \right] \,,
\label{N2g-full}
\end{eqnarray}
with the standard dipole variables -- the dipole impact parameter with respect to the target centre, $\bm{b}=(\bm{x} + \bm{y})/2$, and the dipole separation, $\bm{r}=\bm{x} - \bm{y}$. As the integral above is controlled by predominantly soft (or semi-hard) gluon exchanges, $\Delta_\perp \lesssim 1/R$, one could apply the eikonal approximation where multiple gluon exchanges exponentiate yielding the Glauber approximation for the dipole $S$-matrix, $S=\exp(-N_{2g})$. 

For small dipoles, $r_\perp\ll b_\perp$, corresponding to the perturbative limit with the hard scale $Q^2\sim 1/r_\perp^2 \gg Q_s(b_\perp)$ concerned, the leading-order single-scattering approximation $S \simeq 1 - N_{2g}$ can be used to a good accuracy. Being an even function of both $\bm{r}$ and $\bm{b}$, $N_{2g}$ and hence $S$ are functions of the azimuthal angle $\theta$ only through powers of $(\bm{b}\cdot\bm{r})^2$. Hence, the flow parameters vanish ($v_n=0$) for any odd $n$. This can be readily seen in the single-scattering approximation, where the Fourier-transformed amplitude reads
\begin{eqnarray}
\tilde{N}_{2g}(\bm{p},\bm{b}) &\equiv& \int d^2\bm{r} \, e^{i \bm{p}\cdot\bm{r}} \,N_{2g}(\bm{r},\bm{b}) \nonumber \\
&\simeq& 
-\frac{g^2C_F}{p_\perp^4}\,\mu(b_\perp)\left[1-\frac{b_\perp^2}{8p_\perp^2R^4}\cos(2\phi)\right]\,, \qquad 
p_\perp R \gg 1 \,,
\label{N2g_SSA}
\end{eqnarray}
for an exchanged single hard gluon with momentum (having the same momentum as the final-state quark) $p_\perp\equiv k_\perp$, hence resolving a localized valence substructure in the target. In this regime, we are not sensitive to a gluon mass. The angular correlation emerges at a sub-leading $1/p_\perp^6$ order such that the cross section reaches a maximum at $\pi/2$. As a result, the produced quark is more likely to move perpendicular to its impact parameter vector leading to a negative leading (elliptic) flow coefficient,
\begin{eqnarray}
\label{flow-SS}
v_2(p_\perp,b_\perp) \simeq -\frac{b_\perp^2}{16p_\perp^2R^4} \,,
\end{eqnarray}
while the higher harmonics appear in the higher-order terms in $1/p_\perp$-expansion and hence are suppressed relative to $v_2$. As expected, the correlation for very small dipoles $r_\perp\ll R$ and for central collisions $b_\perp\to 0$ should vanish as the dipole orientation makes no impact on the production cross section in those cases. Besides, the $\cos(2\phi)$ term in the partial dipole amplitude (\ref{N2g_SSA}), and hence the $v_2$ coefficient, are proportional to derivative of the valence color-charge distribution $\mu(b_\perp)$ over the impact parameter $b_\perp$. This illustration provides a clear picture of azimuthal correlations that can be relevant only for soft processes and in the case of non-central impacts, and are generated essentially due to the presence of inhomogeneities in the transverse gluon distribution in the target \cite{Iancu:2017fzn}.

For softer gluon exchanges, i.e. for $p_\perp\lesssim Q_s(b_\perp)$, particularly relevant for understanding the collective phenomena in soft particle production in $pp$ and $pA$ collisions, the single-scattering approximation is no longer valid, and the Glauber-like exponentiated dipole $S$-matrix $S(\bm{b},\bm{r}) = \exp(-N_{2g}(\bm{b},\bm{r}))\to \tilde S(\bm{b},\bm{p})$ should be considered. As the multiple gluon exchange is concerned, the momentum $\bm{p}$ acquired by final-state quark in the course of its propagation through the target can be much larger than the momentum $\bm{k}$ exchanged in a single scattering. Moreover, as has been emphasised in Ref.~\cite{Iancu:2017fzn}, the angular correlations are controlled by soft momentum exchanges, i.e. with $1/r_\perp \gg k_\perp\sim \Delta_\perp$, so one can no longer expand the amplitude in power series over $\Delta_\perp/k_\perp$. Instead, expanding (\ref{N2g-full}) in the limit of small $r_\perp$ and keeping the leading order terms only,
\begin{eqnarray} \nonumber
N_{2g}(\bm{r},\bm{b}) &\simeq&
\frac{g^2C_F}{2}\,r^j r^l \int \frac{d^2\bm{\Delta} }{(2\pi)^2}
\frac{d^2\bm{k} }{(2\pi)^2}  \frac{
\big(k^jk^l-\Delta^j\Delta^l/4\big)\tilde{\mu}(\Delta_\perp)}{
[(\bm{k} + \bm{\Delta}/2)^2 + m_g^2][(\bm{k} - \bm{\Delta}/2)^2 + m_g^2]} \,e^{i\bm{\Delta} \cdot \bm{b}} \\
&=& {\cal N}_0(r_\perp,b_\perp) + {\cal N}_\epsilon(r_\perp,b_\perp) \cos(2\theta) \,,
\label{N2g-small_r}
\end{eqnarray}
where spatial indices $j,l=1,2$ are summed, the dependence on the gluon mass $m_g$ becomes relevant restricting the phase space allowed for soft gluons, $k_\perp\lesssim \Lambda_{\rm QCD}\sim m_g$, and typically $|{\cal N}_\epsilon|\ll |{\cal N}_0|$. As a result, the elliptic flow coefficient (\ref{v2-pointlike}) takes the form,
\begin{eqnarray}
v_2(p_\perp,b_\perp) =\frac{\int r_\perp dr_\perp\,  e^{-\mathcal{N}_0(r_\perp,b_\perp)}\,J_2(p_\perp r_\perp)\,I_1\left(\mathcal{N}_\epsilon(r_\perp,b_\perp)\right)}{\int r_\perp dr_\perp\, e^{-\mathcal{N}_0(r_\perp,b_\perp)}\,J_0(p_\perp r_\perp)\,I_0\left(\mathcal{N}_\epsilon(r_\perp,b_\perp)\right)} \,. 
\label{v2N}
\end{eqnarray}
where $I_0$ and $I_1$ are the modified Bessel functions of the first kind. Due to the inhomogeneities in the target, the quantity $\Delta_\perp$ effectively plays a role of an infrared cutoff in the $k_\perp$-integration, and no infrared divergencies are developed. In the UV limit, a logarithmic divergence is effectively cut off by the dipole separation, i.e. at $k_\perp\sim 1/r_\perp$, such that the leading term at large $k_\perp$ can be subtracted from the integrand in Eq.~(\ref{N2g-small_r}), and then the corresponding contribution to the integral can be replaced by a regularised term,
\begin{eqnarray}
\frac{g^2C_F}{4}\,r_\perp^2 \int \frac{d^2\bm{\Delta} }{(2\pi)^2}\frac{d^2\bm{k} }{(2\pi)^2}  \frac{k_\perp^2 \tilde{\mu}(\Delta_\perp)}{(k_\perp^2+m_g^2)^2} \,e^{i\bm{\Delta}\cdot \bm{b}} = \frac{Q_s^2(b_\perp)r_\perp^2}{4} \ln\left(\frac{1}{r_\perp^2m_g^2}+e\right)\,,
\end{eqnarray}
with the saturation scale defined in Eq.~(\ref{Qsb}). Here, the logarithmic term emerges due to $k_\perp$-integration over $k_\perp\subset [m_g,1/r_\perp]$, while the constant term fixes the renormalization scheme. 
\begin{figure}[h!]
\begin{center}
  \includegraphics[width=0.45\textwidth]{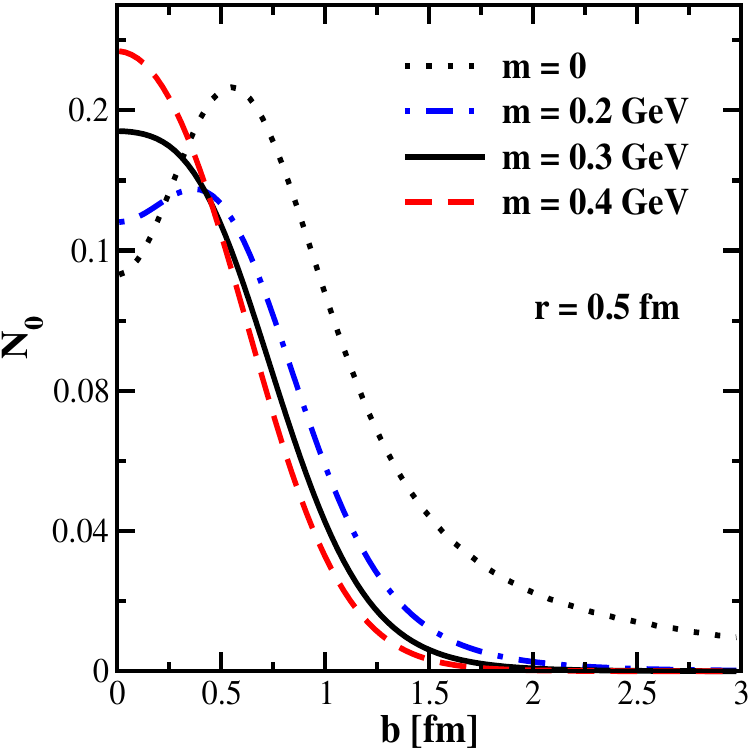}
  \includegraphics[width=0.45\textwidth]{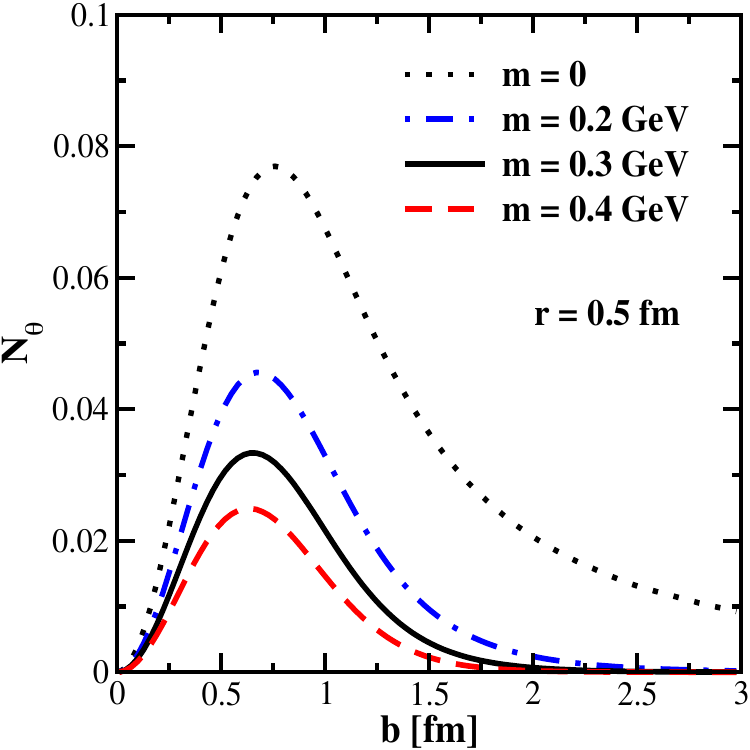}
\end{center}
\caption{Left: isotropic component $N_0\equiv \mathcal{N}_0$ computed in the MV-IR model using Eq.~(\ref{N0}) averaged over the angles; right: elliptic component $N_{\theta}\equiv \mathcal{N}_\epsilon$ encoding the angular dependence of the dipole amplitude computed in the MV-IR model using Eq.~(\ref{N1}), both plotted as a function of $b\equiv b_\perp$ for a fixed dipole size $r\equiv r_\perp = 0.5$~fm and various values of the infrared cutoff $m_g$. All the curves are obtained using $R^2 = 2$~GeV$^{-2}$ and $Q^2_{0s} = 0.165$~GeV$^2$. From Ref.~\cite{Iancu:2017fzn}.}
\label{fig:N0N1-MV}
\end{figure}
As a result of the MV-IR model, the isotropic $\mathcal{N}_0(r_\perp,b_\perp)$ and elliptic $\mathcal{N}_\epsilon(r_\perp,b_\perp)$ components of the amplitude in Eq.~(\ref{N2g-small_r}) for the proton target can be represented in the following integral form \cite{Iancu:2017fzn}:
\begin{eqnarray}
\mathcal{N}_0(r_\perp,b_\perp)= \frac{Q_s^2(b_\perp)r_\perp^2}{4} \ln\left(\frac{1}{r_\perp^2m_g^2}+e\right)  + \frac{g^2C_F}{4(2\pi)^2} r_\perp^2 \int_{0}^\infty d\Delta_\perp\,\Delta_\perp\,  \tilde{\mu}(\Delta_\perp) J_0(\Delta_\perp b_\perp)\nonumber\\
\times \int_{0}^{\infty} dk_\perp\,k_\perp\Bigg[\frac{ k_\perp^2 - \Delta_\perp^2/4 }{\left(k_\perp^2+\Delta_\perp^2/4+m_g^2\right) \left( \left(k_\perp^2+\Delta_\perp^2/4+m_g^2\right)^2-k_\perp^2\Delta_\perp^2\right)^{1/2}} - \frac{k_\perp^2}{(k_\perp^2+m_g^2)^2} \Bigg]\,,\nonumber\\
\label{N0}
\end{eqnarray}
\begin{eqnarray}
\mathcal{N}_\epsilon(r_\perp,b_\perp) &=& \frac{g^2C_F}{4(2\pi)^2} r_\perp^2 \int_{0}^\infty d\Delta_\perp \,\Delta_\perp\, \tilde{\mu}(\Delta_\perp) J_2(\Delta_\perp b_\perp) \nonumber \\
&&\times \int_{0}^{\infty} dk_\perp\,  k_\perp\Bigg[\frac{k_\perp^2 + \Delta_\perp^2/4}
{\left(k_\perp^2+\Delta_\perp^2/4+m_g^2\right) \left( \left(k_\perp^2+\Delta_\perp^2/4+m_g^2\right)^2-k_\perp^2\Delta_\perp^2 \right)^{1/2}}
 \nonumber\\*[0.2cm]
 && \hspace{3.cm}  \ \ +\frac{2}{\Delta_\perp^2} - 
\frac{2\left(k_\perp^2+\Delta_\perp^2/4+m_g^2\right)}{ \Delta_\perp^2 \left(\left(k_\perp^2+\Delta_\perp^2/4+m_g^2\right)^2-k_\perp^2\Delta_\perp^2  \right)^{1/2}} \Bigg] \,, \label{N1}
\end{eqnarray}
where the $k_\perp$-integrals are restricted to $k_\perp\subset [m_g,\Delta_\perp]$. The first term in Eq.~(\ref{N0}) corresponds to one in the original MV model \cite{McLerran:1993ka}. Taking $\Delta_\perp$-integral analytically one arrives at \cite{ReinkePelicer:2018gyh}
\begin{align}
\label{N0p}
\mathcal{N}_0(r_\perp,b_\perp) =& \frac{Q_{0,s}^2 r_\perp^2}{4}  e^{-b_\perp^2/4R^2} \text{ln} \left( \frac{1}{r_\perp^2 m_g^2} + e\right) + Q_{0,s}^2 R^2 r_\perp^2 \int_0^\infty k_\perp d k_\perp e^{-k_\perp^2 R^2} \nonumber \\
& \times J_0(b_\perp k_\perp) \frac{k_\perp \sqrt{k_\perp^2 + 4 m_g^2} - 2 (k_\perp^2 + 2m_g^2) \arctanh{\left( \frac{k_\perp}{\sqrt{k_\perp^2 + 4 m_g^2}} \right)} }{2 k_\perp \sqrt{k_\perp^2 + 4 m_g^2}} \,,
\end{align}
\begin{align}
\label{N1p}
\mathcal{N}_\epsilon(r_\perp,b_\perp) =  Q_{0,s}^2 R^2 r_\perp^2 \int_0^\infty k_\perp dk_\perp & e^{-k_\perp^2 R^2}  J_2(b_\perp k_\perp) \nonumber \\
&\times \frac{k_\perp \sqrt{k_\perp^2 + 4 m_g^2} - 4 m_g^2 \arctanh{\left( \frac{k_\perp}{\sqrt{k_\perp^2 + 4 m_g^2}} \right)} }{2 k_\perp \sqrt{k_\perp^2 + 4 m_g^2}} \,,
\end{align}
which are rather straightforward to use in practical calculations. The functions $\mathcal{N}_0(r_\perp,b_\perp)$ and $\mathcal{N}_\epsilon(r_\perp,b_\perp)$ are illustrated in Fig.~\ref{fig:N0N1-MV} (left and right panels, respectively).

For completeness, we also highlight the MV-IR result of Ref.~\cite{Iancu:2017fzn} for the nucleus target,
\begin{align}
\mathcal{N}^A_0(r_\perp,b_\perp) & = \pi R^2 Q_{0,s}^2 r_\perp^2\text{ln} \left(\frac{1}{r_\perp^2 m_g^2} + e \right) \left[T_A(b_\perp) + R^2 \left(T_A''(b_\perp) + \frac{1}{b_\perp} T_A'(b_\perp) \right) \right] \nonumber \\
& + \frac{\pi R^2}{3 m_g^2} Q_{0,s}^2 r_\perp^2 \left(T_A''(b_\perp) + \frac{1}{b_\perp} T_A'(b_\perp) \right) \,, \label{MV-Wigner-N0A} \\
\mathcal{N}^A_\epsilon(r_\perp,b_\perp) & = \frac{\pi R^2}{6 m_g^2} Q_{0,s}^2 r_\perp^2 \left(T_A''(b_\perp) - \frac{1}{b_\perp} T_A'(b_\perp) \right) \,,
\label{MV-Wigner-N1A}
\end{align}
given in terms of the nuclear thickness function,
\begin{equation}
T_A(b_\perp) = \int dz \rho_A\big(\sqrt{b_\perp^2 + z^2}\big) \,,
\label{thickness}
\end{equation}
where
\begin{equation}
    \rho_A(\bm{r}) = N_A \,\left[1 + \exp\left((r_\perp - R_A)/\delta \right) \right]^{-1} \,, \qquad \int d^3 \bm{r} \rho_A(\bm{r}) = 1 \,,
    \label{WS-density}
\end{equation}
is the normalised nuclear density distribution in the Woods-Saxon parametrization \cite{DeVries:1987atn}, with the nuclear radius $R_A = (1.12$ fm$)A^{1/3}$ and $\delta=0.54$\,fm.

Notably, in the limit of a homogeneous target, $\tilde{\mu}(\bm{\Delta}) \propto \delta^{(2)}(\bm{\Delta})$, only the first term in $\mathcal{N}_0(r_\perp,b_\perp)$ remains while the elliptic component $\mathcal{N}_\epsilon(r_\perp,b_\perp)$ vanishes. Hence, in the considered MV-IR model formulation the azimuthal angle correlations are indeed due to transverse inhomogeneities in the source distribution of the target. Besides, the logarithmic divergence for $m_g\to 0$ in $\mathcal{N}_0(r_\perp,b_\perp)$ is cancelled by the one coming from the second term yielding infrared stability of the result. Yet, $m_g\sim 0.3$ GeV is there on physical grounds effectively cutting off too soft gluon momenta, $k_\perp\lesssim \Delta_\perp$, not consistent with confinement. As was shown in Ref.~\cite{Iancu:2017fzn} starting from Eq.~(\ref{v2N}), multiple scattering effects turn $v_2$ positive and large for peripheral collisions, i.e.~$b_\perp\gtrsim R$, whereas it is negative in the single scattering approximation, see Eq.~(\ref{flow-SS}).
\begin{figure}[h!]
\begin{center}
  \includegraphics[width=0.45\textwidth]{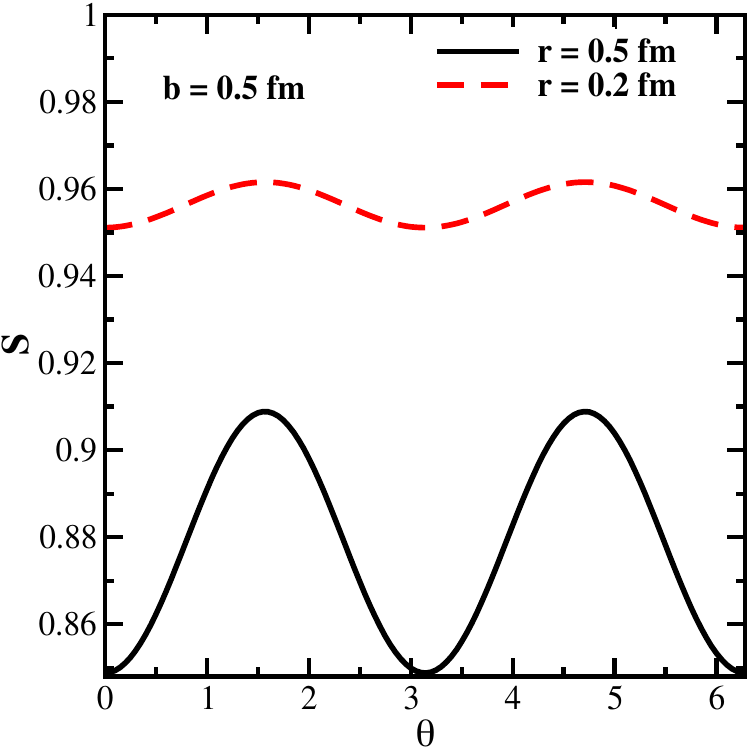}
  \includegraphics[width=0.45\textwidth]{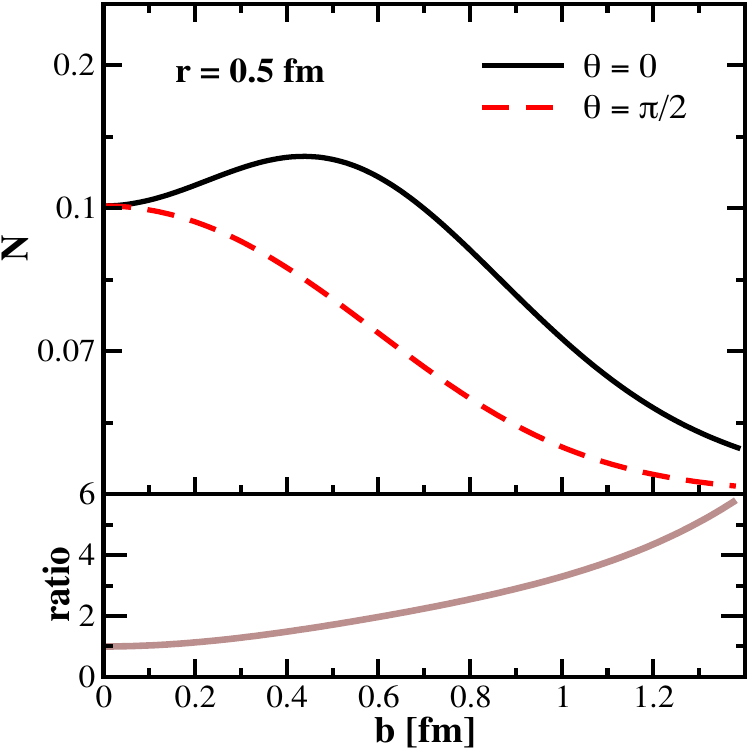}
\end{center}
\caption{Left: The dipole `survival probability' in the MV-IR model as a function of $\theta$ for a fixed $b\equiv b_\perp$ and two values of $r\equiv r_\perp$. Right: the dipole amplitude $N_{2g}(r,b,\theta)$ as a function of $b\equiv b_\perp$ for a fixed value of $r\equiv r_\perp$ and the two extreme possibilities for the orientation: $\theta=0$ and $\theta=\pi/2$. The lower insert shows the ratio $N_{2g}(\theta=0)/N_{2g}(\theta=\pi/2)$. All curves are obtained by numerically evaluating $N_{2g}$ according to Eqs.~\eqref{N2g-small_r}, \eqref{N0} and \eqref{N1}, together  with $m_g=0.25$~GeV, $R^2=2\,\text{GeV}^{-2}$ (i.e. $R\simeq 0.3$~fm), and $Q_{0s}^2=0.165\,\text{GeV}^2$. From Ref.~\cite{Iancu:2017fzn}.}
\label{fig:S-MV}
\end{figure}

The MV-IR model enables one to study quantitatively the dipole orientation phenomena encoded in the dipole $S$-matrix. Using Eqs.~(\ref{N0}) and (\ref{N1}) one obtains the behavior of $S(r_\perp,b_\perp,\theta)$ and $N_{2g}(r_\perp,b_\perp,\theta)$ shown in Fig.~\ref{fig:S-MV} (left and right, respectively). One immediately notices that, as expected, the dipole survival probability is largest for $\bm{b} \perp \bm{r}$ and smallest for $\bm{b} || \bm{r}$, while the target inhomogeneities become irrelevant and the dipole orientation effect disappears for small $b_\perp\lesssim R=0.3$ fm.

For illustration, Ref.~\cite{Linek:2023kga} shows the isotropic part of the gluon GTMD in the forward kinematics, $T_0(k_\perp,\Delta_\perp=0.01\,{\rm GeV})$, for a fixed $x\equiv x_{\Pom}$, obtained using Eq.~(\ref{eq_T-0}) for different selected models for the partial dipole amplitude ${\cal N}$ briefly reviewed above. Such a comparison reveals a significant theoretical uncertainty in modelling of the gluon GTMD in the soft regime of $k_\perp \lesssim m_c$ GeV, even in its isotropic part. Ref.~\cite{Linek:2023kga} further illustrates how such uncertainties in the gluon GTMD modelling are translated into the uncertainties in determination of the differential cross sections of the exclusive diffractive $c\bar c$ photoproduction. Theoretical uncertainties in modelling the elliptic component of the Wigner distribution (or GTMD) are found to be even larger as was discussed extensively in Refs.~\cite{Hagiwara:2017fye,ReinkePelicer:2018gyh,Linek:2023kga} posing a natural question of constraining it directly from the experimental data.
\begin{figure}
  \centering
  \includegraphics[width=.4\textwidth]{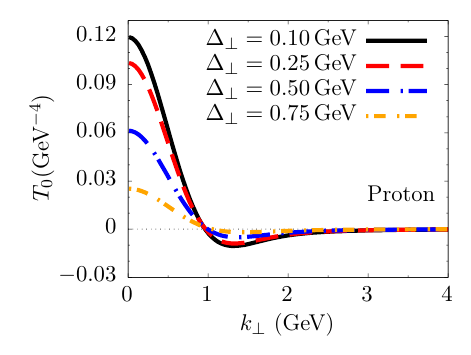}
  \includegraphics[width=.4\textwidth]{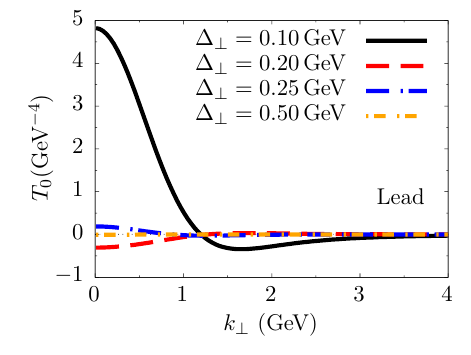}
  \includegraphics[width=.4\textwidth]{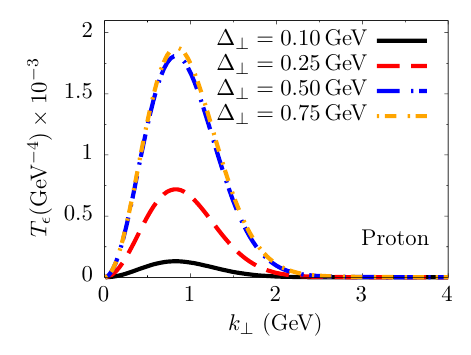}
  \includegraphics[width=.4\textwidth]{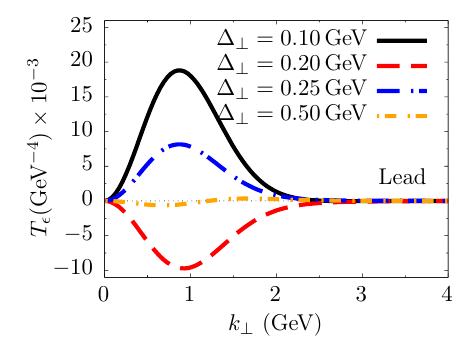}
  \caption{The isotropic $T_0$ (upper panels) and elliptic $T_\epsilon$ (lower panels) components of the dipole $T$-matrix computed in the MV-IR model as functions of $k_\perp$ for different values of $\Delta_\perp$ and for different targets -- the proton target (left panels) and lead nucleus ($A=208$) target (right panels). Here, $R=2$ GeV$^{-1}$, $m_g = 0.25$ GeV have been used. From Ref.~\cite{ReinkePelicer:2018gyh}.}
\label{fig:T0-vs-Teps}
\end{figure}

Fig.~\ref{fig:T0-vs-Teps} illustrates the forward dipole $T$-matrix components -- the isotropic $T_0$ (upper panels) and elliptic $T_\epsilon$ (lower panels) ones -- computed in the MV-IR model using Eqs.~(\ref{eq_T-0}) and (\ref{eq_T-eps}), respectively, as functions of $k_\perp$ for different values of $\Delta_\perp$ and for different targets -- the proton target (left panels) and lead nucleus target (right panels) \cite{ReinkePelicer:2018gyh}. The elliptic effect in nuclear targets is much more pronounced than in the proton in general. Nevertheless, it get quite significant at $k_\perp \sim \Delta_\perp$ even for the proton case, while the ballpark of sensitivity to variations in small $\Delta_\perp$ occurs for $k_\perp\lesssim 2$ GeV.

\begin{figure}[h!]
\begin{center}
  \includegraphics[width=0.9\textwidth]{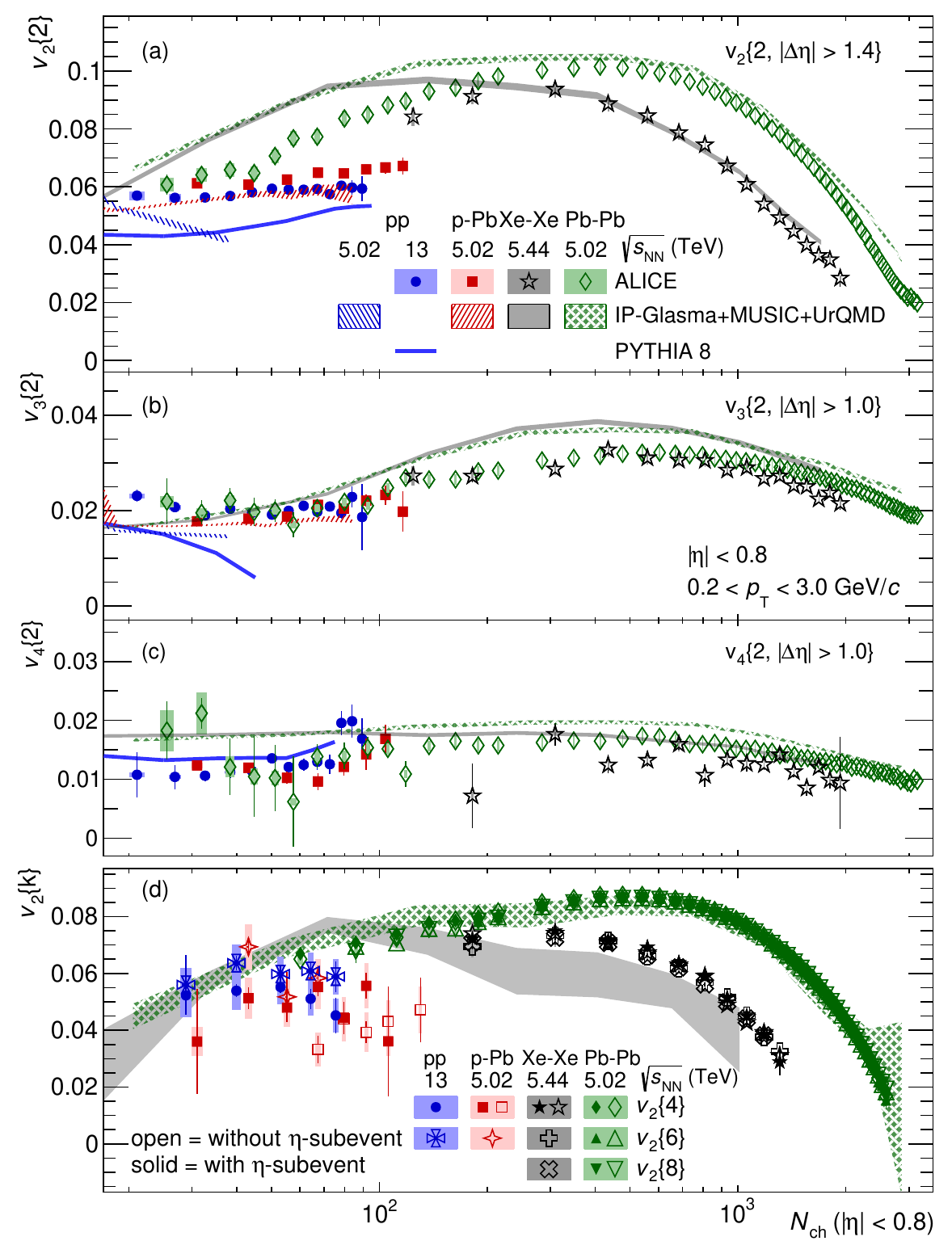}
\end{center}
\caption{Multiplicity dependence of $v_n\{k\}$ for $pp$, pPb, XeXe and PbPb collisions measured by the ALICE experiment and compared with PYTHIA 8.210 Monash 2013~\cite{Sjostrand:2014zea} (solid lines) of $pp$ collisions at $\sqrt{s}$ = 13~TeV and IP-Glasma+MUSIC+UrQMD~\cite{Mantysaari:2017cni,Schenke:2013dpa} calculations of $pp$, pPb, PbPb collisions at $\sqrt{s_{\rm NN}}$ = 5.02~TeV and XeXe collisions at $\sqrt{s_{\rm NN}}$ = 5.44~TeV (filled bands). From Ref.~\cite{ALICE:2019zfl}.}
\label{ALICE-vn}
\end{figure}

To conclude this part, one typically implies that the effects of hydrodynamic evolution become increasingly important and even dominate for large collision systems (especially, at large particle multiplicities), while at low particle multiplicities, often probed in collisions of small systems, initial-state effects are expected to dominate. This is, for example, illustrated in Fig.~\ref{ALICE-vn} summarizing measurements of the particle multiplicity dependence of $v_2$ using 2, 4, 6 and 8-particle cumulants, and $v_3$ and $v_4$ using 2-particle cumulants for $pp$, pPb, PbPb and XeXe collisions. Comparisons of the data with prediction of the hydrodynamic model MUSIC using IP-Glasma for initial conditions and UrQMD for hadronic rescattering \cite{Mantysaari:2017cni,Schenke:2013dpa} may indicate that the overall description could be improved once measurements of the gluon Wigner function become available.

\section{Phenomenological probes for dipole orientation effects}
\label{Sect:pheno}

Here, we would like to elaborate on several phenomenologically important examples of the processes particularly relevant for exploration of the nucleon or nucleus structure through measurement of the azimuthal angle correlations in the particle production processes.

\subsection{Exclusive light-quark dijet photoproduction}
\label{Sect:dijet-photo}

A multitude of new experimental data on various exclusive diffractive processes in hadronic and nuclear collisions are now becoming available from the LHC and making use of these data for detailed studies of the Wigner distribution gains its importance. One of the important classes of diffractive processes that can be utilized for probing non-trivial correlations in phase space parton distributions is the meson-pair photoproduction. For instance, in Ref.~\cite{Hagiwara:2021xkf} it has been shown that the azimuthal $\cos 4\phi$-asymmetry in exclusive di-pion production close to the $\rho^0$-resonance in UPCs recently observed by STAR Collaboration can receive sizeable contributions from the elliptic gluon Wigner distribution. Another important class of processes for probing the gluon tomography concerns diffractive electro- or photoproduction of light- and heavy-quark jets which will be elaborated upon in detail below.

Typically, an extraction of GTMD directly from the data is a rather challenging task due to the fact that a typical cross section, such as that of a semi-inclusive DIS process or an exclusive vector meson production, is computed in terms of an integral convolution of the GTMD with other kinematical functions and an inversion of such an integral is, in general, a very difficult problem. As a way out, one could consider other possible reactions where such an inversion can be realised in a simpler way. A recent proposal to reconstruct both the isotropic and elliptic components of the gluon GTMDs (and hence the gluon Wigner distribution) at low-$x$ through a measurement of exclusive diffractive dijet photoproduction process $\gamma T \to (q\bar q) T$ illustrated in Fig.~\ref{fig:dijet-UPC}, where the target $T$ can be either proton $p$ or a nucleus $A$, has been made in Ref.~\cite{Hatta:2016dxp,Hagiwara:2017fye} (see also a recent work of Ref.~\cite{Boer:2021upt} for more details). In the same way as in the DIS process discussed above, the emerging colorless $q\bar q$ state, the color dipole, is a lowest-Fock fluctuation of the projectile photon $\gamma\to q\bar q$ that plays an important role of the probe of the color background field (and hence substructure) of the target $T$. For comprehensive analysis of this process in the framework of color dipole approach, see e.g. Refs.~\cite{Altinoluk:2015dpi,Mantysaari:2019csc,Salazar:2019ncp} and references therein. 

Such a process can be realised e.g. in $ep$ and $eA$ collisions as well in ultra peripheral $pA$ and $AA$ collisions (UPCs) where an electron, a proton or a heavy projectile nucleus of mass number $A$ is considered as a source of quasi-real photons while another proton or nucleus plays a role of the target whose internal structure is being probed. As will be discussed below, this type of processes can be efficiently studied in UPCs at the LHC, as well as at RHIC and a future EIC providing plausible opportunities for probing the elliptic component theoretically expected to be at the level of only a few percent effect in the perturbative regime. For instance, the EIC \cite{Boer:2011fh,Accardi:2012qut,Aschenauer:2017jsk} is expected to enable proton tomography with unprecedented precision through reconstructing partons' positions in the impact parameter space, their transverse momenta and spins in the proton and nucleus. This may be achieved via precision measurements of diffractive dijet production observables as functions of target's recoil momentum and dijet's total and relative transverse momenta.

The colliding states in UPCs such as relativistic nucleons and/or nuclei scatter at large impact parameters interacting only electromagnetically through an exchange of quasi-real (and hence mostly transversely polarised) WW photons \cite{vonWeizsacker:1934nji,Williams:1934ad}. Let us consider this process in the nucleus-target center-of-mass frame where the WW photon is collinear to the beam axis, with energy $\omega$. In the light-cone decomposition, its momentum reads: $q=(\sqrt{2}\omega,0,\bm{0})$. One then introduces the quark and antiquark $x_{1,2}$ momentum fractions of the light-front plus momentum of the projectile nucleus, respectively, written in terms of their transverse momenta $k_{1,2\perp}$ and center-of-mass rapidities $y_{1,2}$ as
\begin{eqnarray}
x_{1,2} = \sqrt{\frac{k_{1,2\perp}^2 + m_q^2}{s}} \, e^{y_{1,2}}\,,
\end{eqnarray}
while the quark and antiquark light-cone momentum components read
\begin{eqnarray*}
    k_1^+ = z\sqrt{2}\omega\,, \qquad k_2^+=(1-z)\sqrt{2}\omega\,, \\
    \bm{k}_1 = - \bm{P} - \frac{\bm{\Delta}}{2} \,, \qquad \bm{k}_2 = \bm{P} - \frac{\bm{\Delta}}{2} \,.
\end{eqnarray*}
where $z$ is the quark longitudinal momentum fraction of the parent photon in the $\gamma\to q\bar q$ transition. Neglecting the quark mass, it reads
\begin{eqnarray}
    z=\frac{k_{1\perp}e^{y_1}}{ k_{1\perp}e^{y_1} + k_{2\perp}e^{y_2}}\,.
\end{eqnarray}

In the quark massless limit, momentum conservation then provides the standard expressions for the invariant mass squared of the produced dijet and the photon energy,
\begin{eqnarray}
    M^2_{jj} = 2k_{1\perp}k_{2\perp}(\cosh(y_1-y_2)-\cos\phi_{12})\,, \qquad \omega = \frac{1}{2}(k_{1\perp}e^{y_1} + k_{2\perp} e^{y_2}) \,,
\end{eqnarray}
respectively, in terms of the azimuthal angle $\phi_{12}$ between $\bm{k}_1$ and $\bm{k}_2$. Note, $M^2_{jj}$ plays a role of the hard scale of the process enabling the use of QCD perturbation theory for sufficiently large values of quark transverse momenta. The kinematics of the multi-gluon exchange in the $t$-channel is determined in terms of its longitudinal momentum fraction $x$ taken by such an exchange from the target as defined in Eq.~(\ref{xPom}), or equivalently, $x=M^2_{jj}/(x_\gamma s)$. At high energies and in the forward scattering domain, one requires a large rapidity gap $Y \equiv \log(x_0/x) \gg 1$ (with $x_0=0.01$ as often adopted in the literature) stretched between the recoiled target and the produced dijet.

The WW photon carries the fraction $x_\gamma \equiv x_1 + x_2$ whose flux can be written as
\begin{eqnarray}
\frac{dN}{dx_\gamma} = \frac{2Z^2\alpha_{\rm em}}{\pi x_\gamma} 
\left[ \xi K_0(\xi)K_1(\xi)-\frac{\xi^2}{2}(K_1^2(\xi)-K_0^2(\xi))\right]\,, \qquad
\xi = x_\gamma m_p(R_T+R_A) \,, \label{WW-flux}
\end{eqnarray}
in terms of the atomic number (total charge) of the projectile particle $Z$, the fine structure constant $\alpha_{\rm em}$, the radius of the target (projectile) $R_T$ ($R_A$), respectively, $\xi$ is defined in such a way to effectively eliminate an overlap between $T$ and $A$ in the transverse plane, and $m_p$ is the proton mass. The proton and nuclear radii are typically fixed as $R_p = 0.8$ fm and $R_A = (1.12$ fm$)A^{1/3}$, respectively. Consequently, in the c.m.~frame, the differential cross sections of $q\bar q$ production processes in $AT$ UPCs and $\gamma T$ collisions are related as,
\begin{eqnarray}
\frac{d\sigma^{AT\to A(q\bar q)T}}{dy_1dy_2d^2\bm{P}d^2\bm{\Delta}} = x_\gamma \frac{dN}{dx_\gamma}\,\Big(z(1-z)\frac{d\sigma^{\gamma T\to (q\bar q)T}}{dz d^2\bm{P}d^2\bm{\Delta}}\Big)\Big|_{z = \frac{x_1}{x_\gamma}}\,.
\end{eqnarray}

One of the important advantages of using a large nucleus as a source of WW photons is that a suppression due to the electromagnetic coupling $\alpha_{\rm em}$ can be largely eliminated by means of a large prefactor $Z^2$. Moreover, in the target rest frame, the projectile hadron or nucleus appears as a source of WW photons whose spectrum is rather broad, and the peak photon energy scales linearly with the Lorentz $\gamma$-factor in the target rest frame. This process has been studied earlier at the next-to-leading order in QCD in Ref.~\cite{Guzey:2016tek} where it has been shown that the corresponding dijet differential cross section shows a large extension in photon-proton invariant mass and is very sensitive to small momentum fractions of the exchanged gluons.

The shear advantage of $pA$ and $AA$ UPCs is not only a strong reduction of QCD backgrounds and a cleaner hadronic environment due to photon-induced reactions and a large relative separation between the scattered systems, but also an enhancement of the WW photon flux from the projectile nucleus $A$ due to its scaling as the square of the nuclei electric charge, $Z^2$. The measurements in UPCs do not feature complications related to a significant pileup and enable a good subtraction of non-exclusive backgrounds as previously elaborated in e.g.~Refs.~\cite{Bertulani:2005ru,Salgado:2011wc}. The LHC forward proton detectors (FPDs) such as Roman pots in TOTEM~\cite{TOTEM:2008lue}, ALFA~\cite{AbdelKhalek:2016tiv} and AFP~\cite{AFP,Tasevsky:2015xya} at ATLAS, and CT-PPS~\cite{CT-PPS} in the CMS provide the necessary means for precision analysis of exclusive processes in $pA$ and $AA$ UPCs enabling full kinematic reconstruction by detecting the forward final-state protons and possibly nuclei, in addition to reconstruction of the diffractive dijet system. 

A proper extraction of the ``elliptic'' gluon Wigner distribution (or its GTMD counterpart $\tilde{S}(q_\perp,\Delta_\perp)$) requires a detailed reconstruction of kinematics of each final-state jet. Even this may not be sufficient due to a potentially large momentum leakage into the rapidity gap between the dijet and the final-state nucleon (or nucleus). So an extra vetoing of any hadronic activity between the dijet and the forward nucleon at the detector level would enable to suppress the QCD backgrounds related to the Pomeron and/or the nucleon breakup. In order to ensure the full exclusivity of the process and the maximal reduction of the dominant backgrounds, it necessary to also reconstruct the full momentum of the forward nucleon by simultaneously measuring its energy loss and its full transverse momentum. This is in principle available by using forward proton detectors as will be discussed below. In this case, momentum conservation helps veto any additional hadronic activity in the rapidity gap and, hence, reliably probe the Pomeron-exchange contribution relevant for the Wigner distribution measurement. It may be relevant to explore a large class of inclusive and exclusive processes sensitive to the Wigner distribution (or GTMD) to make its extraction more constrained and reliable, and we will further discuss several promising examples below.

As was mentioned above, in order to reconstruct the gluon GTMD (and hence the corresponding dipole $S$-matrix), one has to identify a full set of observables sensitive to both $\bm{q}$ and $\bm{\Delta}$. In the considered example of exclusive dijet photoproduction in the forward region, one aims at precision measurement of the jet transverse momenta, $\bm{k}_1$ and $\bm{k}_2$, enabling to reconstruct the total transverse momentum transfer with the target, $\bm{\Delta}=-(\bm{k}_1 + \bm{k}_2)$. This momentum is transferred across the large rapidity gap in elastic $q\bar q$-dipole scattering off the target. The same $\bm{\Delta}$ is carried by the target in the final state which remains intact and can be, in principle, independently measured in the forward detector. The latter would substantially reduce a possible unobservable momentum leakage, both into the rapidity gap (due to a Pomeron break-up) and off the jets (parton energy loss due to a momentum leakage out of the jet cone), thus ensuring the exclusivity of the process. The full exclusivity then also requires that the projectile nucleus (source of the WW photons) also remains intact and whose kinematics is measured to precisely reconstruct the WW spectrum. The second independent kinematic variable is the relative transverse momentum between jets, $\bm{P} = \frac{1}{2}(\bm{k}_1 - \bm{k}_2)$, whose measurement provides an information about the $\bm{q}$-dependence of the GTMD.

Focusing for simplicity on the transversely polarised projectile WW photon and considering the direct photon (not resolved) process only, the differential cross section of the exclusive di-jet production in $\gamma T$-scattering reads \cite{Hatta:2016dxp},
\begin{eqnarray}
\frac{d\sigma}{dY\,d^2\bm{P} d^2\bm{\Delta}} \propto |\bm{M}|^2\,, \qquad 
\bm{M}(\bm{P},\bm{\Delta}) = \int \frac{d^2\bm{q}}{2\pi}  \frac{\bm{P} - \bm{q}}
{(\bm{P}-\bm{q})^2 + \epsilon_q^2} S_Y(\bm{q}, \bm{\Delta}) \,.
\label{diff-CS}
\end{eqnarray}
Here, the quark energy squared $\epsilon^2_q = z(1-z)Q^2+m_q^2\approx 0$ in the case of photoproduction of light quarks. Turning to the dipole representation \cite{Kopeliovich:1981pz,Nikolaev:1991et}, it is instructive to write down the $\gamma T \to (q \bar q) T$ cross section in terms of the impact parameter space ingredients,
\begin{eqnarray}
    \frac{d\sigma}{dz d^2\bm{P} d^2 \bm{\Delta}} =\overline{\sum_{\lambda_\gamma, \lambda, \bar \lambda}}
    \Big| {\cal A}_{q\bar q}^{\lambda_\gamma\lambda\bar \lambda} \Big|^2 \,, \quad 
    {\cal A}_{q\bar q}^{\lambda_\gamma\lambda\bar \lambda} = \int  \frac{d^2 \bm{b} d^2\bm{r}}{(2\pi)^2}
     e^{-i \bm{\Delta} \cdot \bm{b}} 
    e^{-i \bm{P} \cdot \bm{r}} 
    {\cal N}_Y(\bm{r}, \bm{b}) \, \Psi^{\lambda_\gamma}_{\lambda \bar \lambda}(z,\bm{r}) \,,
    \label{diff_qq-CS}
\end{eqnarray}
namely, the imaginary part of the partial dipole amplitude, ${\cal N}_Y(\bm{r}, \bm{b})$ defined in Eq.~\ref{Imfel-qq}, and the light-front wave function of the $\gamma \to q \bar q$ fluctuation, $\Psi^{\lambda_\gamma}_{\lambda \bar \lambda}(z,\bm{r})$ \cite{Kovchegov.2012}. The cross section is explicitly averaged over the photon polarisation $\lambda_\gamma$ and summed over the quark $\lambda/2$ and antiquark $\bar \lambda/2$ helicities. Using Eq.~(\ref{dipole_N-Y}), one gets \cite{Linek:2023kga}
\begin{eqnarray}
{\cal A}_{q\bar q}^{\lambda_\gamma\lambda\bar \lambda}  &\propto&   \int d^2\bm{\kappa} \, f\Big(Y, \frac{\bm{\Delta}}{2} + \bm{\kappa}, \frac{\bm{\Delta}}{2} - \bm{\kappa}\Big) 
  \, F(\bm{P},\bm{\Delta},\bm{\kappa})  \,,
  \label{A-qq_amp}
\end{eqnarray}
where the impact factor
\begin{eqnarray}
    F(\bm{P},\bm{\Delta},\bm{\kappa}) &=&  \Psi_{\lambda \bar \lambda}^{\lambda_\gamma}\Big(z,\bm{P} + \frac{\bm{\Delta}}{2}\Big) + \Psi_{\lambda \bar \lambda}^{\lambda_\gamma}\Big(z,\bm{P} - \frac{\bm{\Delta}}{2}\Big)
  - \Psi_{\lambda \bar \lambda}^{\lambda_\gamma}(z,\bm{P} + \bm{\kappa}) - 
  \Psi_{\lambda \bar \lambda}^{\lambda_\gamma}(z,\bm{P} - \bm{\kappa}) \nonumber
\end{eqnarray}
describes the coupling of two $t$-channel gluons to the $\gamma \to q \bar q$ amplitude and vanishes when $\bm{\kappa} = \pm \bm{\Delta}/2$.

In Ref.~\cite{Hatta:2016dxp} an extraction of the gluon GTMD has been proposed through a measurement of light quark jets neglecting the quark mass $m_q$ and considering $Q^2\to 0$ (photoproduction limit). In this case, the momentum integral in Eq.~(\ref{diff-CS}) gets dominant contributions for $\bm{q} \sim \bm{P}$ and can be analytically inverted \cite{Hagiwara:2017fye}. Indeed, taking into account that the dominant angular correlation in the Wigner distribution is encoded in its elliptic part as in Eq.~(\ref{GTMD-decomposition}) (see also Refs.~\cite{Hatta:2016dxp,Hagiwara:2016kam}) and performing the angular integral in the amplitude analytically, in the massless quark limit one arrives at \cite{Hagiwara:2017fye}
\begin{eqnarray}
\frac{d\sigma^{AT\to A(q\bar q)T}}{dy_1dy_2 d^2\bm{k}_1 d^2\bm{k}_2} 
&\approx& \omega \frac{dN}{d\omega} 
\frac{2(2\pi)^2 N_c \alpha_{\rm em}}{P_\perp^2}   \nonumber \\
&\times& \sum_q e_q^2 z(1-z) (z^2+(1-z)^2) 
\bigl(A^2+2\cos 2(\phi_P - \phi_\Delta) AB \bigr)\,,
\label{pA->jj_final}
\end{eqnarray}
where the azimuthal angles of $\bm{P}$ and $\bm{\Delta}$ vectors are $\phi_P$ and $\phi_\Delta$, respectively, summation over the light quark flavours $q$ is made explicit, and the following two structure functions
\begin{eqnarray}
&&A(P_\perp,\Delta_\perp)\equiv - \int_0^{P_\perp} dq_\perp  q_\perp {\widetilde S}_{Y,0}(q_\perp,\Delta_\perp)\,, \\
&&B(P_\perp,\Delta_\perp) \equiv -\int_0^{P_\perp} dq_\perp \frac{q_\perp^3}{P_\perp^2}{\widetilde S}_{Y,\epsilon}(q_\perp,\Delta_\perp) +
\int_{P_\perp}^\infty  dq_\perp \frac{P_\perp^2}{q_\perp} {\widetilde S}_{Y,\epsilon}(q_\perp,\Delta_\perp)
\end{eqnarray}
are introduced that contain the basic information about the transverse-momentum dependent isotropic and elliptic gluon densities in the target, respectively. Note that in Eq.~(\ref{pA->jj_final}) only the linear terms for ${\widetilde S}_{Y,\epsilon}\ll {\widetilde S}_{Y,0}$ are kept. A direct reconstruction of $A$ and $B$ structure functions through a precision measurement of the fully differential cross section (\ref{pA->jj_final}) would then enable one to extract the isotropic and elliptic components of the gluon GTMD as follows \cite{Hagiwara:2017fye}
\begin{eqnarray}
{\widetilde S}_{Y,0}(P_\perp,\Delta_\perp) &=& -\frac{1}{P_\perp} \frac{\partial}{\partial P_\perp}A(P_\perp,\Delta_\perp)\,, \\
{\widetilde S}_{Y,\epsilon}(P_\perp,\Delta_\perp) &=& -\frac{\partial B(P_\perp,\Delta_\perp)}{\partial P_\perp^2} +
\frac{2}{P_\perp^2}\int^{P_\perp^2}_0 \frac{dP'^2_\perp}{P'^2_\perp} B(P'_\perp, \Delta_\perp)\,.
\label{WF-extraction}
\end{eqnarray}
It is worth noticing here that, when considering non-zero virtualities of the projectile quasi-real photon, one should also add a contribution from the longitudinally polarized virtual photons~\cite{Chyla:1999pw,Chyla:2000ue}. The latter may somewhat impact the extraction of the Wigner distribution discussed above.

\begin{figure}[!h]
	\includegraphics[width=75mm]{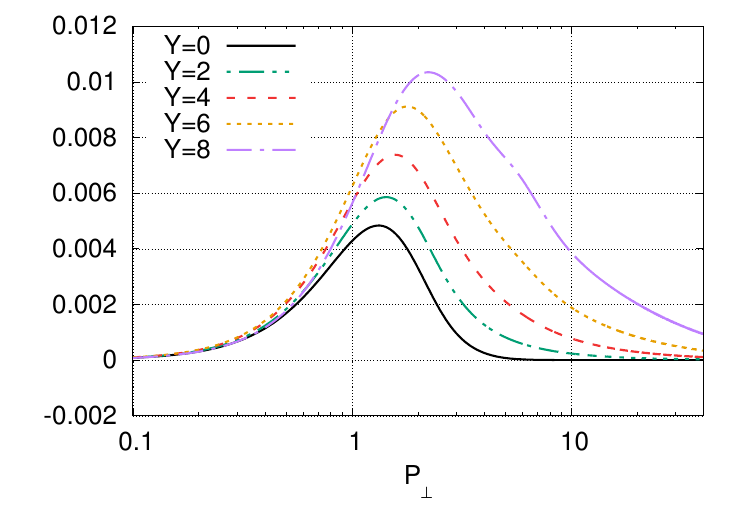}
	\includegraphics[width=75mm]{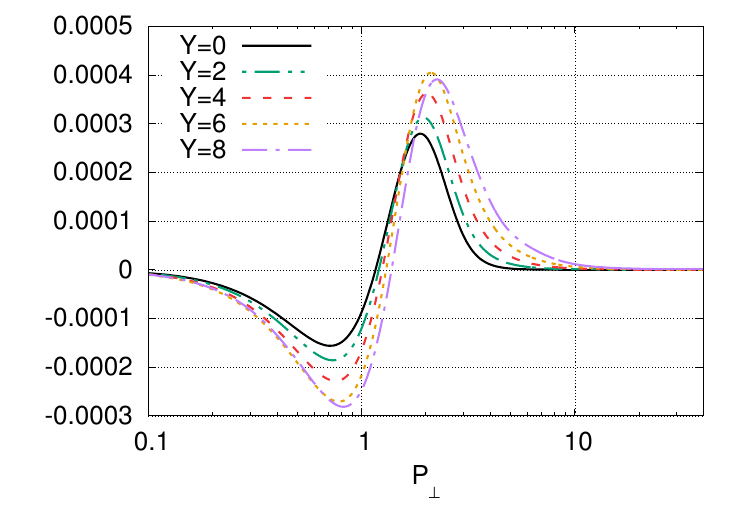}
	\caption{The structure functions $A$ (left) and $B$ (right) versus the transverse momentum $P_\perp$ and rapidity $Y$ for fixed $\Delta_\perp = 1.0$ and the scale parameter $R=0.4$ fm. Both transverse momenta are shown in units of $1/R$. From Ref.~\cite{Hagiwara:2017fye}.} 
\label{fig:A-B_bk-proton}
\end{figure}

With the two commonly used models for the gluon Wigner distribution -- BK and MV-IR models introduced above -- the predictions for the structure functions $A$ and $B$ have been obtained in Ref.~\cite{Hagiwara:2017fye}. In particular, Fig.~\ref{fig:A-B_bk-proton} illustrates the results for rapidity dependence of $A$ (left) and $B$ (right) found as functions of $P_\perp$ using a numerical solution of the impact-parameter dependent BK equation \cite{Balitsky:1995ub,Kovchegov:1999yj} for the proton target following Ref.~\cite{Hagiwara:2016kam} (here, $\Delta_\perp=1$ and the scale parameter $R=0.4$ fm). The transverse momenta $P_\perp$ and $\Delta_\perp$ are defined in units of $1/R$. The function $A(P_\perp)$ exhibits a peak at the saturation scale $Q_s(Y)$ which grows with $Y$ following the geometric scaling, while the peak in $B(P_\perp)$ moves much slower with $Y$.

\begin{figure}[!h]
	\includegraphics[width=75mm]{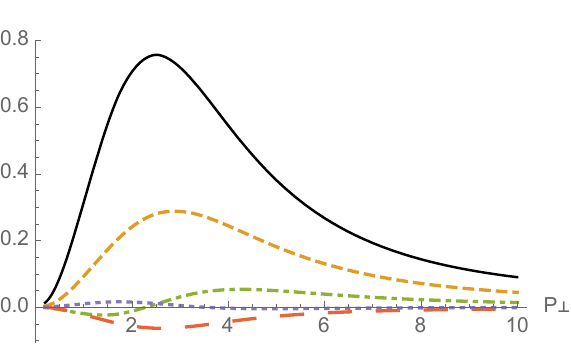}
	\includegraphics[width=75mm]{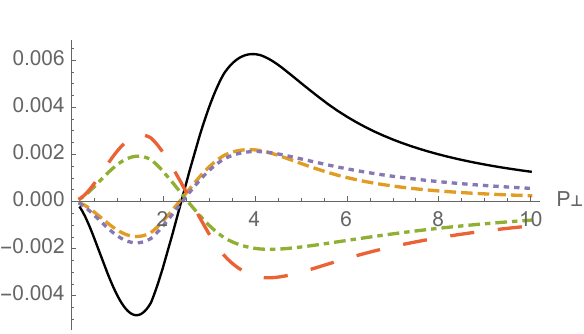}
	\caption{The structure functions $A$ (left) and $B$ (right) versus the transverse momentum $P_\perp$ for lead nucleus target ($A=208$), the scale parameter $R=0.4$ fm, $\Delta_\perp = 0.2$ (solid lines), $\Delta_\perp = 0.25$ (dashed lines), $\Delta_\perp = 0.3$ (dotted-dashed lines), $\Delta_\perp = 0.4$ (long dashed lines), and $\Delta_\perp = 0.5$ (dotted lines). Both transverse momenta are shown in units of $1/R$. Here, the MV-IR model formulation of Ref.~\cite{Iancu:2017fzn} has been employed. From Ref.~\cite{Hagiwara:2017fye}.}
\label{fig:A-B_MV-nucleus}
\end{figure}

\begin{figure}[!h]
	\includegraphics[width=75mm]{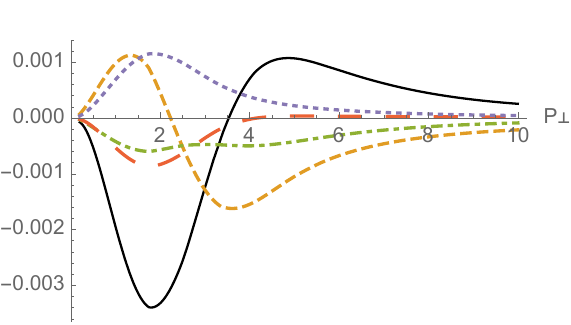}
	\includegraphics[width=75mm]{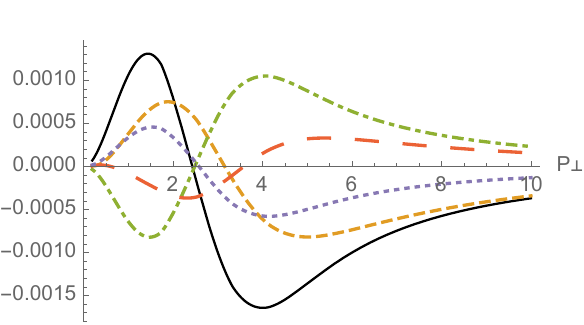}
	\caption{The same as in Fig.~\ref{fig:A-B_MV-nucleus} but for larger $\Delta_\perp$, i.e.~$\Delta_\perp = 0.7$ (solid lines), $\Delta_\perp = 0.8$ (dashed lines), $\Delta_\perp = 0.9$ (dotted-dashed lines), $\Delta_\perp = 1.0$ (long dashed lines), and $\Delta_\perp = 1.1$ (dotted lines). From Ref.~\cite{Hagiwara:2017fye}.}
\label{fig:A-B_MV-nucleus-largeDelta}
\end{figure}

Considering a possibility to probe the Wigner distribution in the nucleus in $AA\to A(jj)A$ UPCs, Ref.~\cite{Hagiwara:2017fye} also shows the predictions for $A$ and $B$ for a large nucleus in the target (such as lead with atomic mass number $A=208$). For this purpose, the MV-IR model for the nuclear gluon distribution has been utilized taking into account its inhomogeneities in the transverse plane as formulated in Ref.~\cite{Iancu:2017fzn} -- see Figs.~\ref{fig:A-B_MV-nucleus} and \ref{fig:A-B_MV-nucleus-largeDelta} for smaller $\Delta_\perp \leq 0.5$ and larger $\Delta_\perp \geq 0.7$, respectively. The corresponding analytical expressions for the isotropic ${\widetilde S}_{Y,0}(r_\perp,b_\perp)$ and elliptic ${\widetilde S}_{Y,\epsilon}(r_\perp,b_\perp)$ densities in the nuclear targets in the small-$x$ limit (not incorporating an explicit $x$-dependence) are found in terms of the nuclear thickness function $T_A(b)$ and its derivatives in Ref.~\cite{Iancu:2017fzn} (see also Ref.~\cite{Zhou:2016rnt}).

We notice that the peaks of $A$ and $B$ in the case large nuclear targets for small $\Delta_\perp<0.3$ exceed those for the proton target evaluated in the BK model by one-two orders of magnitude. While the nuclear $A$ drops very quickly with $\Delta_\perp$ and eventually falls below that of the proton, the MV-IR model predicts the nuclear $B$ being of an order of magnitude larger than that of the proton at $\Delta_\perp \geq 0.7$ where both $A$ and $B$ structure functions feature a node in their $P_\perp$ dependence. This observation of Ref.~\cite{Hagiwara:2017fye} signifies that a measurement of the exclusive dijet production in $AA$ UPCs has a better chance to constrain the elliptic Wigner distribution.

\subsubsection{HERA measurements}

Dijets produced diffractively in photon-induced processes have been measured rather thoroughly at HERA. From comparing data with theory predictions based on the QCD factorization and hence on the universality of diffractive proton PDFs extracted from previous HERA measurements, one usually judges the level of the QCD factorization breaking (the farther from unity, the bigger the breaking is). This factorization breaking is usually explained as a consequence of additional interactions between incoming photon and proton. The H1 Collaboration published results of three analyses, making use of data with large rapidity gap \cite{H1:2007jtx,H1:2010xdi} as well as with a very forward proton spectrometer \cite{H1:2015okx}, the latter allowing one to control better the non-diffractive background and contributions from proton dissociation. The ZEUS Collaboration published one analysis relying on large rapidity gaps \cite{ZEUS:2007uvk}. While all three H1 analyses report the above ratio to be around 0.6, the ZEUS experiment measures it to be close to unity and hence a different level of the factorization breaking, if at all, is claimed by the two collaborations. We conclude that this is rather a small tension given the fact that all three H1 analyses confirm each other after carefully evaluating many sources of uncertainties and using several experimental approaches. As far as the potential information about the Wigner function is concerned, we have to conclude that no $\cos(2\phi)$ or even jet $\phi$ measurement was done. And since the data taking at HERA finished in 2007, chances for any supplemental future measurements of this kind are small.

\subsubsection{LHC measurements}
\label{sec:LHCjj}

One of the main conclusions of Ref.~\cite{Hagiwara:2017fye} is that the Wigner function can be observed at rather low jet transverse momenta $p_{\rm T}$ by reconstructing the full kinematic dependence of the structure functions $A(P_\perp,\Delta_\perp)$ and $B(P_\perp,\Delta_\perp)$ and then inverting them using Eqs.~(\ref{WF-extraction}). As explained above, for LHC experiments the usual $p_{\rm T}$ threshold for jets based on calorimeter information is 20~GeV. As was evident from Figs.~\ref{fig:A-B_bk-proton}-\ref{fig:A-B_MV-nucleus-largeDelta} above, theoretical predictions show a non-trivial dynamics (e.g. the appearance of the node in $P_\perp$ dependence of the structure functions, and a growing relative importance of the elliptic term) in the low $\Delta_\perp$ and $P_\perp < 4$ GeV domains while they are monotonously falling at larger transverse momenta such that their features are very difficult to probe at the LHC. To get to lower hard scales via jet production, it was suggested to use track-based jets at the LHC or to use data from RHIC. Low-$p_{\rm T}$ jets can also be probed at EIC as will be explained in the section~\ref{sec:Future}.

A very useful measurement of the $\cos(2\phi)$ distribution (where $\phi$ is the azimuthal angle between the vector sum and the vector difference of momenta of two highest $E_{\rm T}$ jets) in PbPb collisions has been performed by CMS~\cite{CMS:2022lbi}. It is based on central detector only, so no nucleus tagging is used and exclusivity requirements are reduced to requiring empty regions in the tracker and/or calorimeter, depending on the noise of individual subdetectors. The transverse momenta of the two jets are required to be above 30~GeV for the first jet and 20~GeV for the second jet and at the same time it is required that the absolute value of the difference is larger than the absolute value of the sum. As Fig.~2 (right) of Ref.~\cite{CMS:2022lbi} documents the slow increasing trend for the mean value of $\cos(2\phi)$ as a function of $\Delta_\perp$ observed in the data is described decently by calculations~\cite{Hatta:2020bgy} based on perturbative soft QCD gluon radiation in the final state up to $\Delta_\perp < 15$~GeV, although Wigner function effects are mainly of non-perturbative origin. This is probably not unexpected given the relatively large hard scale given by the jet $p_{\rm T}$ and the absence of forward detectors, more specifically Roman Pots (RPs).

Indeed, the advantage of RPs is manifold: they can be used to i) tag nuclei, ii) measure their fractional energy loss, $\xi$, iii) time-of-flight (ToF) and iv) transverse momenta. The first three would allow one to use more stringent requirements on exclusivity than those based on rapidity gaps used in Ref.~\cite{CMS:2022lbi}, by requiring that the value of an observable measured in the central detector is identical to the value measured by FPDs within resolutions of subdetectors used in the above measurements. In this way, quantities based on $\xi$ (dijet mass or rapidity) obtained in the central detector are required to match those measured in Silicon trackers of FPDs, while $z_{\rm vtx}$ of a primary vertex obtained in the central tracker is required to match that obtained by the ToF detector in FPDs. Finally, the measurement of the nucleus transverse momentum would then be significantly more precise than the measurement of the sum of the two jet momenta.

\subsection{Diffractive photoproduction of heavy quarks}
\label{Sect:heavy-QQ}

As was elaborated above, the elliptic component of the the gluon Wigner distribution becomes relatively more pronounced compared to its isotropic component when the transverse momenta of final-state quark and antiquark approach the saturation scale of the gluon density in the target. As there is no other hard scale in the $\gamma+(gg)\to q\bar{q}$ matrix element for light quarks, with $gg$ in the color singlet state, except for the (anti)quark transverse momentum (being translated into the jet-$p_T$), pushing into the soft domain poses natural questions about the validity of perturbation theory. Besides, the existing experimental capabilities at ATLAS are limited by the detector acceptance imposing with a lower cut-off on jet at 20 GeV or so. These constraints, together with substantial backgrounds, severely restrict one's capability to probe the elliptic density in the kinematics domain where it is typically enhanced.

As a way out of this problem, over the past years it has been suggested in Refs.~\cite{ReinkePelicer:2018gyh,Linek:2023kga} (see also Ref.~\cite{Mantysaari:2019csc}) that exclusive photoproduction of heavy ($Q=c,b$) quark pairs in high-energy diffractive $eT$ (with the target $T=p,A$) collisions as well as in $pT$ and $AA$ UPCs offers a powerful tool for probing the gluon Wigner distribution in the target $T$ in both perturbative and soft kinematics domains. This is due to the fact that the quark mass $m_Q$ provides a naturally hard scale (instead of a large $P_\perp$ in the case of light-quark jets). The latter enables us to go effectively to much lower transverse momenta for final-scale heavy-quark jets (i.e.~a jet that contains an open heavy flavour meson reconstructed in a detector) and, hence, to probe the gluon GTMD at transverse momenta $P_\perp \gtrsim \Delta_\perp$ down to a GeV scale, or even lower. The analysis of Refs.~\cite{ReinkePelicer:2018gyh,Linek:2023kga} was inspired by an earlier work of Refs.~\cite{Hatta:2016dxp,Hagiwara:2017fye} on light-quark jets discussed above in the previous subsection, and below we outline some of its basic details.

Besides providing a natural hard scale for the perturbation theory to more reliably work at low quark transverse momenta, this process is also advantageous from the experimental point of view as the reconstruction of open heavy-flavor states (such as $B$ and $D$ mesons) can be achieved at lower transverse momenta than those for light-quark jets. Finally, despite of a smaller cross section, this process is expected to benefit from a better control of QCD backgrounds. However, there is a price to pay for these gross advantages: the integral in the diffractive amplitude is not analytically invertible for a large $m_Q$ with respect to the components of the Wigner distribution as was shown above for massless quarks. Thus, so far in the literature only predictions for the differential observables were given, for a set of distinct theoretical models, and no efficient reconstruction procedure for the elliptic density directly from the data has been advised yet in this case.

The formalism of exclusive heavy quarks' photoproduction in $A+T$ ($T=p,A$) UPCs is very similar to that of light quarks, upon an appropriate introduction of the finite quark mass terms~\cite{ReinkePelicer:2018gyh,Linek:2023kga}. The differential cross section for the diffractive $\gamma T \to (Q \bar Q) T$ process can be represented as
\begin{eqnarray}
\frac{d\sigma^{\gamma T\to (Q\bar Q)T}}{dzd^2\bm{P}d^2\bm{\Delta}} = e_f^2 \alpha_{\rm em}\,2N_c (2\pi)^2 \Big\{ \big(z^2 + (1-z)^2\big)|\bm{M}_0|^2 + m_Q^2 |M_1|^2 \Big\} \,.
\end{eqnarray}
where the matrix elements can be represented in terms of the gluon GTMD in the target, $T(Y,\bm{k},\bm{\Delta})$, absorbing all the $\bm{\Delta}$-dependence as follows \cite{Linek:2023kga}
\begin{eqnarray}
\bm{M}_0 &=& \int\frac{d^2\bm{k}}{2\pi}\,T(Y,\bm{k},\bm{\Delta})\, 
\Big\{\frac{\bm{P} - \bm{k}}{(\bm{P} - \bm{k})^2 + m_Q^2} - 
\frac{\bm{P}}{P_\perp^2 + m_Q^2}\Big\} \,, \nonumber \\
M_1 &=& \int\frac{d^2\bm{k}}{2\pi}\,T(Y,\bm{k},\bm{\Delta})\, 
\Big\{\frac{1}{(\bm{P} - \bm{k})^2 + m_Q^2} - \frac{1}{P_\perp^2 + m_Q^2}\Big\}\,.
\end{eqnarray}
In this representation, the last $\bm{k}$-independent term in both integrands effectively drops out upon the use of the ``sum rule'' (\ref{sumrule}). Then, one ends up with the ``formfactor representation'' for the matrix elements \cite{ReinkePelicer:2018gyh} leading to a generalization of Eq.~(\ref{pA->jj_final}):
\begin{align}
\frac{d\sigma^{AT\to A(Q\bar Q)T}}{dy_1dy_2d^2\bm{P}d^2\bm{\Delta}} \simeq\, & 
\omega \frac{dN}{d\omega} 2(2\pi)^2 N_c \alpha_{\rm em} e_Q^2 z(1-z) \frac{1}{P_\perp^2} \nonumber \\
&\left\{(z^2 + (1-z)^2) \; \left[ A(P_\perp, \Delta_\perp) + B(P_\perp, \Delta_\perp) \cos 2(\phi_P - \phi_\Delta) \right]^2 \right.  \nonumber \\ 
& \left. + \frac{m_Q^2}{P_\perp^2} \; \left[ C(P_\perp, \Delta_\perp) + D(P_\perp, \Delta_\perp)\cos 2(\phi_P - \phi_\Delta) \right]^2 \right\}\,,
\end{align}
where the incident WW photon flux is defined in Eq.~(\ref{WW-flux}). Note that two additional formfactors $C$ and $D$ emerge in the heavy quark case whose expressions are given explicitly in Ref.~\cite{ReinkePelicer:2018gyh}.
\begin{figure}
  \centering
  \includegraphics[width=.4\textwidth]{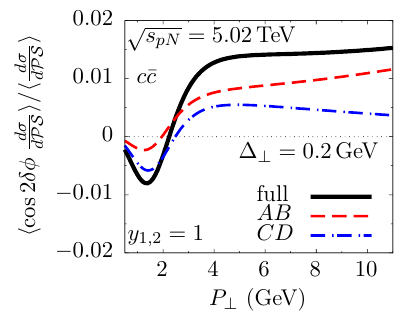}
  \includegraphics[width=.4\textwidth]{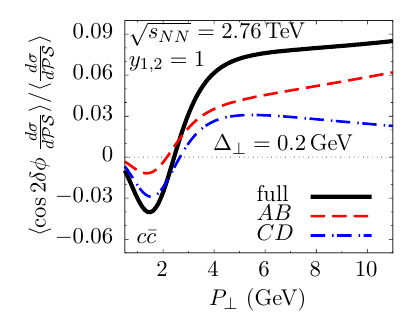}
  \includegraphics[width=.4\textwidth]{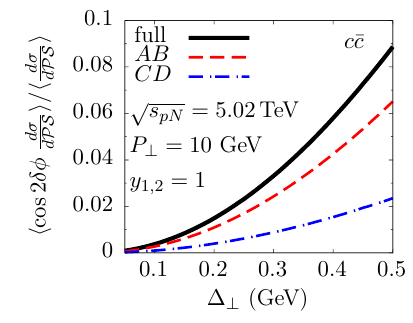}
  \includegraphics[width=.4\textwidth]{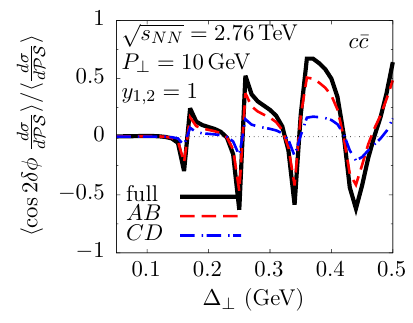}
  \caption{The angular cosine-weighted average cross section normalised to the angle integrated one for exclusive photoproduction of $c\bar c$ pairs in the MV-IR model as a function of $P_\perp$ with fixed $\Delta_\perp = 0.2$ GeV (upper panels) and of $\Delta_\perp$ with fixed $P_\perp = 10$ GeV (lower panels). This quantity is shown for two different types of targets -- the proton (left panels) and the lead nucleus (right panels). From Ref.~\cite{ReinkePelicer:2018gyh}.}
\label{fig:CS_ratio}
\end{figure}

The impact of the additional $C$ and $D$ formfactors due to the finite-mass corrections on the angular correlation of the differential cross section is illustrated in the case of exclusive $c\bar c$-pair photoproduction in Fig.~\ref{fig:CS_ratio}. Here, the ratio of the angular cosine-weighted average cross section defined as
\begin{equation} 
\label{av-cosine} 
\left\langle \frac{d\sigma}{d\mathcal{PS}}  \cos 2(\phi_P - \phi_\Delta) \right\rangle  = \int_0^{2\pi} d\phi_{P}  \int_0^{2\pi} d\phi_{\Delta}  \frac{d\sigma^{AT\to A(Q\bar Q)T}}{dy_1dy_2d^2\bm{P}d^2\bm{\Delta}} \cos 2(\phi_P - \phi_\Delta) 
\end{equation}
to the angle-integrated one is shown as a function of $P_\perp$ at fixed $\Delta_\perp = 0.2$ GeV (upper panels) and of $\Delta_\perp$ at fixed $P_\perp = 10$ GeV (lower panels). Here, MV-IR model discussed above has been used. Such an observable appears to be a good probe for the angular correlation since, indeed, it tends to more positive the more collinear (parallel or antiparallel) $\bm{P}$ and $\bm{\Delta}$ become (while negative case corresponds to the vectors turning perpendicular). Naturally, the quark mass effect diminishes at larger transverse momenta but becomes quite important for softer $P_\perp$ and in the forward limit $\Delta_\perp \sim \Lambda_{\rm QCD}$. For more details on the quark mass effects, see Ref.~\cite{ReinkePelicer:2018gyh}, while a comparison of exclusive photoproduction cross section of charm quarks computed for different models for the dipole $T$-matrix can be found in Ref.~\cite{Linek:2023kga}. 

In the literature, there are also other relevant exclusive diffractive processes involving heavy quarks that are sensitive to the elliptic component of the gluon Wigner distribution. An earlier complementary analysis of exclusive charm production in $ep$ collisions at the EIC (at large $Q^2$) has been performed in Ref.~\cite{Mantysaari:2019csc} in the framework of CGC. Here, the gluon Wigner distribution has been modelled by solving the JIMWLK equations for Wilson lines using the initial conditions constrained through the HERA data. Besides, the quarkonia pair production process and its connection to the gluon Wigner distribution has been studied in Ref.~\cite{Bhattacharya:2018lgm}.

\subsection{Photon bremsstrahlung}
\label{Sect:direct-photon}

A particularly clean and important phenomenological source of new information on the nucleon and nucleus structure is direct (or prompt) production of photons. Such a reaction is typical, for instance, in DIS processes well explored at the HERA collider. As photons do not take part in strong interactions, they transmit unaffected information about the hard scattering in $pp$ or $pA$ collisions, and can be measured in the final state, either real or virtual, through a detection of a Drell-Yan lepton pair. 

\begin{figure*}[!h]
\begin{minipage}{1.0\textwidth}
 \centerline{\includegraphics[width=0.75\textwidth]{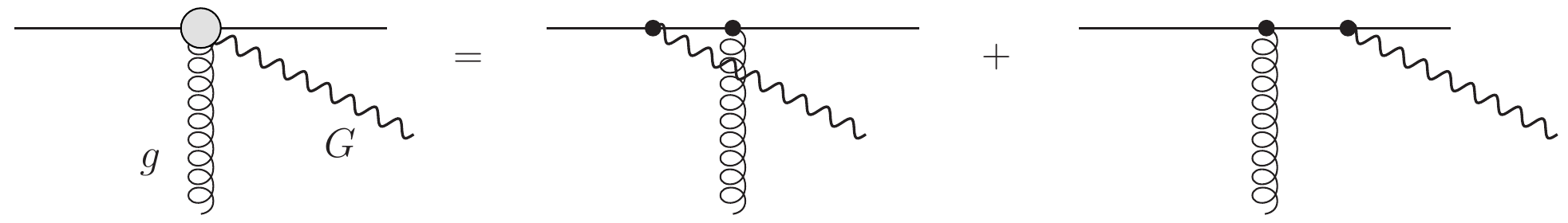}}
\end{minipage}
   \caption{Photon radiation by a projectile quark in the target rest frame. From Ref.~\cite{Pasechnik:2015fxa}.}
 \label{fig:G-vertex}
\end{figure*}

In the parton model, the direct photon emission can be thought of as a Compton scattering $gq\to\gamma q$. On the other hand, in the target rest frame the same process should be considered as photon bremsstrahlung off the projectile quark scattered off the target $T$ (nucleon, $T=N$, or nucleus, $T=A$). Its amplitude is represented by an interference of two simple diagrams at tree-level as is shown in Fig.~\ref{fig:G-vertex}. The corresponding color-dipole description in the case of direct production of real photons has been thoroughly discussed e.g. in Refs.~\cite{Kopeliovich:1998nw,Jalilian-Marian:2005tod,Kopeliovich:2007fv,Goncalves:2020tvh,Benic:2022ixp}, while the process of virtual (massive) photon production with dilepton (Drell-Yan) final states have been analysed both in $pp$ and $pA$ collisions e.g.~in Refs.~\cite{Kopeliovich:2000fb,Kopeliovich:2001hf,Baier:2004tj,Kopeliovich:2007yva,Basso:2015pba,Schafer:2016qmk,Basso:2016ulb,Goncalves:2016qku,Ducloue:2017zfd}.
As was discussed above, the elliptic gluon Wigner distribution in the target encodes a non-trivial dependence on the dipole orientation with respect to the color background field of the target. The latter can be probed by measuring the azimuthal angular correlation of the final-state photon and a jet or a leading hadron originating from the recoil quark after hadronisation. Here, we briefly elaborate on the main results and implications of this process for phenomenology of the elliptic gluon distribution.

In order to highlight the relevance of this process for gluon tomography of the nucleon, we follow the analysis of Ref.~\cite{Kopeliovich:2007fv} based on phenomenological dipole framework. Following the discussion of Ref.~\cite{Kopeliovich:1998nw}, the factorized dipole formula for the transverse momentum distribution of the direct (unpolarised) photons radiated by the projectile quark at a certain impact distance $\bm{b}$ from the centre of the target $T$ reads
\begin{eqnarray}
&&\frac{d \sigma^{qT\to\gamma X}}
{d(\ln \alpha)\,d^2{\bm p}\,d^2\bm{b}}=\frac{1}{(2\pi)^2}
\int d^2\bm{r}_1d^2\bm{r}_2
e^{i \bm{p}\cdot (\bm{r}_1-\bm{r}_2)}
\sum_{i,f} \phi^{\star}_{\gamma q}(\alpha, \bm{r}_1)
\phi_{\gamma q}(\alpha, \bm{r}_2)
F_{\rm T}(\bm{b},\alpha\bm{r}_1,\alpha\bm{r}_2,x) \,, \nonumber\\
&&\sum_{i,f}\phi^{\star}_{\gamma q}(\alpha, \bm{r}_1)\phi_{\gamma q}(\alpha, \bm{r}_2) = \frac{\alpha_{\rm em}}{2\pi^{2}}m^2_q\alpha^2 \Bigl\{\alpha^2 K_0(\alpha m_q r_{1\perp})
K_0(\alpha m_q r_{2\perp}) \label{CS-gammaX} \\
&& \qquad \qquad +\,[1+(1-\alpha)^2] \frac{\bm{r}_1\cdot\bm{r}_2}{r_{1\perp}r_{2\perp}} K_{1}(\alpha m_q r_{1\perp})K_{1}(\alpha m_q r_{2\perp})
\Bigr\} \nonumber
\end{eqnarray}
in terms of the constituent quark mass $m_q\approx 0.2\,{\rm GeV}$ that can be viewed as an adjustable IR cutoff parameter \cite{Kopeliovich:1999am,Kopeliovich:2000ra}, the final-state photon transverse momentum $\bm{p}$, its LC momentum fraction $\alpha \equiv p_\gamma^+/p_q^+$, the Bjorken $x$ variable of soft gluons in the target, the LC wave function of the lowest-Fock $q\to q\gamma$ transition, $\phi_{\gamma q}(\alpha,\bm{r})$, at a transverse separation $\bm{r}$, and the effective dipole amplitude off a target $T$, $F_{\rm T}(x,\alpha\bm{r}_{1},\alpha\bm{r}_{2},\bm{b})$. At high energies, the latter can be represented as a superposition of partial dipole amplitudes accounting for dipole-target scatterings of colorless dipoles with different separations at impact parameter $b_\perp$, namely,
\begin{eqnarray*}
F_{\rm T}(x,\alpha\bm{r}_{1},\alpha\bm{r}_{2},\bm{b}) = 
{\rm Im}f^T_{q\bar{q}}(x,\alpha \bm r_{1},\bm{b})+
{\rm Im}f^T_{q\bar{q}}(x,\alpha \bm r_{2},\bm{b}) -
{\rm Im}f^T_{q\bar{q}}(x,\alpha(\bm{r}_{1} - \bm{r}_{2}),\bm{b}) \,. 
\end{eqnarray*}

As was elaborated above, in order to quantify the azimuthal asymmetry in the direct photon production one could track the correlation in azimuthal angle $\phi_p$ of the transverse momentum $\bm{p}$ with respect to a fixed impact parameter vector $\bm{b}$ represented by the elliptic flow coefficient,
\begin{eqnarray}
&& v_2 = 2 \left\langle \left(\frac{\bm{p}\cdot\bm{b}}{p_\perp b_\perp}\right)^2\right\rangle_{\phi_p} - 1 , \left\langle \left(\frac{\bm{p}\cdot\bm{b}} {p_\perp b_\perp}\right)^2\right\rangle_{\phi_p}\propto \int\limits_0^{2\pi} d\phi_p\,
\left(\frac{\bm{p}\cdot\bm{b}}{p_\perp b_\perp}\right)^2\,
\frac{d\sigma^{qA\to\gamma X}}
{d(\ln \alpha)\,d^2\bm{p}\,d^2\bm{b}}\,
\label{gamma-v2}
\end{eqnarray}
where the averaging can be performed analytically. For this purpose, it is convenient to switch from the integration of the differential cross section (\ref{CS-gammaX}) over direction of vector $\bm{p}$ at fixed $\bm{b}$ to the integration over direction of vector $\bm{b}$ at fixed $\bm{p}$, such that $\phi_p\to\phi_b$. This is a helpful step due to a particularly simple (Gaussian) form of the phenomenological $b_\perp$-dependence in the effective amplitude $F_{\rm T}$ stemming from the dipole parameterisations in Eqs.~(\ref{2gluon-dipole-N-x}) and (\ref{approx-nuc-fel}) for the nucleon and nucleus targets, respectively.

Now, we have all the ingredients to write down the azimuthal asymmetry of direct photons emitted in quark-target collisions as \cite{Kopeliovich:2007fv}
\begin{eqnarray}
v_2\,\int\limits_0^{2\pi} d\phi_b\,\frac{d \sigma^{qA\to\gamma X}}
{d(\ln \alpha)\,d^2\bm{p}\,d^2\bm{b}} &=&
\frac{1}{2\pi}\sum_{i,f}\int d^2\bm{r}_1d^2\bm{r}_2 e^{i \bm{p}\cdot(\bm{r}_{1}-\bm{r}_{2})}
\phi^{\star}_{\gamma q}(\alpha, \bm{r}_{1})\phi_{\gamma q}(\alpha, \bm{r}_{2})
\tilde F_{\rm T}(x,\alpha\bm{r}_{1},\alpha\bm{r}_{2},b_\perp) \,, \nonumber\\
\tilde F_{\rm T}(x,\alpha\bm{r}_{1},\alpha\bm{r}_{2},b_\perp) &=&
{\rm Im} \tilde f^T_{q\bar{q}}(x,\alpha r_{1\perp},b_\perp)+
{\rm Im} \tilde f^T_{q\bar{q}}(x,\alpha r_{2\perp},b_\perp)-
{\rm Im} \tilde f^T_{q\bar{q}}\bigl(x,\alpha|\bm{r}_{1}-\bm{r}_{2}|,b_\perp\bigr) \,, \nonumber \\
\label{gamma-asymmetry}
\end{eqnarray}
where
\begin{eqnarray}
{\rm Im}\tilde f^T_{q\bar{q}}(x,r_\perp,b_\perp) \equiv
\frac{1}{2\pi}\int\limits_0^{2\pi} d\phi_b\,\cos(2\phi_b){\rm Im} f^T_{q\bar{q}}(x,\bm{r},\bm{b}) \,.
\end{eqnarray}
In particular, for a heavy nuclear target $T\equiv A$, using Eq.~(\ref{approx-nuc-fel}) one obtains \cite{Kopeliovich:2007fv}
\begin{eqnarray}
{\rm Im}\tilde f^A_{q\bar{q}}(x,r_\perp,b_\perp) = \frac{1}{4}\,e^{-\frac{1}{2}\sigma_{q\bar{q}}(r_\perp,x)T_A(b_\perp)}\Bigl[T_A^{\prime\prime}(b_\perp)\,g(r_\perp)-T_A^{\prime 2}(b_\perp)\,h(r_\perp)\Bigr]\,,
\label{phi-int-A}
\end{eqnarray}
where
\begin{eqnarray*}
g(r_\perp)&=&\int d^2\bm{s}\,{\rm Im} f^N_{q\bar q}(\bm{r},\bm{s})\,
\left[2\,\frac{(\bm{p}\cdot\bm{s})^2}{p_\perp^2}-s_\perp^2\right] \,, \\
h(r_\perp)&=&\int d^2\bm{s}_1 d^2\bm{s}_2\,{\rm Im} f^N_{q\bar q}(\bm{r},\bm{s}_1)\,
{\rm Im} f^N_{q\bar q}(\bm{r},\bm{s}_2)\left[2\,\frac{(\bm{s}_1\cdot\bm{p})
(\bm{s}_2\cdot\bm{p})}{p_\perp^2}-(\bm{s}_1\cdot\bm{s}_2)\right] \,.
\end{eqnarray*}
This result explicitly demonstrates that the azimuthal asymmetry is non-zero only if a non-trivial correlation between the transverse vectors $\bm{b}$ and $\bm{r}$ is present in the partial dipole amplitude $f^N_{q\bar q}(\bm{r},\bm{b})$ (otherwise, $g(r_\perp)=h(r_\perp)=0$). Besides, the asymmetry is strongly enhanced at the periphery of the nucleus, $b_\perp\sim R_A$, as expected.
\begin{figure}[!h]
\begin{minipage}{0.48\textwidth}
\centerline{\includegraphics[width=1.0\textwidth]{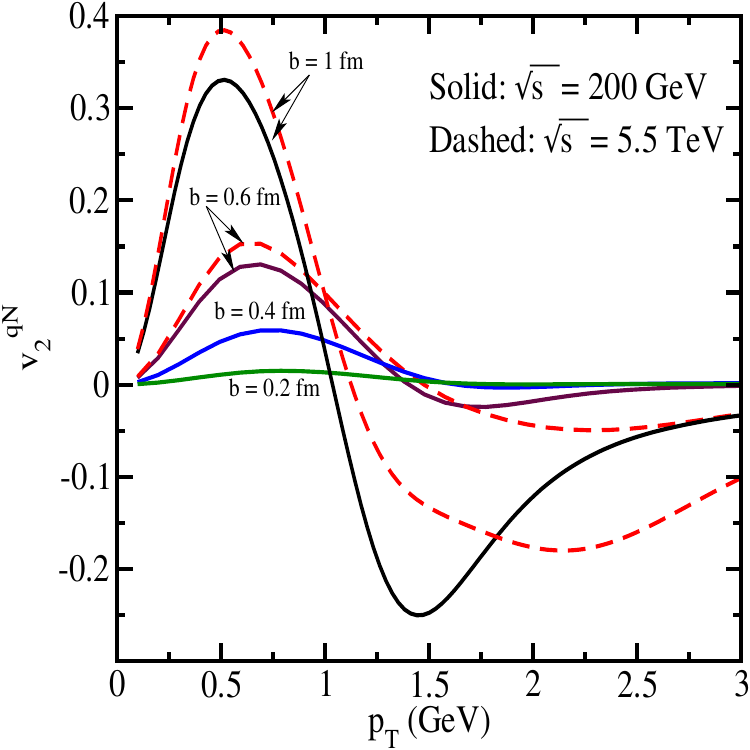}}
\end{minipage}
\begin{minipage}{0.48\textwidth}
\centerline{\includegraphics[width=1.0\textwidth]{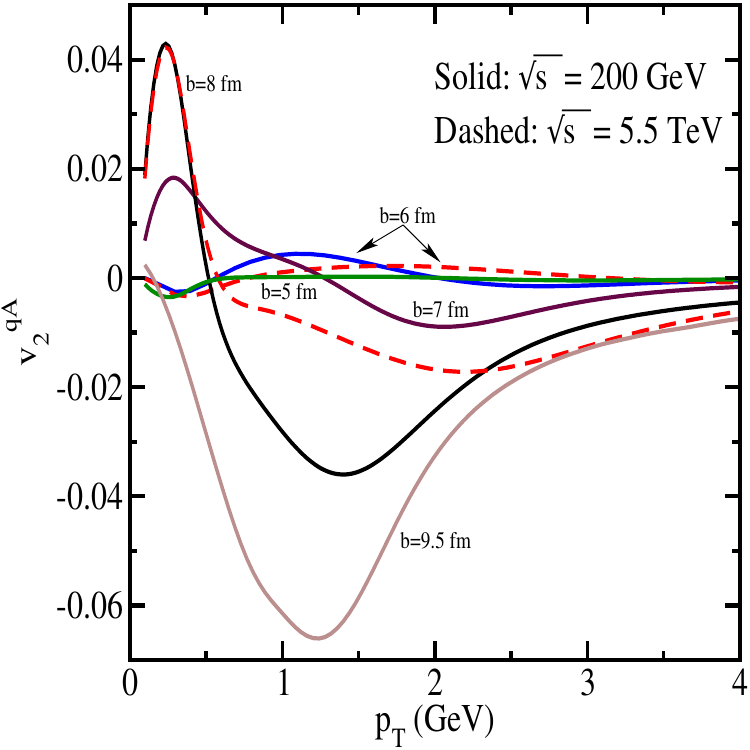}}
\end{minipage}
\caption{In the left panel, the elliptic flow coefficient of the photon production in $qN$ collisions, $v_2^{qN}(p_T\equiv p_\perp,b\equiv b_\perp,\alpha=1)$, is shown as a function of the photon transverse momentum $p_T$ at different values of $b$ and c.m.s.~collision energies $\sqrt{s}$. In the right panel, the same quantity is shown for $qA$ collisions, with $A=Pb$, computed with the use of the Woods-Saxon model for nuclear density (\ref{WS-density}). From Ref.~\cite{Kopeliovich:2007fv}.
\label{fig:v2_qN-qA}}
\end{figure}

Fig.~\ref{fig:v2_qN-qA} illustrates the sensitivity of the azimuthal angular correlations of the prompt photon production in quark-nucleon $qN$ (left panel) and quark-nucleus $qA$ (right panel) collisions represented by the corresponding elliptic flow parameter, $v_2^{qN}(p_\perp,b_\perp,\alpha)$, to the impact parameter $b_\perp$ and the photon transverse momentum $p_\perp$ for a fixed $\alpha=1$ \cite{Kopeliovich:2007fv}. Here, a few cases with different center-of-mass collision energies (at RHIC and LHC) and impact parameters $b_\perp$ are shown. One notices an increase of the azimuthal anisotropy with the rise of $b_\perp$ while it vanishes with $p_\perp$. As can be anticipated on general grounds, the azimuthal anisotropy reaches maximal values at the nucleon and nucleus peripheries where the sensitivity to the dipole orientation effects is largest, while the latter diminishes for smaller dipoles. It was demonstrated in Ref.~\cite{Kopeliovich:2007fv} that the azimuthal asymmetry of direct photons is strongly sensitive to the behavior of the nuclear thickness function at the nuclear edge. When the quark propagates through the nucleus at small values of $b_\perp$, it effectively interacts with nucleons placed at different azimuthal angles with respect to the quark path such that their contributions largely cancel each other in $v_2^{qA}$ making it smaller compared to $v_2^{qN}$. This cancellation is not exact, hence the azimuthal symmetry is broken, due to a non-trivial $b_\perp$-dependence of $T_A$.

\subsubsection{Exclusive and semi-exclusive mesons and photons at HERA}

The quasi-elastic production of vector mesons, production of photons in the Deeply Virtual Compton Scattering (DVCS) or prompt photon production belong to the simplest 
processes at HERA. The vector meson production was thoroughly measured in DIS as well as in inclusive or diffractive photoproduction. Thanks to the wealth of data and the level of details, the measurements form a legacy. Properties of vector mesons of the photoproduction origin are mapped in Refs.~\cite{H1:1996prv,ZEUS:1997rof,H1:1999pji,ZEUS:1998xpo,ZEUS:1999wqh,ZEUS:2000rhm,H1:2002laa,H1:2003ksk,H1:2009cml,H1:2005dtp,ZEUS:2009ixm,H1:2013okq,ZEUS:2002wfj,ZEUS:2011spj}. 
The vector meson species including $\rho, \omega, \phi, \rho^{\prime}, J/\Psi, \Psi^{\prime}$ and $\Upsilon$ had been detected via mainly two-prong decays by reconstructing charged decay products and requiring no further activity beyond the noise levels in the relevant subdetectors. Due to this requirement of exclusivity, we have $|t - t_{\rm min}| = p_{\rm T}^2$, where the vector meson transverse momentum $p_{\rm T}$ is obtained from its decay products and $t_{\rm min}$ is the kinematically allowed minimum which is usually negligible. Thanks to the high-precision trackers, the $t$ variable is measured precisely without the need to tag the intact proton. Other variables whose distributions were usually measured are $Q^2$ and $W$. 
The $\Phi$ angles have not been measured, so the elliptic part can not be constrained but all the HERA results can be used in global analyses to extract the radial part (i.e. TMD).  

The two processes with the photon in the final state are attractive due to the fact that hadronization effects do not need to be considered and thanks to a high precision of the energy measurement of an isolated object in the calorimeter. This, however, is balanced by a necessity to demand a good level of isolation whose aim is to separate the signal photon from background photons (e.g. from decays of $\pi^0$ and $\eta^0$). 

Photoproduction of the prompt photon in association with a jet was a subject of numerous HERA analyses, see Refs.~\cite{H1:2010pfw,ZEUS:2013xjr,H1:2004jab,ZEUS:1997edw,ZEUS:1999zvn,ZEUS:2001lmk,ZEUS:2006uxx}. Besides the distributions of $E_{\rm T}^{\gamma}, \eta^{\gamma}$ and $\xi^{\gamma}$ of the prompt photon shown in all the above papers, in Refs.~\cite{H1:2010pfw,ZEUS:1997edw} there are distributions of the difference between the azimuthal angle of the jet and prompt photon. 

\subsubsection{Inclusive $\gamma$+jet production}

Globally speaking the inclusive production of prompt photons in association with jets in proton–proton collisions is measured routinely and with a high precision. Precise measurements are useful for many reasons. They provide a testing ground for perturbative QCD (for example, the $t$-channel quark exchange) and serve to tune MC models or validate generators used to estimate backgrounds in searches for BSM physics signals which involve photons. The results include also measurements of angular correlations between the photon and the jet which, on one hand, can be used to probe the dynamics of the hard-scattering process but on the other, have a potential to give crucial information about the elliptic part of the WF, as we stress throughout this review. To separate the prompt-photon in a final state with accompanying hadrons one has to require a tight isolation of the photon to avoid the large contribution from neutral-hadron decays into photons. The production of isolated photons in association with jets in pp collisions at 7, 8 and 13~TeV was studied by the ATLAS~\cite{ATLAS:2013hsm,ATLAS:2016jxf,ATLAS:2012ar,ATLAS:2017xqp} and CMS~\cite{CMS:2013jnr,CMS:2013myp,CMS:2015onn,CMS:2021yzl} Collaborations.
This process is also extremely useful in the procedure of determining jet energy scales at hadron colliders (see detailed descriptions by ATLAS~\cite{ATLAS:2019oxp} and CMS~\cite{CMS:2016lmd}). Once the reconstructed jets are corrected to the particle level, the so called in-situ calibration needs to be done whose purpose is to account for differences between the jet response in data and simulations. The jet response is evaluated by balancing the $p_{\rm T}$ of a jet against that of a well-calibrated reference object, the $Z^0$-boson or a photon.  

\subsection{Deeply virtual Compton scattering}
\label{Sect:DVCS}

In the previous subsection we discussed the prompt photon production as a bremsstrahlung process off the projectile quark in the dipole picture where the angular correlation between the produced photon and the leading hadron or jet (emerging from the recoil final-state quark) has been shown to probe the elliptic Wigner contribution. In lepton-nucleon (and lepton-nucleus) colliders, the real photon can be produced either in the elecromagnetic bremsstrahlung (Bethe-Heitler) process, which dominates in fixed-target experiments (such as COMPASS) for not-so-hard photon virtualities, or through the deeply-virtual DVCS process \cite{H1:2005gdw,Aschenauer:2013hhw}.

Let us discuss how the elliptic gluon Wigner distribution impacts the DVCS process providing an important complementary way to analyse the inner structure of the proton, both theoretically and experimentally, being of particular relevance for lepton-nucleon colliders (see e.g.~Refs.~\cite{Goeke:2001tz,Diehl:2003ny,Belitsky:2005qn,Boffi:2007yc,Boer:2011fh,Accardi:2012qut}). Among particularly important implications of DVCS, it enables to access the information about orbital angular momentum of partons inside a nucleon~\cite{Ji:1996ek,Muller:1994ses,Ji:1996nm,Radyushkin:1997ki}. A thorough evaluation of the DVCS cross section accounting for the $\cos 2\phi$ term in the dipole picture at small $x$ realised in the framework of CGC formalism~\cite{McLerran:1993ni} has been performed in detail in Ref.~\cite{Hatta:2017cte}, and here we briefly outline the basic results. 
\begin{figure*}[!h]
\begin{minipage}{1.0\textwidth}
 \centerline{\includegraphics[width=0.9\textwidth]{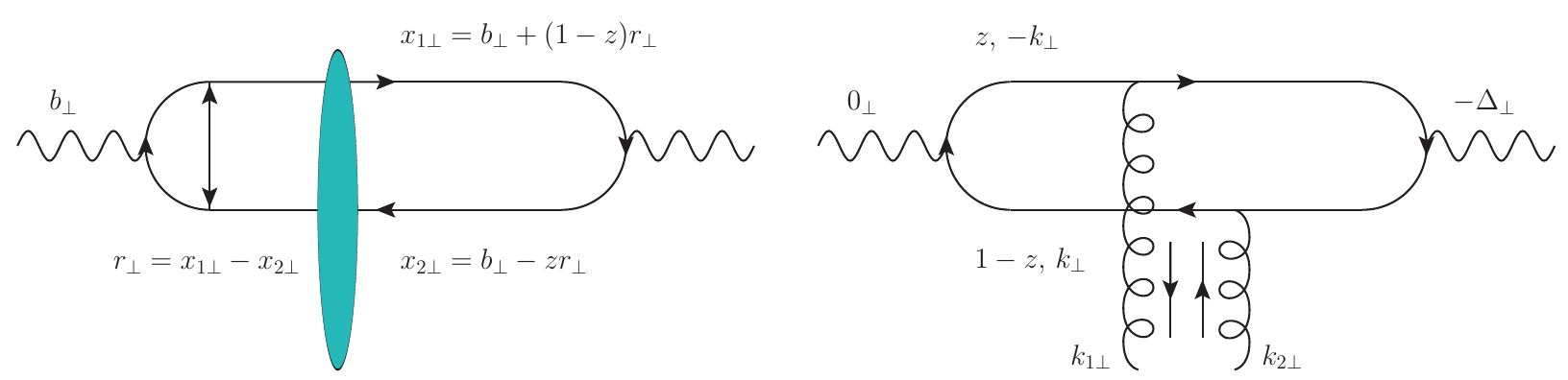}}
\end{minipage}
   \caption{DVCS amplitude in the dipole picture in the impact-parameter (left) and momentum (right) representations. From Ref.~\cite{Hatta:2017cte}.}
 \label{fig:DVCS-amplitude}
\end{figure*}

The differential DVCS cross section in $ep$-scattering takes the following standard form:
\begin{eqnarray}
\frac{d\sigma(ep\to e'\gamma p')}{dx_{\rm Bj}dQ^2d^2\bm{\Delta}}=
\frac{\alpha_{\rm em}^3 x_{\rm Bj}y^2}{4\pi Q^4}\frac{L_{\mu\nu}{\cal M}^{\mu\nu}}{Q^4} \,, 
\label{DVCS-CS}
\end{eqnarray}
in terms of the standard leptonic and hadronic tensors, $L_{\mu\nu}$ and ${\cal M}^{\mu\nu}$, respectively. The DVCS kinematics in the dipole approach is analogical to that of exclusive dijet production discussed above (and depicted in Fig.~\ref{fig:DVCS-amplitude}). In particular, denoting the lepton momenta in the initial and final states as $l$ and $l'$, while keeping the same notation for initial and final proton momenta, $p$ and $p'=p+\Delta$, respectively, one finds the momentum of the collinear initial virtual photon, $q = l - l'$, $q^2 =-Q^2$. The standard DIS variables then read, $x_{\rm Bj}= {Q^2/ (2q\cdot p)}$, $y= {q\cdot p /(l\cdot p)}$, $t=-\Delta^2_\perp$ and $W^2=(q+p)^2 \approx Q^2/x_{\rm Bj}$, such that the lepton transverse momentum, $l_\perp=l'_\perp$, satisfies $l_\perp^2 = \frac{1-y}{y^2}Q^2$.

In the dipole factorization approach valid at small $x$, the DVCS amplitude, illustrated in Fig.~\ref{fig:DVCS-amplitude} in impact-parameter (left) and momentum (right) representations, can be found  as~\cite{Donnachie:2000px,Bartels:2003yj,Favart:2003cu,Kowalski:2006hc}
\begin{equation}
{\cal A}_{\rm DVCS}\sim \int d^2\bm{b} e^{i\bm{b} \cdot \bm{\Delta}} 
\int dzd^2\bm{r} \Psi_{\gamma^*}(z,\bm{r})\Psi_{\gamma}^*(z,\bm{r})
T_Y(\bm{r},\bm{b}) \ ,
\end{equation}
in terms of the dipole $T$-matrix $T_Y\equiv 1-S_Y$ discussed above valid at $x\ll1$. In this process, $x$ is related to the Bjorken variable as $x\simeq x_{\rm Bj}/2$ being the same as the skewness $\xi = {(p^+ - p'^+)/( p^+ + p'^+)}$. Indeed, at high energies, or small $x$, the dipole amplitude is dominated by its imaginary part which to the leading order is found in terms of GPDs at $\xi=x$. Besides, $\Psi_{\gamma^*}(z,\bm{r})$ and $\Psi_{\gamma}^*(z,\bm{r})$ are the light-cone wave functions for the virtual ($T,L$-polarised) photon in the initial state and for real ($T$-polarised) photon in the final state, respectively (found e.g.~in Refs.~\cite{Bjorken:1970ah,Kowalski:2006hc,Hatta:2017cte}).

In the considered light-front kinematics, the gluon GPDs, in particular, are defined as
\begin{eqnarray}
&&\frac{1}{P^+} \int \frac{d\zeta^-}{2\pi}e^{ix P^+\zeta^-} \langle p'|F^{+i}(-\zeta/2)F^{+j}(\zeta/2)|p\rangle \nonumber\\
&& \qquad \quad =\frac{\delta^{ij}}{2}xH_g(x,\Delta_\perp)+ \frac{xE_{Tg}(x,\Delta_\perp)}{2M^2}\left( \Delta^i_\perp \Delta^j_\perp- \frac{\delta^{ij}\Delta_\perp^2}{2}\right)+\cdots \,, 
\label{gluon-GPDs} \\
&& \qquad \quad H_g(x,\Delta_\perp)\equiv H_g(x,\xi=x,\Delta_\perp)\,, \quad E_{Tg}(x,\Delta_\perp)\equiv E_{Tg}(x,\xi=x,\Delta_\perp) \,, \nonumber
\end{eqnarray}
such that the conventional unpolarised gluon PDF is recovered in the forward limit as $xH_g(x,\Delta_\perp\to 0)\equiv xG(x)$, while $E_{Tg}$ stands for the gluon transversity (or helicity-flip) GPD normalised as in Ref.~\cite{Hoodbhoy:1998vm}. Then, using the relation of the above matrix element to the dipole $S$-matrix found in Ref.~\cite{Hatta:2016dxp} in the form (c.f.~Eq.~(\ref{xWg-F})) 
\begin{eqnarray*}
\frac{1}{P^+} \int \frac{d\zeta^-}{2\pi}e^{ix P^+\zeta^-} \langle p'|F^{+i}F^{+j}|p\rangle
\approx  \frac{2N_c}{\alpha_s} \int d^2\bm{q}   \left(q_\perp^i -\frac{\Delta_\perp^i}{2}\right)\left(q_\perp^j+\frac{\Delta_\perp^j}{2}\right)  {\widetilde S}_Y(q_\perp,\Delta_\perp) \,,
\end{eqnarray*}
and utilising the gluon GTMD ${\widetilde S}_Y$ decomposition in Eq.~(\ref{GTMD-decomposition}), one arrives at the following important relations connecting the gluon GPDs to the isotropic and elliptic components of the small-$x$ gluon Wigner distribution as follows \cite{Hatta:2017cte}
\begin{eqnarray}
xH_g(x,\Delta_\perp) &=& \frac{2N_c}{\alpha_s}  \int d^2 \bm{q}\, q_\perp^2 {\widetilde S}_{0,Y}(q_\perp,\Delta_\perp) \,, 
\label{xH_g-S0} \\
xE_{Tg}(x,\Delta_\perp) &=&   \frac{4N_c M^2}{\alpha_s \Delta_\perp^2} \int d^2 \bm{q}\, q_\perp^2 {\widetilde S}_{\epsilon,Y}(q_\perp,\Delta_\perp)\,,
\label{xE_{Tg}-S1}    
\end{eqnarray}
such that the elliptic gluon Wigner distribution is responsible for the helicity-flip contribution to the DVCS amplitude. Hence, the azimuthal $\cos 2\phi$-correlation in DVCS induced by the helicity-flip gluon GPDs as initially predicted in Refs.~\cite{Diehl:1997bu,Hoodbhoy:1998vm,Belitsky:2000jk,Diehl:2001pm,Belitsky:2001ns} can be considered as an important probe for the CGC dynamics and the gluon tomography at small $x$.

The full result for the differential DVCS cross section in the dipole formulation reads \cite{Hatta:2017cte}
\begin{eqnarray}
\frac{d\sigma (ep \to e'\gamma p')}{dx_{\rm Bj} dQ^2d^2\Delta_\perp} = 
\frac{\alpha_{\rm em}^3 }
{\pi x_{\rm Bj}Q^2}\Biggl\{ \left(1-y+\frac{y^2}{2}\right)({\cal A}_0^2+{\cal A}_2^2)+
2(1-y){\cal A}_0{\cal A}_2 \cos(2\phi_{\Delta l}) \nonumber \\
+(2-y)\sqrt{1-y} ({\cal A}_0+{\cal A}_2){\cal A}_L \cos \phi_{\Delta l} 
+(1-y){\cal A}_L^2\Biggr\}\,,
\end{eqnarray}
where the correlation in the azimuthal angle $\phi_{\Delta l}$ of the final state photon with respect to the lepton plane is concerned. Here are the UV-finite leading (in the collinear $Q^2\gg q_\perp^2$ limit) contributions to the helicity-conserving
\begin{eqnarray}
{\cal A}_0(\Delta_\perp) \simeq \sum_q e_q^2 N_c \int_0^1 dz \left[z^2+(1-z)^2\right] \int d^2 \bm{q} \ln \left[1+\frac{(q_\perp+\delta_\perp)^2}{z(1-z)Q^2}\right]  {\widetilde S}_{0,Y}(q_\perp,\Delta_\perp)\,,\nonumber\\
\label{helicity-conserv}
\end{eqnarray}
and helicity-flip
\begin{eqnarray}
{\cal A}_2(\Delta_\perp) \simeq -\sum_q \frac{e_q^2 N_c}{Q^2} \int d^2\bm{q} \, q_\perp^2 {\widetilde S}_{\epsilon,Y}(q_\perp, \Delta_\perp)  \,, 
\label{helicity-flip}
\end{eqnarray}
amplitudes for $T$-polarised photons, as well as the finite amplitude for $L$-polarised photons
\begin{eqnarray}
&& {\cal A}_L(\Delta_\perp) \simeq -\sum_q e_q^2 N_c Q\,\Delta_\perp \int_0^1 dz z(1-z)(1-2z)^2 \int d^2\bm{q} \nonumber\\
&& \qquad \quad \times \left[ \frac{{\widetilde S}_{0,Y}(q_\perp,\Delta_\perp)}{\epsilon_q^2 + q_\perp^2} + {\widetilde S}_{\epsilon,Y}(q_\perp,\Delta_\perp) \left(\frac{1}{\epsilon_q^2+q_\perp^2}-\frac{1}{q_\perp^2} \ln \left(1+\frac{q_\perp^2}{\epsilon_q^2}\right) \right) \right] \,, \label{L-polarised} 
\end{eqnarray}
where $\epsilon_q^2=z(1-z)Q^2$ in the massless quark limit. In fact, as was pointed out in Ref.~\cite{Hatta:2017cte} the latter amplitude describing $\gamma^*_L \to \gamma$ transition proportional to $\cos \phi$ is often neglected in the dipole model calculations and vanishes unless one incorporates the correct phase factor $e^{-i\bm{\delta}\cdot \bm{r}}$ \cite{Bartels:2003yj}. The results of the CGC dipole-model analysis have demonstrated the full consistency with those of the collinear factorisation framework at $Q^2\gg q_\perp^2$, in particular, for the $\cos \phi$ and $\cos 2\phi$ correlations \cite{Diehl:2003ny,Belitsky:2005qn}. Due to a sensitivity to the transverse-momentum gluon distribution in the target at small $x$, the dipole approach becomes more appropriate for lower $Q^2$.

Hence, a possible reconstruction of ${\cal A}_2(\Delta_\perp)$ and ${\cal A}_L(\Delta_\perp)$ from the measurement of $\cos \phi_{\Delta l}$ and $\cos 2\phi_{\Delta l}$ angular correlations in the fully-differential DVCS cross section would potentially enable us to extract the elliptic gluon Wigner distribution ${\widetilde S}_{\epsilon,Y}$. Since the latter can also source the $\cos 2\phi$ correlation in inclusive hadron production, hence, giving rise to the elliptic flow in high energy $pp$ 
and $pA$ collisions~\cite{Dusling:2015gta,Hagiwara:2017ofm}, a combination of both measurements would certainly help understand the physical origin of the elliptic flow and the role of hydrodynamic evolution in inclusive hadron production in $pA$ collisions as well as better constrain the nucleon and nucleus structure (gluon tomography) at high energies \cite{Zhou:2016rnt}. The DVCS analysis can  straightforwardly be generalised to the diffractive vector meson (such as $J/\psi$, $\rho$ and $\phi$) production in DIS \cite{Brodsky:1994kf,Accardi:2012qut,Kowalski:2006hc,Kowalski:2003hm,Lappi:2014foa}. For more recent works along these lines, see e.g.~Refs.~\cite{Mantysaari:2016ykx,Mantysaari:2016jaz,Mantysaari:2018zdd,Krelina:2018hmt,Cepila:2019skb,Henkels:2020kju,Kopeliovich:2020has,Henkels:2020qvo,Kopeliovich:2021dgx,Kopeliovich:2022jwe,Wang:2022jwh,Boer:2023mip} and references therein.

As discussed in Ref.~\cite{Newman:2013ada}, HERA data are sensitive to GPDs at low $x$. With respect to the vector meson production, the advantage of DVCS is the absence of the wave function but the disadvantage is a lower production cross section (due to the coupling of the final-state photon). Published results from HERA about DVCS~\cite{H1:2007vrx,H1:2009wnw,ZEUS:2008hcd} show a good precision of the $Q^2, W$ and $t$ measurements. In Ref.~\cite{H1:2009wnw} the azimuthal $\Phi$ angle between the plane of incoming and outgoing lepton and the plane of virtual and real photon was measured (the incoming lepton is the beam electron/positron and the virtual/real photon is the incoming/outgoing photon).

\section{Future measurements}
\label{sec:Future}

In this section, we briefly discuss proposals for future measurements, with a potential to probe the gluon Wigner function.

Very large integrated luminosities assumed to be achievable at future hadron colliders promise to give high-statistics data samples that would allow us to map as many as all five kinematical dependencies in the Wigner distribution with a relatively fine binning. We stress the importance of measuring the $\phi$-angle of the objects in the pair required to be present in the final state, respectively their difference, if we want to access the elliptic part of the Wigner distribution. Thanks to precise trackers present in large experiments, the precision of the $\phi$-angle measurement is usually rather high. The $t$-variable can also be potentially measured using the central tracker if the final-state objects are individual tracks (and not jets) and if the tracker stays more-or-less empty. This is impossible to reach at future hadron colliders that are designed to run at high instantaneous luminosities and hence with large amounts of pile-up events that lead to overwhelming combinatorial backgrounds. 

There are two approaches to cope with the pile-up problem: i) to take data using the so-called special runs, and ii) to utilize forward proton detectors equipped with the so-called time-of-flight (ToF) detectors and to try to impose exclusivity criteria. The advantage of the first approach is low pile-up rates, while the disadvantage is low statistics of the collected data. The asset of the second approach (besides the measurement of the $t$-variable directly from forward protons, see Section~\ref{sec:LHCjj}) is an access to very-high-statistics samples but we pay for that by a limited acceptance of forward detectors and a limited efficiency of the exclusivity criteria including a finite resolution of the ToF detector. The requirement of exclusivity practically means requiring a matching between measurements of the same variable by the forward proton detectors and by the central detectors, within resolutions of the considered detectors. 

At the HL-LHC with an assumed pile-up rate of 200 events per bunch crossing and assumed integrated luminosity of 4~ab$^{-1}$ (or 140 pile-up events and 3~ab$^{-1}$), a 10~ps ToF resolution with a fine granularity (of the order of 1~mm) seems to be imperative. The latest Run~2 results indicate about 25--30~ps resolution achieved with a 3--5~mm granularity of quartz bars, obtained at relatively low pile-up rates, but it is believed that the design characteristics above are conceivable if proper R\&D studies are performed, especially with respect to performance and life-time of photo-multipliers expected to be exposed to extremely high radiation doses during the nominal HL-LHC running. 

In the case of Central Diffraction process of interest (all (semi)exclusive processes considered in this review), the dominant part of the combinatorial background comes from a sum of three independent events: one hard-scale event and two soft SD events, each giving a proton on one side of the forward proton detector. The virtue of the ToF detector lies in the capability to suppress this combinatorial background. If our signal is selected using the double-tag signature (tagging nucleon/nucleus on each side of the IP), combining the time information from both ToF detectors enables us to constrain the longitudinal position of the IP, \zvtx. From a direct comparison of such measurements with a very precise \zvtx\ measurement from the central tracker (the latter of the order of 100~$\mu$m), we can suppress the combinatorial background coming from vertices more distant from the primary vertex than the resolution of the ToF method. It is thus evident that the better resolution of the ToF detector, the better suppression of the combinatorial backgrounds can be expected (for more performance studies, see e.g.~Ref.~\cite{Staszewski:2019yek}. The role of ToF detector granularity and $\mu$-dependence is studied in Ref.~\cite{Cerny:2020rvp}). The impact of the ToF resolution on significance to extract various CD processes from the backgrounds can be found e.g.~in Refs.~\cite{Harland-Lang:2018hmi,Martins:2022dfg}.

At very high levels of pile-up rate, for example at the HL-LHC, more handles will most likely be necessary. One possibility is to make use of timing detectors in the central part of the main detector. Such detectors (HGTD~\cite{ATLASHGTD} in ATLAS, MTD~\cite{Butler:2019rpu} in CMS) have been approved as upgrades for the HL-LHC period and are currently being built. With a design resolution for both of about 30~ps, one can get an additional information about time-of-flight of tracks in the central tracker and hence suppress tracks and vertices belonging to pile-up. Another handle is the so-called track veto (or $z$-vertex veto). By requiring no (or very few) tracks and vertices in a very narrow region around the primary vertex, one can significantly reduce the effect of pile-up. This requirement is an alternative to requiring large rapidity gaps in an environment where the gaps are obscured by tracks from pile-up. One should, however, be aware that the efficiency of all suppressors of the combinatorial background (exclusivity (or matching including ToF of forward protons), ToF of central tracks and track veto) decrease with the increasing level of pile-up. It will therefore be advisable to use some combination of them, with the exact composition depending on the signal and background processes under study. In this context it is useful to note that not necessarily the double-tag signature should be required to get the optimum statistical significance. Requiring both forward detector sides to be tagged leads to a better control over the combinatorial background but unavoidably leads to smaller signal yields and hence sometimes to smaller significances. A single-tag requirement naturally gives more signal but leads to a more dangerous background contamination. This should not discourage us from considering this option, if we are able to model the combinatorial background properly, for example by data-driven methods such as event-mixing or machine learning. The event mixing technique has successfully been used in modeling the combinatorial background in searches for Axion-like particles with AFP based on the single-tag signature~\cite{ATLAS:2023zfc}, or in searches for New Physics with CT-PPS based on the double-tag signature~\cite{CMS:2023roj}.

The experimental environment in detectors to be used at the future lepton colliders, such as FCC-ee \cite{FCC:2018evy} and EIC \cite{Accardi:2012qut}, is believed to be more friendly compared to future hadron colliders due to a much lower pile-up and underlying event contaminations. This promises to provide very ``clean'' jets with access to low $p_T$ and masses which, together with a sensitivity to fragmentation reduced compared to final-state hadrons, makes them an excellent probe of the gluon Wigner function properties~\cite{Abir:2023fpo}. On the other hand, the jets are going to contain significantly fewer particles than at hadron colliders, and also the fragmentation contributions are more visible at low jet masses, which may entail jet-algorithm and fragmentation-model dependencies. The ability to polarize beams at the EIC allows one to access spin-dependent observables in which jets can play a leading role. Diffractive dijets (as elaborated above) and jet-lepton correlations in DIS~\cite{Liu:2018trl,Arratia:2020nxw} probe the 3D unpolarized and polarized TMD, and serve as important roads to the nucleon tomography at the EIC.

\section{Summary and conclusions}
\label{Sect:summary}

In summary, in this review we have collected some of the basic theoretical and phenomenological developments and experimental considerations regarding the fundamental properties and principal possibilities of eventual measuring the 5D gluon Wigner distribution in QCD through scattering processes at high energies. The characteristic $\cos 2\phi$-correlations in the azimuthal angle of the produced particles, in both inclusive and (semi)exclusive reactions in high-energy $pp$, $pA$ and $AA$ collisions provide important means for probing the nucleon and nucleus structure in the transverse plane and at $x\ll 1$, i.e.~the gluon tomography, encoded in the elliptic Wigner distribution. This is especially pronounced in the dipole scattering picture in the target rest frame where multi-particle correlations are influenced directly by the dipole orientation relative to the color field of the target nucleon or nucleus, and can be theoretically estimated e.g. in the Color Glass Condensate framework. 

While the fully differential cross sections for (semi)exclusive processes are most sensitive to the kinematical dependencies of the Wigner distribution, their measurement is very challenging as the statistics are very low, and no reliable measurement directly accessing the elliptic density is yet available. Multi-particle correlations in inclusive processes, that manifest themselves through, for instance, the elliptic flow effect, have been measured in a number of different reactions for both small ($pp$ and $pA$) and large ($AA$) systems. However, information about the initial-state phenomena encoded in the elliptic Wigner distribution is convoluted with hydrodynamics effects impacting the hadronic final states, especially for high-multiplicity events and large collision systems. A comprehensive global analysis combining the yet-to-be-measured azimuthal-angle correlations in (semi)exclusive reactions with the elliptic flow measurements in inclusive processes in several different channels (such as DVCS, diffractive dijet and prompt photon production) would be mandatory for the reliable reconstruction of the elliptic Wigner distribution and for constraining the genuine effect of hydrodynamic evolution in the flow measurements. We believe that our review will be useful for QCD theory and phenomenology research in these interconnected areas.

\section*{Acknowledgments}
This work was supported by ESIF and MEYS (Project FZU researchers, technical and administrative staff mobility - CZ.02.2.69/0.0/0.0/18\_053/0016627).
R.P.~is supported in part by the Swedish Research Council grants, contract numbers 621-2013-4287 and 2016-05996, as well as by the European Research Council (ERC) under the European Union's Horizon 2020 research and innovation programme (grant agreement No 668679). M.T.~is supported by MEYS of the Czech Republic within the project LTT17018.

\bibliography{references}{}
\bibliographystyle{h-physrev}

\end{document}